\documentclass[11pt]{article}

\usepackage{amsfonts} 
\usepackage{amscd}
\usepackage{amsbsy}
\usepackage{url}
\usepackage{graphicx}
\usepackage{color}
\usepackage{graphics}
\usepackage{amsmath}
\usepackage{rotating}
\usepackage{float} 
 \usepackage{rotating}
\usepackage{amsthm}
\usepackage{float}
\usepackage{ulem}
 \usepackage{url}
\usepackage{subfigure}

\usepackage{enumerate}
\usepackage{enumitem}
 
 \usepackage{algorithm}
\usepackage{algorithmic}
\usepackage{array,arydshln}

\definecolor{amethyst}{rgb}{0.6, 0.4, 0.8}

\usepackage[margin=1.5in]{geometry}% http://ctan.org/pkg/margin
\usepackage{booktabs}% http://ctan.org/pkg/booktabs
\usepackage{array}% http://ctan.org/pkg/array

\newcolumntype{M}{>{\centering\arraybackslash}m{\dimexpr.4\linewidth -2\tabcolsep}}
\usepackage{soul}

\newcommand\ba {\mathbf a}
\newcommand\bb {\mathbf b}

\newcommand\be {\mathbf e}

\newcommand\bu {\mathbf u}

\newcommand\bx {\mathbf x}

\newcommand\bA {\mathbf A}
\newcommand\bB {\mathbf B}

\newcommand\bM {\mathbf M}

\newcommand\bV {\mathbf V}
\newcommand\bW {\mathbf W}
\newcommand\bX {\mathbf X}

\newcommand\indica {\mathbb{I}}

\newcommand\wa {\widehat{{a}}}
 
\newcommand\wb {\widehat{{b}}}
\newcommand\wbb {\widehat{\bb}}

\newcommand\wese {\widehat{s}}

\newcommand\wbx {\widehat{\bx}}

\newcommand\wX {\widehat{X}}

\newcommand\wtX {\widetilde{X}}

\newcommand\itA {{\mathcal{A}}}
\newcommand\itB {{\mathcal{B}}}

\newcommand\itE {{\mathcal{E}}}
\newcommand\itF {{\mathcal{F}}}

\newcommand\itH {{\mathcal{H}}}
\newcommand\itI {{\mathcal{I}}}

\newcommand\itK {{\mathcal{K}}}
\newcommand\itL {{\mathcal{L}}}

\newcommand\itN {{\mathcal{N}}}

\newcommand\itS {{\mathcal{S}}}

\newcommand\itU {{\mathcal{U}}}

\newcommand\bphi {\mbox{\boldmath $\phi$}}

\newcommand\bmu {\mbox{\boldmath $\mu$}}

\newcommand\bxi {\mbox{\boldmath $\xi$}}

\newcommand\bSi {\mbox{\boldmath $\Sigma$}}

\newcommand\wbeta {\widehat{\beta}}

\newcommand\wphi {\widehat{\phi}}
\newcommand\wbphi {\widehat{\bphi}}
\newcommand\wgamma {\widehat{\gamma}}

\newcommand\wlam {\widehat{\lambda}}

\newcommand\wmu {\widehat{\mu}}
\newcommand\wbmu {\widehat{\bmu}}

\newcommand\wsigma {\widehat{\sigma}}

\newcommand\wxi {\widehat{\xi}}

\newcommand\wGamma {\widehat{\Gamma}}

\newcommand\wbSi {\widehat{\bSi}}

\newcommand\wtbeta {\widetilde{\beta}}

\newcommand\wtgamma {\widetilde{\gamma}}

\def\real{\mathbb{R}}
\def\natu{\mathbb{N}}

\newcommand{\esp}{\mathbb{E}}
\newcommand{\prob}{\mathbb{P}}

\newcommand{\var}{\mbox{\sc Var}}

\newcommand{\convdist}{ \buildrel{D}\over\longrightarrow}

\newcommand{\trasp}{^{\mbox{\footnotesize \sc t}}}

\newcommand\bcero {{\bf{0}}}

\def\dst{\displaystyle}

\def\median{\mathop{\mbox{median}}}

\def\argmax{\mathop{\mbox{argmax}}}

\def\argmin{\mathop{\mbox{argmin}}}

\newcommand{\identidad}{\mbox{\bf I}}

\def\dst{\displaystyle}

\def\square{\ifmmode\sqr\else{$\sqr$}\fi}
\def\sqr{\vcenter{
         \hrule height.1mm
         \hbox{\vrule width.1mm height2.2mm\kern2.18mm
\vrule width.1mm}
         \hrule height.1mm}}

\newcommand{\rob}{\mbox{\footnotesize \sc rob}}

\newcommand\smooth {\mbox{{\footnotesize\sc s}}}

\newcommand\Signo {\mbox{\footnotesize \sc s}}
\newcommand\gm {\mbox{\footnotesize \sc gm}}

\newtheorem{proposition}{Proposition}[section]

\usepackage{authblk}

\begin{document}
\title{Robust functional principal components for sparse longitudinal data} 

\author[1]{Graciela Boente}
\author[2]{Mat\'{\i}as Salibi\'an-Barrera}
\affil[1]{Universidad de Buenos Aires and CONICET, Argentina \\
gboente@dm.uba.ar }
\affil[2]{University of British Columbia, Canada\\
matias@stat.ubc.ca } 

%
%\author{Graciela Boente \and Mat\'{\i}as Salibi\'an-Barrera}
%\institute{G. Boente \at
%              Universidad de Buenos Aires and CONICET, Argentina \\
%              Tel.: +54-11-5285-8373\\
%              Fax: +54-11-5285-8378\\
%              \email{gboente@dm.uba.ar}          
%           M. Salibi\'an-Barrera \at
%            University of British Columbia, Canada\\
%            Tel.: +1-604-822-3410\\
%              Fax: +1-604-822-6960\\
%              \email{matias@stat.ubc.ca} 
%}

\date{\today}

\maketitle

%%%%%%%%%%%%%%%%%% ABSTRACT%%%%%%%%%%%%%%%%%%%%%%%%%%%%%%%%
\begin{abstract}
In this paper we review existing methods for robust 
functional principal component analysis (FPCA) and 
propose a new method for FPCA that 
can be applied to longitudinal data where only a few observations
per trajectory are available. This method is robust against the presence
of atypical observations, and can also be used to derive 
a new non-robust FPCA approach for 
sparsely observed functional data.  
We use local regression to estimate
the values of the covariance function, taking advantage of the 
fact that 
for elliptically distributed random vectors the conditional location parameter
of some of its components given others is a linear function of the conditioning
set.  This observation allows us to
obtain robust FPCA estimators by using robust local regression methods. 
The finite sample performance of our proposal is 
explored through a simulation study that shows 
that, as expected, the robust method outperforms 
existing alternatives when the data are contaminated. Furthermore, we also see that for samples that do 
 not contain outliers the 
 non-robust variant of our proposal compares favourably to 
 the existing alternative in the literature.  
A real data example is also presented. 

\end{abstract}

\normalsize

\section{Introduction}

Functional Data Analysis refers to a collection of statistical
models that apply to data that can be 
%methods  used to model data that may be 
naturally represented as observations taken from
(mostly unobserved) underlying functions. 
Instead of simply treating the data as vectors, 
models with a functional structure naturally 
incorporate into the analysis important characteristics of the random
process generating the observations, such as smoothness. 
Functional models represent the partially observed
functions as random elements on a functional space, such 
as $L^2({\itI})$, with $\itI\subset \real$  a finite interval. Given the
infinite dimension of this type of sample spaces, some form of dimension
reduction or regularization is typically required. Functional Principal
Components Analysis (FPCA) is a commonly used 
approach to 
obtain optimal lower-dimensional representations of the observations 
(Boente \textsl{et al.}, \cite{BST}). 
Moreover, 
FPCA can also be used to describe the main characteristics of 
the process generating the functions, similarly 
to what is done in the finite-dimensional case
with the interpretation of PCA 
loadings. Specifically, 
the trajectories of each subject can be represented by their coefficients 
(scores) on a few elements of the basis of eigenfunctions (see for instance, 
Ramsay and Silverman \cite{RS} and Hall and Horowitz \cite{HH}.) 
Other bases can also be used to represent the trajectories, e.g.
a sufficiently rich $B$-splines basis (James \textsl{et al.}, \cite{JHS}).

In this paper, we study the problem of  
reliably estimating the principal functions when the data may 
contain a small proportion of atypical observations. 
It is well-known that, similarly to what happens in the finite dimensional 
case, most FPCA methods can be seriously affected in such 
a situation. The outliers need not be ``extreme'' data points, but might
consist of curves that behave differently from the others, or 
that display a persistent behaviour either in shift, amplitude
and/or shape (see, e.g., Hubert \textsl{et al.} \cite{HRS}).
 
Robust FPCA methods in the literature can be 
classified in three groups, depending on the specific property of 
principal components on which they focus. Some of them
rely on performing the eigenanalysis of a
robust estimator of the covariance or scatter 
operator. Others estimate the principal functions by searching for
directions that maximize a robust estimator of the 
spread or scale of the corresponding projections. 
A third group of 
robust methods estimate the principal subspaces by minimizing 
a robust measure of the reconstruction error
(the discrepancy between the observations and their
orthogonal projections on the eigenspace). 
A more detailed
discussion can be found in Section 2 below.

Most functional data analysis methods assume that each trajectory 
is observed on a relatively fine grid of 
points, often equally spaced. The corresponding approaches to
analyze such functional data can be found, for instance, in Ramsay and Silverman \cite{RS}, Ferraty and  
Vieu  \cite{FV} and Ferraty and Romain \cite{FR}. 
Recent reviews of functional data analysis methods include:
Horv\'ath and Kokoszka \cite{HK},  
Hsing and Eubank \cite{HE}, 
Cuevas  \cite{C}, Goia and Vieu \cite{GV},
and Wang \textsl{et al.} \cite{WCM}.  

To the best of our knowledge, most of the 
robust FPCA methods proposed in the literature also require that the curves be 
observed in a relatively dense grid. 
However, there are many applications in which 
trajectories are measured only a few times for each sampling unit. 
Such longitudinal data sets
with only a few available observations per curve (possibly recorded at irregular 
intervals), are relatively 
common in applications, and in many cases it is sensible to consider 
an underlying functional structure 
(of smooth trajectories, for example). 
Only a few FPCA methods exist in the literature
for this type of data (see for
example, James \textsl{et al.} \cite{JHS},   Yao \textsl{et al.} \cite{YMW} and  Li and Hsing \cite{LHS}). 
%, and 
%the review of Wang \textsl{et al.} \cite{WCM}). 

The approach of James \textsl{et al.}, \cite{JHS} consists of 
assuming a finite-dimensional expansion for the underlying random process
(i.e. a Karhunen-Loeve expansion with only finitely many terms), and 
approximating the mean function and eigenfunctions with splines. 
An alternative proposal was given by 
Yao \textsl{et al.} \cite{YMW} 
(see also Staniswalis and Lee \cite{SL}). It consists of 
estimating the covariance function 
using a bivariate smoother over the cross-products of the 
available observations. In this way, information about the
covariance function of the process at different points is ``borrowed''
from different curves. 
Principal directions are constructed from the estimated covariance function, 
and scores are estimated using best linear predictors. 
 Similarly, 
Li and Hsing \cite{LHS} estimate the mean and covariance functions
with local linear smoothers, but use weights
to ensure that the effect that each curve has on the optimizers is not overly affected by the 
number of points at which they were observed.

To  construct  robust FPCA estimators, 
Gervini \cite{G09} modified the 
reduced rank proposal of
James \textsl{et al.}, \cite{JHS} assuming that  
the finite-dimensional standardized scores and the measurement errors have a joint multivariate Student's T 
distribution. The resulting estimators for the centre function and eigenfunctions are Fisher-consistent. 
 A related method using $MM$-estimators for the basis coefficients and scores
has been recently proposed by Maronna \cite{M19}. 
 
An intuitively straightforward approach to obtain a robust version of the method 
proposed by Yao \textsl{et al.} \cite{YMW} would be to 
replace the bivariate smoother of the cross-products with a robust alternative. 
However, the regression approach of  Staniswalis and Lee \cite{SL}, and
Yao \textsl{et al.} \cite{YMW}
works because the smoother estimates the conditional mean of the 
cross-products, which is the covariance function. 
The main challenge with using a robust smoother in this setting is that, 
to the best of our knowledge, there are no robust estimators for the
expected value of asymmetrically distributed random variables  
without imposing distributional assumptions. Hence, 
a robust smoother of the cross-products will typically not be able to estimate 
the covariance function.

In this paper, we provide an alternative approach to 
obtain a robust estimate of the covariance function in the case of 
sparsely observed functional (longitudinal) data. 
Furthermore, this proposal can naturally be adapted to 
offer another way to perform FPCA for longitudinal data 
in the settings considered by Yao \textsl{et al.} \cite{YMW}. 
Our proposal is based on a well-known property of the conditional distribution  of 
elliptically distributed random vectors: the conditional 
location parameter
of one of its components given others is a linear function of the set
of conditioning values. We first show that for an elliptically distributed
random process, the vector of its values at a fixed set of points has
a multivariate elliptical distribution. Combining this property with the previous one, 
we propose to use a robust local regression method to estimate
the values of the covariance function when outliers may be present in the data. 
Our approach can also be used with a non-robust local regression method, 
and our numerical experiments show that this method compares favourably 
to that of Yao \textsl{et al.} \cite{YMW}.

The rest of the paper is organized as follows. 
In Section \ref{sec:review} we review previously
proposed robust FPCA methods which are applicable when either the entire trajectories
were observed, or they were recorded on a dense grid of points. 
Section \ref{sec:proposal} describes our robust estimator of the principal directions for sparsely recorded data. 
We discuss the results of our numerical experiments studying 
the robustness and finite sample performance of the different methods 
in Section \ref{sec:simu}. An illustration is discussed
in Section \ref{sec:realdata}, 
while Section \ref{sec:conclusion} contains a discussion and 
further recommendations.

\section{Robust methods for FPCA}{\label{sec:review} }

As mentioned in the Introduction, atypical observations 
in a functional setting need not be ``extreme'' points and may 
occur in several different ways. Often they 
correspond to individuals that follow a different pattern 
than that of the majority of the data.  
The detection of such outliers is generally a difficult problem, and some
proposals exist in the literature. Among others, we can 
mention the procedures described in Febrero \textsl{et al.} \cite{FGG07,FGG08}, 
Hyndman  and Ullah \cite{HU}, Hyndman  and  Shang \cite{HS}, Sawant \textsl{et al.} \cite{SBS}, 
and Sun and Genton \cite{SG}. 

In this paper we are concerned with methods that do not require the preliminary
detection of potential atypical observations in the data. 
We need to introduce some notation and definitions. 
Although we are motivated by situations where data that can be represented 
as random elements $X$ on a functional space, like $L^2({\itI})$,  
with $\itI\subset \real$ a finite interval,
most of our discussion can be generalized to 
the case where $X$ is a random element on 
an arbitrary separable Hilbert space $\itH$. 
Denote the corresponding inner
product and norm with $\langle \cdot,\cdot\rangle$ and  
$\|u\|=\langle u,u\rangle^{1/2}$, respectively. 
In classical FPCA one assumes that 
$\esp\|X\|^2<\infty$, which ensures the existence of the mean   
and the covariance operator (denoted by 
$\mu_X  =  \esp(X )$ and $\Gamma_X :{\itH}\to{\itH}$, respectively). 
The covariance operator can be written as 
$\Gamma_X = \esp\{ (X -\mu_X ) \otimes (X -\mu_X ) \}$, where
$\otimes$ denotes the tensor product on $\itH$, i.e., 
the operator $u\otimes v:{\itH}\to \itH$ given by 
$(u\otimes v)w= \langle v,w\rangle u$ for $w \in \itH$. 
When $\itH = L^2({\itI})$, the mean function $\mu_X$ satisfies
$\mu_X(t) = E(X(t))$ for $t \in {\itI}$,  
and the covariance operator 
$\Gamma_X$ is associated with a symmetric, 
 non-negative  definite covariance function $\gamma_X : 
\itI \times \itI \to \real$ such that 
$\int_{\itI} \int_{\itI} \gamma_X^2(t,s) \, ds\, dt <\infty$, as follows $(\Gamma_X f)(t)=\int_{\itI}\gamma_X(s,t)\, f(s) \, ds$.

The covariance operator
$\Gamma_X $ is linear, self-adjoint, continuous, and Hilbert-Schmidt.
Hilbert-Schmidt operators have a countable number of
eigenvalues, 
all of them real when the operator is self-adjoint. 
Let $\{\phi_{\ell}:\ell\,\ge 1\}$ be the eigenfunctions of
$\Gamma_X$ associated with its eigenvalues 
$\lambda_1 \ge \lambda_2 \ge \cdots$ in decreasing order. 
The  Karhunen-Lo\`eve expansion of the process $X$ is given by  
the representation
$X \, = \, \mu_X \, + \, \sum_{\ell=1}^\infty  \xi_{ \ell}\,\phi_{\ell}$, 
where the convergence of the series is in mean square norm, 
that is, $\lim_{q\to \infty} \esp \| X- \mu_X\, -\;\sum_{\ell=1}^q \,\xi_{\ell}\,\phi_{\ell} \|^2 = 0$. 
The random variables $\{\xi_{\ell}:\ell\,\ge 1\}$ are the  coordinates of $X -\mu_X$ on the basis
$\{\phi_{\ell}:\ell\,\ge 1\}$, that is, $ \xi_\ell=\langle X -\mu_X ,\phi_\ell \rangle$. Note that
$\esp (\xi_\ell)=0$, 
$\esp (\xi_\ell^2)=\lambda_\ell$, and 
$\esp (\xi_\ell \xi_{s})=0$ for $\ell\ne s$.

In multivariate analysis, elliptical distributions provide an important 
framework for the development of robust
principal component procedures. 
Elliptical processes play a similar role for the development of robust FPCA methods, 
allowing for random elements that may not have finite second moments. 
Given $\mu \in  \itH$, usually known as the location parameter, and $\Gamma$, 
a self-adjoint, positive
semi-definite and  Hilbert-Schmidt  operator on $\itH$ (the scatter  operator), 
we say that the process $X$ has an elliptical distribution $\itE(\mu, \Gamma, \varphi)$, 
if for any linear bounded operator $A : \itH \to \real^d$ 
the random vector $AX$ has a multivariate elliptical 
distribution $\itE_d( A \mu, A \Gamma A^*, \varphi)$, where 
$A^*$ denotes the adjoint operator of $A$ and $\varphi$ stands for the characteristic generator. It is easy to see that Gaussian processes 
satisfy the above definition with $\varphi(a) = \exp( -a/2)$.  
  When second moments exist, the mean of $X$ is $\mu$ and
the covariance operator is proportional to $\Gamma$, i.e., 
$\mu_X=\mu$ and $\Gamma_X = \alpha \, \Gamma$ for some $\alpha>0$
(see Lemma 2.2 in Bali and Boente  \cite{BB}). 
As noted in Boente \textsl{et al.}  \cite{BST},  the scatter operator $\Gamma$ is confounded with the function $\varphi$ meaning that for any $c > 0$,
$\itE(\mu, \Gamma,\varphi) \sim \itE(\mu, c\,\Gamma , \varphi_c)$ where
$\varphi_c(w) = \varphi(w/c)$. For that reason, without loss of generality, when second moments exist, we will assume that $\Gamma_X =  \Gamma$.
We refer the interested reader to 
Bali and Boente  \cite{BB} and Boente \textsl{et al.}  \cite{BST}.

Like their finite-dimensional counterparts, functional
principal components satisfy the
following three equivalent properties: 
\begin{itemize}
\item \textbf{Property 1:}  The first principal directions correspond   to the eigenfunctions of 
the covariance operator $\Gamma_X$ associated with its largest eigenvalues. 
\item \textbf{Property 2:}  The first principal direction maximizes $\var\left(\langle\alpha,X\rangle\right)$ over 
the unit sphere $\itS=\{\alpha: \|\alpha\|=1\}$, and subsequent ones solve the same optimization problem 
subject to the condition that $\alpha$ be orthogonal to all the previous principal directions.
\item \textbf{Property 3:} The first $q$ principal directions provide the best
$q$-dimensional linear approximation to the random element in terms of mean
squared error. Specifically, let $\pi(x, {\itL})$ denote the orthogonal
projection of $x \in {\itH}$ onto an arbitrary $q$-dimensional closed linear space $\itL$, and $\itL_0$
be the linear space spanned by the eigenfunctions of $\Gamma_X$ associated with its 
$q$ largest eigenvalues $\lambda_1 \ge \cdots \ge \lambda_q$. 
If $\lambda_q>\lambda_{q+1}$, then 
we have 
$\esp(\|(X-\mu_X)-\pi((X-\mu_X),{\itL}_0)\|^2) \le \esp(\|(X-\mu_X)-\pi((X-\mu_X),{\itL})\|^2)$. 
\end{itemize}

As mentioned in the Introduction, robust functional principal 
components estimators proposed in the
literature can be classified in three groups, according to the property 
on which they focus. The first approach 
is based on obtaining a robust counterpart 
to the sample covariance operator, 
and use its eigenfunctions to perform robust FPCA. 
The second group consists of methods that 
sequentially search for
directions that maximize a robust estimator of the 
spread or scale of the projections. The last class of 
robust FPCA methods estimate\textcolor{blue}{s} principal subspaces
by searching the closed linear subspace $\itL$ that minimizes a 
robust measure of scale of the reconstruction error 
$\|(X-\mu)-\pi((X-\mu),{\itL})\|$ (the distance between the data and their
orthogonal projections onto the subspace), 
where $\mu$ denotes a location parameter.  

In the rest of this section we review the robust FPCA methods
that are available in each of these three groups. Our proposal 
for the case of sparsely observed trajectories is discussed in
the next Section.

\subsection{Methods based on Property 1}{\label{sec:property1}}

Probably the earliest robust functional principal components 
estimators in the literature are those introduced in 
Locantore \textsl{et al.} \cite{Letal}. They 
%Based on \textbf{Property 1} of the principal components, 
%Locantore \textsl{et al.} \cite{Letal} 
proposed the so-called spherical principal components, which were 
further studied in Gervini \cite{G08} and Boente \textsl{et al.} \cite{BRS}. 
The basic idea is to control potential outliers in the data by 
normalizing the observations (forcing all curves to 
lie on the unit sphere). 
Formally, given a centre $\mu\in \itH$, the spatial or sign covariance
operator of 
$X$ centered at  $\mu$ is defined as
\begin{equation} \label{eq:spherical.cov}
\Gamma^{\Signo}(\mu) \, = \, \esp \left[
\left( \frac{X-\mu}{ \| X - \mu\|}  \right) \otimes 
\left( \frac{X-\mu}{ \| X - \mu\|} \right)
\right] \, ,
\end{equation}
which can be estimated through its sample version:
$$
\wGamma^{\Signo}(\mu) \, = \, \frac{1}{n}
\sum_{i=1}^n \left( \frac{X_i-\mu}{\| X_i-\mu \| } \right) \otimes 
\left( \frac{X_i-\mu}{ \| X_i-\mu \|} \right) \, .
$$
Several location parameters (centres) 
have been considered in the literature. When using the
sign covariance operator \eqref{eq:spherical.cov} 
the standard choice for the centre $\mu$
is the functional spatial 
or geometric median: 
$\mu_{\gm}=\argmin_{u\in \itH}  \esp\left(\|X-u\|-\|X\|\right)$
(see Locantore \textsl{et al.} \cite{Letal} and 
Gervini \cite{G08}).   
%is the usual choice to center the data when using the spatial operator. 
It can be estimated by its sample version 
$\wmu_{\gm}=\argmin_{u\in \itH}  \sum_{i=1}^n \|X-u\|$ 
leading to the usual spatial operator estimator 
$\wGamma^{\Signo}(\wmu_{\gm})$. 
The spherical principal direction estimators 
correspond to its eigenfunctions. % of the spatial sign operator estimator above. 
We refer to Gervini \cite{G08} and Cardot \textsl{et al.} \cite{CCZ} for  
consistency results of the geometric median estimator and to 
Boente \textsl{et al.} \cite{BRS} for results regarding the asymptotic 
distribution of $\wGamma^{\Signo}(\wmu)$
and the spherical principal directions.
Other robust location estimators for $\mu$ above may be considered. 
For example, the $\alpha$-trimmed mean introduced in  
 Fraiman and Mu\~niz \cite{FM},  the deepest point (e.g. Cuevas \cite{C}, 
Cuevas \textsl{et al.} \cite{CFF} and L\'opez-Pintado and  Romo \cite{LPR})
and the $M$-location estimator in Sinova \textsl{et al.} \cite{SGRVA}.   

In addition to being easy to compute, another 
appealing property of the spherical principal directions
is that, under certain conditions they can be shown to be Fisher-consistent.  
Specifically, 
%%%%%%%%%%%%%%%%%
%\st{recall that the 
%the operator
%$\Gamma_X $ is linear, self-adjoint, continuous and Hilbert-Schmidt.
%Hilbert-Schmidt operators have a countable number of
%eigenvalues, all of them being real when the operator is self-adjoint. 
%Let $\lambda_1 \ge \lambda_2 \ge \cdots$ be the eigenvalues of
%$\Gamma_X$ in decreasing order, and $\{\phi_{\ell}:\ell\,\ge 1\}$ the
%associated eigenfunctions. 
%If $\mu = \esp( X)$, the 
%the Karhunen-Lo\`eve expansion of the process $X$ is given by  
%the representation
%$X \, = \, \mu \, + \, \sum_{\ell=1}^\infty  \xi_{ \ell}\,\phi_{\ell}$, 
%where the convergence of the series is in mean square norm, 
%that is, $\lim_{q\to \infty} \esp \| X- \mu\, -\;\sum_{\ell=1}^q \,\xi_{\ell}\,\phi_{\ell} \|^2 = 0$. 
%The random variables $\{\xi_{\ell}:\ell\,\ge 1\}$ are the  coordinates of $X -\mu$ on the basis
%$\{\phi_{\ell}:\ell\,\ge 1\}$, that is, $ \xi_\ell=\langle X -\mu ,\phi_\ell \rangle$. Note that
%$\esp (\xi_\ell)=0$, 
%$\esp (\xi_\ell^2)=\lambda_\ell$, and 
%$\esp (\xi_\ell \xi_{s})=0$ for $\ell\ne s$.}
%%%%%%%%%%%%%%%%%%%%%%%%%%%
assume that either $X$ has an
 elliptical distribution, as defined above, %in Bali and Boente \cite{BB}, 
with location $\mu$ and scatter operator $\Gamma$, or
that $X$ has a finite-rank representation  
$X \, = \, \mu + \sum_{\ell=1}^q \lambda_\ell ^{1/2}\, f_{ \ell}\,\phi_{\ell}$, 
where the random vector $(f_1, \dots,f_q)$ has exchangeable symmetric marginal
distributions  (in this case, we denote 
$\Gamma = \sum_{j=1}^q \lambda_j \,   \phi_j \otimes \phi_j  $). 
Then, the eigenfunctions of $\Gamma^{\Signo}(\mu)$ 
are the same as those of $\Gamma$ and in the same order 
(see Boente \textsl{et al.} \cite{BST} and Gervini \cite{G08}). 

Other robust scatter operator estimators have been proposed in the literature. 
For example, the $\rho$-scatter operator of Kraus and Panaretos \cite{KP},
 and the geometric median covariation studied in 
 Cardot and Godichon-Baggioni \cite{CGB}. 
However, the eigenfunctions of these operators are not guaranteed to 
 remain in the same order as those of the true  covariance operator when the latter exists, or of the true scatter 
  operator for elliptical  processes.

 Sawant \textsl{et al.} \cite{SBS} proposed to estimate the 
 principal directions and their size indirectly
 using the following approach. Ramsay and Silverman \cite{RS} showed that if the 
 observed trajectories $X_i$ and the eigenfunctions $\phi_\ell$ of the
 covariance operator are
 represented on a common known basis (e.g. splines or Fourier), then
 FPCA can be performed using standard finite-dimensional PCA 
 on the matrix of coefficients representing the observations on the chosen basis. 
 The robust version of this procedure in Sawant \textsl{et al.} \cite{SBS}
 consists of replacing the PCA step above with a robust PCA alternative. 
 In particular, they used  ROBPCA (Hubert \textsl{et al.} \cite{HRB})
 and BACONPCA (Billor \textsl{et al.} \cite{BHV}).
 
 \subsection{Methods based on Property 2}{\label{sec:property2}}

Proposals to perform robust FPCA relying on Property 2 above are 
typically referred to as projection-pursuit approaches. 
Hyndman and Ullah \cite{HU} discussed a robust projection-pursuit 
method applied to smoothed observed trajectories. 
Bali \textsl{et al.} \cite{BBTW} generalized this approach 
combining it with penalization and basis reduction. 
In order to impose smoothness on the estimated eigenfunctions, 
regularization is included via a penalization operator. Consider 
the subset ${\itH}_{\smooth}$,  of \textsl{smooth elements} of ${\itH}$ 
and $D:{\itH}_{\smooth} \rightarrow {\itH}$  a linear operator, 
usually called the \lq\lq differentiator\rq\rq, 
since when $\itH=L^2(\itI)$ 
one typically takes $D\alpha= \alpha^{\prime\prime}$. 
Associated with $D$ one can construct the 
following symmetric positive semi-definite bilinear form 
$\lceil \cdot, \cdot \rceil:{\itH}_{\smooth}\times {\itH}_{\smooth} \rightarrow \real$, 
where $\lceil \alpha, \beta \rceil = \langle D\alpha, D\beta \rangle$, and
the functional 
$\Psi: {\itH}_{\smooth} \rightarrow \real$ as $\Psi(\alpha) = \lceil \alpha,\alpha\rceil$.
Finally, a penalized inner product is defined as 
$\langle\alpha,\beta\rangle_{\tau}=\langle\alpha,\beta\rangle+\tau \lceil \alpha,\beta \rceil$, 
and  the associated norm is $\|\alpha\|_{\tau}^2=\|\alpha\|^2+\tau \Psi(\alpha)$.

Let $s_{n}(\alpha)$  be a robust univariate scale estimator $\sigma_n$
%\st{$s_n$} \textcolor{blue}{$\sigma_n$} 
(e.g. an M-scale estimator)
computed on the sample projections 
$\langle\alpha,X_1\rangle$, $\dots$, $\langle\alpha,X_n\rangle$. 
The intuitive idea is to estimate the first principal direction as the element $\alpha$ that 
maximizes $s^2_n(\alpha) - \rho \Psi(\alpha)$, over the unit sphere 
$\left\{ \|\alpha\|_{\tau}=1 \right\} \subset \itH$. 
 In addition, Bali \textsl{et al.} \cite{BBTW}
used a sieves approach to approximate the elements of $\itH$ with an increasing 
sequence of known bases (e.g. splines).  Formally, let 
$\{\delta_i\}_{i\ge 1}$ be a basis of ${\itH}$, and 
${\itH}_{p_n}$   the linear space spanned by $\delta_1,\dots, \delta_{p_n}$, with 
$p_n \to \infty$ as $n \to \infty$. 
The robust projection-pursuit estimator for the first principal direction
in  Bali \textsl{et al.} \cite{BBTW} is given by
$$
\wphi_{1} =
\argmax_{\alpha \in {\itH}_{p_n},\|\alpha\|_{\tau}=1}\left\{s_n^2(\alpha)- \rho \Psi(\alpha)\right\} \, .
$$
The other principal directions are defined sequentially 
adding the condition that they be orthogonal to the previous ones
with respect to
the inner product $\langle\cdot,\cdot\rangle_{\tau}$.
Bali \textsl{et al.} \cite{BBTW} showed that the estimators are qualitative robust. 
Furthermore, the procedure is Fisher-consistent 
when the robust univariate scale estimator $\sigma_n$ 
is the empirical version of a functional $\sigma_{\rob}$
such that there exist a constant $c > 0$ and a 
self-adjoint, positive semidefinite and 
Hilbert-Schmidt operator $\Gamma_0$ satisfying
%%%%%%%%%%%%%%%%%%%%%%%%%%%%%%%%%%%%%%%%%%%%%%%%%%%%%%%%%%%%%%%%%
%\st{$\sigma_{\rob}^2(\alpha)={c}\langle\alpha, \Gamma_0\alpha\rangle$ for
%all $\alpha \in {\itH}$, 
%where $\sigma_{\rob}^2(\alpha)$ denotes the functional 
%$\sigma_{\rob}$ computed on the distribution of $\langle \alpha, X \rangle$.}
%%%%%%%%%%%%%%%%%%%%%%%%%%%%%%%%%%%%%%%%%%%%%%%%%%%%%%%%%%%%%%%
$\sigma_{\rob}^2(P[\alpha])={c}\,\langle\alpha, \Gamma_0\alpha\rangle$ for
all $\alpha \in {\itH}$,  where $P[\alpha]$ stands for the distribution of 
$\langle \alpha, X \rangle$. 
This condition is satisfied when $X$ has an 
elliptical distribution. 

Note that the approaches described here 
and in Section \ref{sec:property1} require that one 
is able to compute or approximate well the 
norms $\|X_i-\mu\|$ and inner products $\langle \alpha, X_i \rangle$. 
For example, when $\itH=L^2(\itI)$ this typically means that 
the trajectories need to have been observed
over a relatively dense grid of points. When the data
are measured sparsely (and only a few points per trajectory are available) these
methods become infeasible.

 \subsection{Methods based on Property 3}

The last group of robust FPCA methods estimate the eigenspaces 
directly, for example by minimizing a robust measure of the 
distance between the observations and their orthogonal projections
on finite-dimensional subspaces. 
 Lee \textsl{et al.} \cite{LSB} proposed a sequential algorithm fitting linear spaces of dimension 1
to the observed data.
Their proposal assumes that all trajectories 
$X_i$, $1 \le i \le n$, 
were observed on a  
common grid $t_1, \ldots, t_m$. 
The goal is to find a smooth function $v$ and a vector 
$\ba=(a_1,\dots, a_n)^\top$ that minimize the size of the residuals $X_i(t_j) - a_i v(t_j)$. 
Because of the potential presence of outliers, the method uses a bounded loss function 
$\rho$. Given an initial estimator for $v$, 
let $a_i= \sum_{j=1}^m X_i(t_j) v(t_j)$, $1 \le i \le n$, and 
for each $j = 1, \ldots, m$, 
let $\wsigma_j$ be a robust scale
estimator of the residuals $X_i(t_j) - a_i v(t_j)$, $1 \le i \le n$. 
The estimators for $v$ and $\ba$ are defined as the minimizers of 
 $$
 \sum_{i=1}^n \sum_{j=1}^m \wsigma_j^2 \rho\left(\frac{X_i(t_j)-\,a_i\, v(t_j)}{\wsigma_j}\right) +\frac{\lambda}{2}\int \left[ v^{\prime\prime}(t)\right]^2 \! dt \, , 
 $$
 subject to the constraint $\int  v^{2}(t) \, dt = 1$, 
 where $\lambda > 0$ is a user-chosen regularization parameter. 
This problem can be solved iteratively as follows. 
Given $\ba=(a_1,\dots, a_n)$
let $\phi_\ba$ be the minimizer of the objective function above, which can be
obtained using iterative reweighted penalized least-squares, and 
given $\phi_\ba$, the entries of $\ba$ are 
the projections of each trajectory along the direction $\phi_\ba$:  
 $a_i=\sum_{j=1}^m X_i(t_j) \phi_\ba(t_j)$, $1 \le i \le n$. These steps are then iterated until convergence. 
 Once the first principal direction $\wphi_1$ is obtained, we can compute the corresponding 
  estimated scores $\wxi_{i,1}=\sum_{j=1}^m X_i(t_j) \wphi_1(t_j)$.  
  The other estimated principal directions are computed sequentially as follows: assume 
  that $\wphi_s$,  $s=1,\dots, \ell-1$,  are estimators of the first $\ell-1$ principal directions, 
  then, the $\ell-$th principal direction  and the related scores  are computed applying the 
  previous procedure to   $X(t_j)-\,\sum_{s=1}^{\ell-1} \wxi_{i,s} \wphi_s(t_j)$. 
Lee \textsl{et al.} \cite{LSB} also discussed a data-dependent and  
resistant procedure to select $\lambda$.

 In some applications each curve $X_i$ may be observed on a different 
 (but dense) grid, and in that case we can, in principle, proceed as recommended in
 Ramsay and Silverman \cite{RS}, namely: 
 smooth each curve $X_i$ and apply the method above to the 
 values of the smoothed trajectories on a fixed grid. 
A drawback of this strategy is that the errors introduced by the data smoothing 
step cannot be included easily in the analysis.  

A different approach to estimate functional principal subspaces is as follows. 
Assume that each trajectory is observed on a dense grid of points (that may
be specific to each curve), and let 
$\{\delta_i\}_{i\ge 1}$ be an orthonormal basis of ${\itH}$. 
The basic idea is to identify each curve $X_i$ with the vector 
of its coefficients on a finite-dimensional basis, apply a
robust multivariate method to estimate principal subspaces
in this finite-dimensional space, and then map the results 
back to $\itH$. 
Specifically, fix $p_n$, the basis size, and let 
$x_{ij}=\langle X_i,\delta_j\rangle$ be the coefficient of the $i$-th 
trajectory on the $j$-th element of the basis, $1 \le j \le p_n$.
The grids where each $X_i$ are observed need to be sufficiently dense
so that these inner products can be approximated well using 
finite Riemman  sums. For each $1 \le i \le n$, let $\bx_i$ be 
the vector
of coefficients of $X_i$ on the basis $\{ \delta_j \}_{1 \le j \le p_n}$:
$\bx_i=(x_{i 1},\dots,x_{i p_n})\trasp$.
%Both Boente and Salibian-Barrera \cite{BS} and Cevallos-Valdiviezo \cite{CV}, allow the observations to be recorded in different dense grids, situation that appears in some applications. To be more precise,  assume that, for each $1\le i\le n$, the trajectory $X_i$ is observed at  design points $t_{ij}$, $1 \le j \le m_i$. These authors assume  that  as the sample size $n$ increases, so does the number of points $m_i$  and that, in the limit, these points cover the interval $[0,1]$.  The basic idea in these papers is to identify each observed point in $\itH$ with the vector
%formed by its coordinates on a finite-dimensional basis applying to them a robust multivariate procedure. Let $\{\delta_i\}_{i\ge 1}$ be an orthonormal basis of ${\itH}$ and fix $p=p_n$ the dimension of the linear space where to project the data. Let $x_{ij}=\langle X_i,\delta_j\rangle$ be the coefficient of the $i-$th trajectory on the $j-$th element of the basis, $1 \le j \le p$  and
%form the $p-$dimensional vector $\bx_i=(x_{i 1},\dots,x_{i p})\trasp$. The
%inner products $\langle X_i, \delta_j \rangle $ can be numerically computed using a Riemann
%sum over the design points for the $i$th trajectory $\{t_{ij}\}_{1\le j\le m_i}$. 
We can now apply robust 
multivariate methods on the sample 
$\bx_1, \ldots, \bx_n$ to estimate a $q$ dimensional 
principal subspace 
$\widehat{\itL} \subset \real^{p_n}$. 
Once the principal subspace 
$\widehat{\itL}$ is found, 
the functional principal direction estimators 
can be reconstructed by ``mapping back'' the results in
$\real^{p_n}$ onto $\itH$. 
To fix ideas, let $\wbb^{(1)}, \dots, \wbb^{(q)}$ be an orthonormal basis for 
$\widehat{\itL}$ 
and let $\wbx_i = \wbmu + \sum_{\ell=1}^q \wa_{i\ell} \wbb^{(\ell)}$ 
be the $q$-dimensional approximation to $\bx_i$, 
where $\wbmu = (\wmu_1, \ldots, \wmu_p )\trasp$.    
The functional location parameter can be reconstructed as 
$\widehat{\mu}=\sum_{j=1}^{p_n} \wmu_j \delta_j$, while the
 $q$-dimensional principal direction basis in $\itH$ is 
$\wphi_\ell=\sum_{j=1}^{p_n}\wb_{\ell\,j}\delta_j/\|\sum_{j=1}^{p_n}\wb_{\ell\,j}\delta_j\|$, for $1\le \ell\le q$.
Furthermore, the ``fitted values'' in $\itH$ are 
$\wX_i=\wmu + \sum_{\ell=1}^q \wa_{i\ell} \wphi_{\ell}$.  

The proposals of 
Boente and Salibian-Barrera \cite{BS} and Cevallos-Valdiviezo \cite{CV}
are variants of this approach that differ on how the principal subspace $\widehat{\itL} 
\subset \real^{p_n}$ and its orthonormal basis are 
estimated. 
Let $\bB=\left(\bb^{(1)}, \dots, \bb^{(q)}\right) \in \real^{p_n \times q}$ 
and $\bb_j\trasp$  the $j$-th row of $\bB$. 
For a given $\bmu \in \real^{p_n}$ and 
$\ba_i\in \real^q$, the corresponding ``fitted values'' are
$\wbx_i(\bmu, \bB, \bA) 
= \bmu + \bB \ba_i$, % =(\wx_{i1}, \dots, \wx_{ip})\trasp$, $1 \le i \le n$, 
%where $\wx_{ij} = \mu_j + \ba_i\trasp \bb_j$, and
where $\ba_i$ are 
the columns of the matrix $\bA \in \real^{q \times p_n}$.

Boente and Salibian-Barrera \cite{BS} estimated the 
principal subspace minimizing 
\linebreak $\sum_{j=1}^p\wsigma_j^2 (\bmu, \bB, \bA)$, 
where 
$\wsigma_j(\bmu, \bB, \bA)$
is a robust scale of the $j$-th coordinate of the 
residual vectors ${\mathbf r}_i = \bx_i - \wbx_i(\bmu, \bB, \bA)$, 
$1 \le i \le n$. 
%$r_{ij}(\bmu, \bB, \bA)=x_{ij}-\wx_{ij}$, 
 Although in principle one can use any robust scale estimator, 
 Boente and Salibian-Barrera \cite{BS} discussed in detail the 
 case of $M$-scale estimators  (see  Maronna \textsl{et al.} \cite{MMYS}), 
 for which an iterative algorithm can be derived. 
Cevallos-Valdiviezo \cite{CV} proposed using a multivariate $S$-estimator for 
PCA, minimizing 
$\wsigma (\bmu, \bB, \bA)$, where $\wsigma (\bmu, \bB, \bA)$ is a
robust scale of the distances
$d_{i}(\bA,\bB,\bmu) = \| \bx_i -\wbx_i(\bmu, \bB, \bA)\|$, 
using $M$-scale and least-trimmed scale estimators. 
All these proposals are Fisher-consistent for elliptically distributed random processes.

\section{FPCA for longitudinal (sparse) data}{\label{sec:proposal} }

In this section we describe a new robust FPCA method that is applicable 
to the case where few observations per trajectory may be available. Moreover,
when robustness is not a concern, our proposal results in a novel method for
non-robust FPCA that compares favourably with existing methods in the literature. 

\subsection{Model}

We will assume the following framework. 
Let $\{X(t)\, : \, t\in  \itI \}$ be a stochastic process defined on a probability
space $(\Omega,{\itA},\prob)$ with continuous trajectories,
where $\itI \subset \real$ is a finite interval, 
which can be assumed to be $\itI =[0,1]$
without loss of generality. 
To allow for atypical observations, we will include processes 
$X$ for which finite second moments may not exist. 
Specifically, we will only assume that the process $X$ has an 
elliptical distribution $\itE(\mu, \Gamma, \varphi)$, 
where $\mu \in  L^2({\itI})$ and  $\Gamma$  is a self-adjoint, positive
semi-definite and  Hilbert-Schmidt  operator on $L^2({\itI})$. 
We will denote as $\gamma:L^2({\itI})\times L^2({\itI})$ the 
kernel defining $\Gamma$, that is, $(\Gamma f)(s)= \int_{\itI} \gamma(t,s) f(t) dt$. 
Recall that $\gamma$ is symmetric and 
$\int_{\itI} \gamma^2(t,s)\,ds\,dt <\infty$, 
since $\Gamma$ is a Hilbert-Schmidt  operator. 
Finally, $\left\{ \phi_k \right\}_{k \ge 1}$ will denote an orthonormal basis of 
$L^2({\itI})$ consisting of eigenfunctions of the 
  scatter  operator $\Gamma$ ordered
according to the associated non-zero eigenvalues 
$\lambda_1 \ge \lambda_2 \ge \ldots$.
 
%\st{in the following sense. Consider the separable Hilbert space $L^2({\itI})$ endowed with 
%the usual inner product $\langle f, g \rangle = \int f(t) g(t) \, dt$. Given a function 
%$\mu \in  L^2({\itI})$ and  a self-adjoint, positive
%semi-definite and  Hilbert-Schmidt  operator $\Gamma$ on $L^2({\itI})$ dispersion (the dispersion operator), 
%we say that the process $X$ has an elliptical distribution $\itE(\mu, \Gamma, \varphi)$, 
%if for any linear bounded operator $A : L^2({\itI}) \to \real^d$ 
%the random vector $AX$ has a multivariate elliptical 
%distribution $\itE_d( A \mu, A \Gamma A^*, \varphi)$, where 
%$A^*$ denotes the adjoint operator of $A$ (see Bali and Boente,} \cite{BB}).
% \st{It is easy to see that Gaussian processes 
%satisfy the above definition with $\varphi(a) = \exp( -a/2)$ (see 
%Boente \textsl{et al.},} \cite{BST}).  
% 
%\st{When second moments exist, $\gamma_X$ will 
%denote the covariance function of $X$ (i.e. 
%$\gamma_X(t,s) = \cov\left(X(t),X(s)\right)$), and 
%we will use $\Gamma_X$ for the corresponding 
%covariance operator that satisfies
%$\Gamma_X (f) = \int \gamma_X(t, \cdot ) f(t) \, dt$. 
%In this case we also have $\Gamma_X = \alpha \, \Gamma$ for some $\alpha>0$. }

Note that elliptically distributed random processes as defined in 
Section \ref{sec:review}
accept a Karhunen-Lo\`eve representation 
when $\Gamma$ has finite rank  or when the kernel $\gamma$ is continuous. 
Consider first the case where
the scatter operator $\Gamma$ has finite rank (i.e. a finite number of
non-zero eigenvalues). It follows that 
$X \, \sim  \, \mu  + \sum_{k=1}^q  \xi_k \, \phi_k$, 
where $q=\mbox{rank}(\Gamma)$ and  
$\bxi=\left(\xi_1, \dots, \xi_q\right)\trasp$ has a 
multivariate elliptical distribution  $\itE_q(\bcero_q, \mbox{diag}(\lambda_1, \dots, \lambda_q), \varphi)$ 
(see the proof of Proposition \ref{lemma:proposal}).
When there are infinitely many positive eigenvalues, 
Proposition 2.1 in Boente \textsl{et al.} \cite{BST} shows that 
$X \, \sim \,  \mu + S \,  V $, where $S \ge 0$ is a random variable independent from the 
zero-mean Gaussian random element $V$. The standard Karhunen-Lo\`eve expansion
for Gaussian processes implies that $V =  \sum_{k=1}^\infty  \eta_k \, \phi_k(t)$
with  $\eta_k \sim {\cal N} (0, \lambda_k)$, $k \ge 1$. The continuity of the covariance
function $\gamma$ and the process $V$ implies that 
the convergence of the series is uniform over $\itI$ with probability 1. 
Thus, defining $\xi_k=S\; \eta_k$, we obtain that $\bxi=\left(\xi_1, \dots, \xi_q\right)\trasp$ 
has a multivariate elliptical distribution  
$\itE_q(\bcero_q, \mbox{diag}(\lambda_1, \dots, \lambda_q), \varphi)$ and 
$ X \, \sim \,  \mu + \sum_{k=1}^\infty  \xi_k \, \phi_k$
with covariance operator $\Gamma$. 
Hence, we can write 
\begin{equation} \label{eq:repr}
X(t) = \mu(t) + \sum_{k=1}^\infty  \xi_k \, \phi_k(t) \, , \quad t \in \itI \, ,  
\end{equation}
where the scores $\xi_{k} \, = \, \langle X - \mu \, , \phi_k \, \rangle $ 
are such that, for any $q\ge 1$, the random vector $\bxi=\left(\xi_1, \dots, \xi_q\right)\trasp$ 
has a multivariate elliptical distribution  $\itE_q(\bcero_q, \mbox{diag}(\lambda_1, \dots, \lambda_q), \varphi)$.
 When second moments exist, the scores are uncorrelated random variables.

The following Proposition provides the main 
motivation for our approach. It shows that, similarly to what holds
for Gaussian processes, the vector obtained by evaluating an elliptically 
distributed random process on a fixed finite set of $k$ points is an elliptical random 
vector on $\real^k$. 
Furthermore, it also shows that the conditional distribution of the scores $\xi_j$
in \eqref{eq:repr}
given a set of $k$ observations of the process is elliptical. This 
result suggests a natural estimator for the $\xi_j$'s which 
can be used to reconstruct the full trajectories in the sample. 
The proof can be found in Section \ref{sec:appendix}. 

\begin{proposition} \label{lemma:proposal}
Let $X \, \sim \, \itE\left( \mu, \Gamma, \varphi \right)$
be a random element on $L^2({\itI})$, with $\itI\subset \real$,
and assume that
the kernel $\gamma$  associated with $\Gamma$  is continuous. 
Let $\lambda_1\ge \lambda_2\ge \dots$ be the non-null eigenvalues of $\Gamma$ and as $\phi_k$ the eigenfunction of $\Gamma$ associated with $\lambda_{k}$ chosen so that the set
$\{\phi_{k}, k \in \natu \}$ is an orthonormal set in $L^2({\itI})$. Then, 
\begin{enumerate}[label=\alph*)]
\item For any fixed $m$ and 
$t_1$, $t_2$, \ldots, $t_m$ in $\itI$,
the random vector $( X(t_1),  \ldots, X(t_m) )\trasp$
has an elliptical distribution in $\real^m$ with location
$\bmu_m = (\mu(t_1),   \ldots, \mu(t_m))\trasp$ and 
scatter matrix $\bSi$ with elements $\Sigma_{(\ell, j)} = \gamma(t_\ell, t_j)$, $1 \le \ell, j  \le m$. 
\item For any $s_0\ne t_0\in \itI$ we have that
 $\bigl. X(t_0) \, \bigr| \, X(s_0)  \ \sim \ 
\itE_{1} \left( \mu_{t_0|s_0}, \, \sigma_{t_0|s_0} ,\varphi^{\star}_{s_0}\right)\,,
 $
 where the conditional location is given by
\begin{equation}
 \mu_{t_0|s_0} \, = \,  \mu(t_0) +  \frac{\gamma(t_0, s_0) }{\gamma(s_0,s_0)} 
 \, \left( X(s_0) - \mu(s_0) \right) \, .
 \label{eq:mut0dados0}
\end{equation}
\item For any fixed $m$ and $t_1$, $t_2$, \ldots, $t_m$ in $\itI$, 
%Let $\phi_k$ stand for the orthonormal eigenfunction of $\Gamma$ related to the 
%eigenvalue $\lambda_k$ and denote as $\xi_k=\langle X-\mu, \phi_k\rangle$. 
%Given  $m$ and any  $t_1$, $t_2$, \ldots, $t_m$ in $\itI$,   
let $\bX_m=(X(t_1),\dots, X(t_m))\trasp$.
We have that  $\xi_k \, \bigr| \, \bX_m \sim \itE_{1} \left( \mu_{k}, \, \sigma_{k}, \varphi^{\star}_{\bX_m} \right)$ 
where 
 \begin{equation}
  \mu_{k} = \lambda_k\; \bphi_k\trasp \bSi_{\bX_m}^{-1} \left(\bX_m-\bmu_m \right) \, ,
 \label{eq:mediaescores}
\end{equation}
 with  $\bphi_k=(\phi_k(t_1), \dots, \phi_k(t_m))\trasp$, $\bmu_m =(\mu(t_1), \dots, \mu(t_m))\trasp$ 
 and the $(\ell, j)$-th element of $ \bSi_{\bX_m}$ equals $\gamma(t_\ell\,,\, t_j)$. 
\end{enumerate}
\end{proposition}
 
 \subsection{Method} \label{sec:ourmethod}
In what follows we will use $\{X_i: 1 \leq i \leq N\}$ to denote independent realizations of the stochastic process $X$.
We assume that each trajectory $X_i$ 
is observed at independent random ``times'' $t_{ij}$, $1 \le j \le n_i$, where
the $n_i$'s, $1 \le i \le N$, are usually assumed to be
random variables, independent from all others.  
Then, our model is:
\begin{equation}
X_{ij} \, = \, X_i(t_{ij}) \, = \, \mu(t_{ij}) + \sum_{k=1}^\infty \xi_{ik} \phi_k(t_{ij})
\, , \quad j = 1, \ldots, n_i, \quad i = 1, \ldots, N \, .
\label{model}
\end{equation}
 The procedure has three steps:  
 (1) estimating the center function $\mu$; (2) estimating 
 the diagonal elements of the scatter function $\Gamma$, and (3) estimating 
 the off-diagonal entries of $\Gamma$. 

\subsubsection{Step 1}
We start by estimating the center function $\mu$ using a robust local $M$-estimator. For
example, one can consider the  local $M-$smoothers proposed in Boente and Fraiman \cite{BF}, 
H\"ardle and Tsybakov \cite{HT}, H\"ardle \cite{H}  and
Oh \textsl{et al.} \cite{ONL} or   robust local linear smoothers as defined in Welsh \cite{W}. 
Here we use the latter option.

Consider a $\rho$-function $\rho_1$ as defined in  
Maronna \textsl{et al.}  \cite{MMYS}. That is, $\rho_1$ is even, 
nondecreasing as a function of  $|x|$,  $\rho_1(0)=0$, 
and increasing for $x>0$ when $\rho_1(x)< \sup_t \rho_1(t)$.  
For each $t_0 \in \itI$, let
\begin{equation}
\left( \wbeta_0(t_0),\wbeta_1(t_0) \right)\trasp \, = \, \argmin_{\beta_0, \beta_1} \,
\sum_{i=1}^N \sum_{j=1}^{n_i} w_{ij}(t_0) \, \rho_1\left(\frac{X_{ij}-\beta_0-\beta_1(t_0-t_{ij})}{\wsigma(t_0)}\right) \,,
\label{eq:Mpolylocalmuest}
\end{equation}
where 
$\wsigma(t)$ is a preliminary robust consistent   estimator of scale and the 
weights $w_{ij}(t_0)$ are given by 
\begin{equation} \label{eq:weights}
w_{ij}( t_0 )  \, = \,  \itK \left( \frac{ t_{ij} -  t_0 }{h} \right) \left\{
  \sum_{k=1}^N \, \sum_{\ell=1}^{n_k} \itK \left( \frac{ t_{k \ell} -  t_0 }{h} \right) \right\}^{-1} \, , 
\end{equation}
where $h=h_n$ is a bandwidth parameter and 
$\itK:\real \to \real$ is a kernel function, i.e. continuous, non-negative, and integrable. 
Then, the estimate for $\mu(t_0)$ is
$$
\wmu(t_0) \, = \, \wbeta_0(t_0) \, .
$$
The preliminary scale estimator
$\wsigma(t_0)$ in \eqref{eq:Mpolylocalmuest} 
%or \eqref{eq:Mlocalmuest} 
can, for example, be a ``local \textsc{MAD}'', that is: the MAD of the 
observations in a neighborhood of $t_0$. Formally:
$$
\wsigma(t_0) \  = \ \kappa^{-1} \, \median_{|t_{ij}-t_0|\le h} \left| X_{ij} - 
\median_{|t_{ij}-t_0|\le h}(X_{ij}) \right| \, ,
$$
where $\kappa \in \real$ is a constant ensuring Fisher-consistency at 
a given distribution. 
If we consider a gaussian distribution as the central model, then 
$\kappa=\Phi^{-1}(3/4)$, where $\Phi$ is the cumulative
distribution function of the standard normal.

\subsubsection{Step 2}

To estimate the diagonal elements $\gamma(t_0, t_0)$ of the 
scatter function we use
a robust $M-$scale estimator of the residuals. 
Although in principle one can use any robust scale estimator, we expect 
smooth functionals (like M-scales) to provide more stable estimators.  
More precisely, 
let $\rho_2 : \real \to \real$ be 
a bounded 
$\rho$-function, such that $\sup_t \rho_2(t) =1 $, 
and let $b \in (0, 1)$ be a fixed constant
which defines the robustness of the scale estimator
(Maronna \textsl{et al.} \cite{MMYS}). Then 
$\wgamma( t_0, t_0) $ satisfies
\begin{equation} \label{eq:diagonal}
\sum_{i=1}^N \sum_{j=1}^{n_i} \, w_{ij}( t_0 ) \, 
\rho_2 \left( \frac{ X_{ij} - \wmu(t_{ij})  }{ \wgamma ( t_0, t_0) } \right)
 \, = \, b \, ,
\end{equation}
where the weights $w_{ij}(t_0)$ are as in \eqref{eq:weights}. 
We choose $\rho_2$ so that 
$E( \rho_2(Z) ) = b = 1/2$, where $Z \sim {\cal N}(0, 1)$ 
to ensure that in 
the case of i.i.d. observations with Gaussian errors, the resulting scale estimator 
is consistent and has maximum breakdown point. 
 
\subsubsection{Step 3}

To estimate the off-diagonal elements   
$\gamma(t_0, s_0)$, $s_0\ne t_0$, we 
take advantage of  Proposition \ref{lemma:proposal}, which shows that
the conditional mean of $X(t_0)$ given $X(s_0)$ is a
linear function of the conditioning value, and furthermore, that
the slope is proportional to $\gamma(t_0, s_0)$. 
This suggests that we can use local regression methods to 
estimate $\gamma(t_0, s_0)$. 
 Specifically, we  
first estimate the slope $\beta(  t_0, s_0)$ in \eqref{eq:mut0dados0} with
a local regression estimator
and then use that $\gamma(t_0, s_0) = \beta(  t_0, s_0) \, \gamma(s_0, s_0)$, 
where $\gamma(s_0, s_0)$ can be estimated as in (\ref{eq:diagonal}).  
More precisely, let $\wtX_{ij} = X_{ij} - \wmu(t_{ij})$ be the 
centred observations, and let
$\wbeta(t_0, s_0)$ be the local $M$-regression estimator satisfying
\begin{equation}
 \wbeta(  t_0, s_0)=\argmin_{ \beta \in \real}\sum_{i=1}^N\sum_{j\ne  \ell}   
 \rho\left(\frac{\wtX_{ij}-\beta \wtX_{i\ell}}{\wese(  t_0, s_0)} \right) 
 \; \itK \left( \frac{ t_{ij} -  t_0  }{h} \right)  \itK \left( \frac{ t_{i \ell} -   s_0  }{h} \right) \, ,
\label{eq:wbeta}
\end{equation}
where $\wese( t_0, s_0)$ is a preliminary robust scale estimator 
and $\rho$ is a bounded $\rho$-function. 
To compute $\wese( t_0, s_0)$ we use a 
local MAD of the residuals from an initial robust estimator. 
Specifically, 
let  $Z_{ij\ell}=\wtX_{ij}/\wtX_{i\ell}$,  and
noting that the model does not include an intercept, let 
%the preliminary estimate of $\beta(t_0,s_0)$ as 
$\wtbeta(t_0,s_0)$ be the local median of the slopes: 
$\wtbeta(t_0,s_0) =\median_{|t_{ij}-t_0|\le h\;, |t_{i\ell}-s_0|\le h} Z_{ij\ell}$. 
Construct residuals 
$r_i(t_0,s_0)=\wtX_{ij}-\wtbeta(t_0,s_0)\,\wtX_{i\ell}$ and let 
$\wese (  t_0, s_0)$  be the corresponding local MAD:
% of $r_i(t_0,s_0)$, 
%where $r_i(t_0,s_0)=\wtX_{ij}-\wtbeta(t_0,s_0)\,\wtX_{i\ell}$, that is, 
$$
\wese (  t_0, s_0)=\median_{|t_{ij}-t_0|\le h\;, |t_{i\ell}-s_0|\le h}\, |r_i(t_0,s_0)- m(t_0,s_0) | \, ,
$$ 
with $m(t_0,s_0)=\median_{|t_{ij}-t_0|\le h\;, |t_{i\ell}-s_0|\le h} r_i(t_0,s_0)$.

With $\wbeta_0(  t_0, s_0)$ computed in \eqref{eq:wbeta} and 
$\wgamma( s_0,  s_0)$ from Step 2, we define 
$$
\wtgamma( {t_0}, {s_0}) = \,  \wbeta_0(  t_0, s_0) \; \wgamma( s_0,  s_0) \, .
$$
To ensure that the estimated function $\wgamma$ 
is symmetric and smooth, we also 
compute $\wbeta_0( s_0, t_0)$ and use a two-dimensional smoother 
(e.g. a bivariate $B$-spline  as described 
in Section \ref{sec:simu}) on each of them, 
resulting in 
$\widetilde{\wtgamma}(s_0,t_0)$ and $\widetilde{\wtgamma}(   t_0, s_0)$. 
Finally, the estimated scatter function is:
$$
\wgamma (  t_0, s_0) \, = \, \wgamma (s_0, t_0) \, = \, \left. \left( 
\widetilde{\wtgamma}(   t_0, s_0) + 
\widetilde{\wtgamma}(  s_0, t_0)  \right) \right/ 2 \, .
$$
Note that even though $\wgamma (t, s)$ defines a 
symmetric kernel, it may not be semi-positive definite. 
If necessary, 
after computing its eigenvalues 
$\wlam_1\ge \wlam_2\ge \ldots$,  and corresponding eigenfunctions $\wphi_1,\wphi_2,\ldots$, 
we set $\wgamma$ as 
$\wgamma(t,s)=\sum_{j: \wlam_j\ge 0} \wlam_j \wphi_j(t)\; \wphi_j(s)$.

\subsubsection{Reconstructing trajectories}

Once we have estimated the eigenfunctions and eigenvalues 
of the scatter function, 
Proposition \ref{lemma:proposal}(c) suggests a natural 
way to reconstruct the trajectories. 
Let $\xi_{i,k} = \langle X_i - \mu, \phi_k \rangle$ the 
$k$-th score of the $i$-th observation. 
Recall that 
 the conditional distribution of $\xi_{i,k}$ given 
 $\bX_i=\left(X_i(t_{i1}), \dots X_i(t_{i n_i})\right)\trasp$ is 
 elliptical with location parameter 
%has location 
%
%For a given number of principal directions $q$, to reconstruct the trajectories as
%$\wX_i(t)=\wmu(t) + \sum_{k=1}^q \wxi_{i,k} \wphi_{k}(t)$, we need 
%estimates of the scores $\xi_{i,k}$. Denote as  $\bX_i=\left(X_i(t_{i1}), \dots X_i(t_{i n_i})\right)\trasp$. 
%According to Proposition \ref{lemma:proposal}c), the conditional distribution of $\xi_{i,k}$ given $\bX_i$ 
%has location 
\begin{equation}
%\mu_{i,k} = 
\lambda_k \, 
\phi_{ik}\trasp \, 
%\left(\phi_k(t_{i1}), \dots, \phi_k(t_{i \,n_i}\right)\trasp 
\bSi_i^{-1} \, \left(\bX_i-\bmu_i \right) \, ,
\label{eq:muik}
\end{equation}
where 
$\phi_{ik} = \left(\phi_k(t_{i1}), \dots, \phi_k(t_{i \,n_i}) \right)\trasp$, 
$\bmu_i=\left(\mu(t_{i1}), \dots, \mu(t_{i \,n_i}) \right)\trasp $ and $\bSi_i \in \real^{n_i \times n_i}$ is the 
matrix with $(\ell, j)-$th element  equal to $\gamma(t_{i\ell}\,,\, t_{ij})$. 
Note that if the process has finite mean then \eqref{eq:muik} equals 
$\esp \left(\xi_{i,k} \, \bigr| \, \bX_i \right)$, 
which is the best predictor for $\xi_{i,k}$ based on the 
observed trajectory. 
%In the case of Gaussian processes, this
%approach leads to the estimated scores as in Yao \textsl{et al.} \cite{YMW}.

A natural estimator for \eqref{eq:muik} consists of replacing the unknown quantities with
their estimators obtained in Steps 1 - 3 above. Specifically:
%This suggest that, using the estimated directions $\wphi_k$ and their size $\wlam_k$ as well as the 
%estimated location $\wmu$, the scores may predicted  by plugging-in the unknown quantities in \eqref{eq:muik}, i.e., 
\begin{equation}
\label{eq:wxi}
\wxi_{i,k} \, = \, 
\wlam_k \, \wbphi_{ik}\trasp \, 
(\wbSi_i+\delta\identidad_{n_i})^{-1} \, ( \bX_i - \wbmu_i )\,,
\end{equation}
 where $\wbmu_i=\left(\wmu(t_{i1}), \dots \wmu(t_{i n_i})\right)\trasp$,  
 $\wbphi_{ik}=\left(\wphi_k(t_{i1}), \dots \wphi_k(t_{i n_i})\right)\trasp$, 
  $\wbSi_i$ is the matrix with $(j,\ell)-$th element $\wgamma(t_{ij},t_{i\ell})$,
  and $\delta > 0$ is  a small regularization constant
  to ensure non-singularity (see Yao \textsl{et al.} \cite{YMW}).
 Finally, for a fixed $q$, the reconstructed curves are
 $$
 \widehat{X}_i \, = \, \sum_{k = 1}^q \wxi_{i,k} \, \wphi_k \, , \qquad i = 1, \ldots, n \, .
 $$
 
\subsection{The non-robust case} \label{sec:non-robust}

Note that using $\rho(u) = \rho_1(u) =\rho_2(u) = u^2$ in the method 
described above naturally yields a 
non-robust version of our proposal, which is different from that of 
Yao \textsl{et al.} \cite{YMW}. Moreover, our numerical experiments 
(see Section \ref{sec:simu}) 
indicate that when the data do not contain outliers
the non-robust version of 
our proposal compares favourably to PACE (Yao \textsl{et al.} \cite{YMW}). 

%It is worth noticing that the classical counterpart of the robust procedure described 
%above is not the proposal given in Yao \textsl{et al.} \cite{YMW}, but corresponds to 
%the choice $\rho(x)=x^2$ in \eqref{eq:Mlocalmuest}, \eqref{eq:wbeta} and in \eqref{eq:diagonal}, 
%taking in this last case $b=1$.

\section{Monte Carlo study}{\label{sec:simu}}

In this section we report the results of a Monte Carlo study carried out to investigate the
finite-sample performance and robustness of the proposed robust FPCA estimator. 
Our experiments compared the robust and non-robust
versions of our proposal and PACE, the approach in Yao \textsl{et al.} \cite{YMW}.  
We report results for random processes with two covariance functions and different 
atypical observations. In each case we  generated 500
samples of size $N=100$, and focused on the 
behaviour of the estimated eigenfunctions, eigenvalues, 
predicted scores, and accuracy of the 
scatter function estimator for both clean and contaminated samples.

\subsection{Simulation settings}{\label{sec:settings}}
For clean samples, the data sets are generated from the following model
\begin{equation} \label{eq:clean-simu}
X_i \ = \ 
%\mu \, + \, \sum_{j=1}^q \, \xi_{ij}  \, \phi_j = 
\mu \, + \, \sum_{j=1}^q \ \sqrt{\lambda}_j  \, Z_{ij} \, \phi_j 
\, , \qquad i = 1, \ldots N \, , 
\end{equation}
with  $Z_{ij}$ i.i.d. $Z_{ij}\sim \itN(0, 1)$ and $\lambda_1 \ge \lambda_2 \ge \cdots \lambda_q > 0$.
We considered two models for the location function $\mu$ 
and the principal directions $\phi_j$:
\begin{itemize}
\item \textbf{Model 1:}  $\itI=(0, 10)$, $\mu(t) = t \, + \, \sin(  t  )$
$q = 2$, $\lambda_1 = 4$, $\lambda_2 = 1$ and $\phi_j$ 
two elements of the Fourier basis:
\begin{eqnarray*}
\phi_1(t) = -\dst\frac{\cos( t \, \pi / 10 )}{ \sqrt{5}} \, , \quad \quad
\phi_2(t)  = \dst\frac{\sin( t \, \pi / 10 ) }{ \sqrt{5}}\,.
\end{eqnarray*}
The design points were generated as follows. First 
build an equally spaced grid of $51$ points $\{c_\ell\}_{\ell=1}^{51}$  on $[0, 10]$ with $c_1 = 0$ and $c_{51} = 10$, and 
 define $ s_\ell = c_\ell + \epsilon_\ell$, where $\epsilon_\ell$ are i.i.d. with 
 $\epsilon_\ell\sim N(0,0.1)$ (set $s_\ell=0$ if $s_\ell < 0$, 
 and $s_\ell = 10$ when  $s_\ell > 10$). 
 Each curve was sampled at a random number of points $n_i$, 
 chosen from a discrete uniform distribution on $\{2,3,4\}$, and 
 the locations $t_{ij}$ were randomly chosen 
 from $\{s_\ell\}_{\ell=2}^{50}$ without replacement.  
 This is the same setting as in Yao \textsl{et al.} \cite{YMW}. 

\vskip .1in

\item  \textbf{Model 2:} In this case, $ \itI=(0, 1)$, 
$\mu(t) = 10 \, \sin( 2 \, t \, \pi ) \, \exp( -3 \, t ) $, 
$q = 4$, 
$\lambda_1 = 0.83$, $\lambda_2 = 0.08$, $\lambda_3 = 0.029$ and 
$\lambda_4 = 0.015$, and the eigenfunctions 
$\phi_j$ are the first $q$ eigenfunctions of the ``Mattern'' covariance  
function:
$$
\gamma(s, t)  \, = \, C\left( \frac{\sqrt{2 \nu}\,|s - t|}{\rho}\right) \, , \quad 
C(u)= \frac{\sigma^2 \, 2^{1 - \nu}}{\Gamma(\nu)} \, u^\nu
K_\nu \left( u \right)  \, ,
$$
where $\Gamma(\cdot)$ is the Gamma function
and $K_{\nu }$ is the modified Bessel function of the second kind. 
We set $\rho = 3$, $\sigma = 1$ and $\nu = 1/3$.   
%We choose  $q = 4$, $\lambda_1 = 0.83$, $\lambda_2 = 0.08$, 
%$\lambda_3 = 0.029$ and $\lambda_4 = 0.015$. 
The eigenvalues above were chosen to have very similar
ratios $\lambda_j / \lambda_{j+1}$, $j=1, 2, 3$ 
to those of the first 4 eigenvalues of this 
Mattern covariance function. 
To obtain sparsely observed trajectories each curve was observed at a random
number of times $n_i$ chosen from a discrete uniform distribution on 
$\{3, 4, 5\}$. The observed times $t_{ij}$ 
satisfy $t_{ij} \sim \itU(\itI)$, i.i.d., $j = 1, \ldots, n_i;  i = 1, \ldots, N$. 

\end{itemize}
These models have different characteristics that may
affect the performance of the FPCA estimators on finite samples. 
For example, while the ratio between
the first and second eigenvalues is 4 for Model 1, it is larger than 10 for Model 2. This 
means that in the second model a single principal direction 
already explains a  larger  proportion (87\%) of the total variation. 
In addition, although the first two principal directions are similar in both models 
(with the order reversed), the third and fourth eigenfunctions of Model 2 will tend
to produce less smooth trajectories, and a slightly more complex covariance function.

\begin{figure}[ht!] 
\begin{center}
 %  \newcolumntype{G}{>{\centering\arraybackslash}m{\dimexpr.36\linewidth-1\tabcolsep}}
\begin{tabular}{cc }
Model 1 & Model 2 \\
 \includegraphics[scale=0.4]{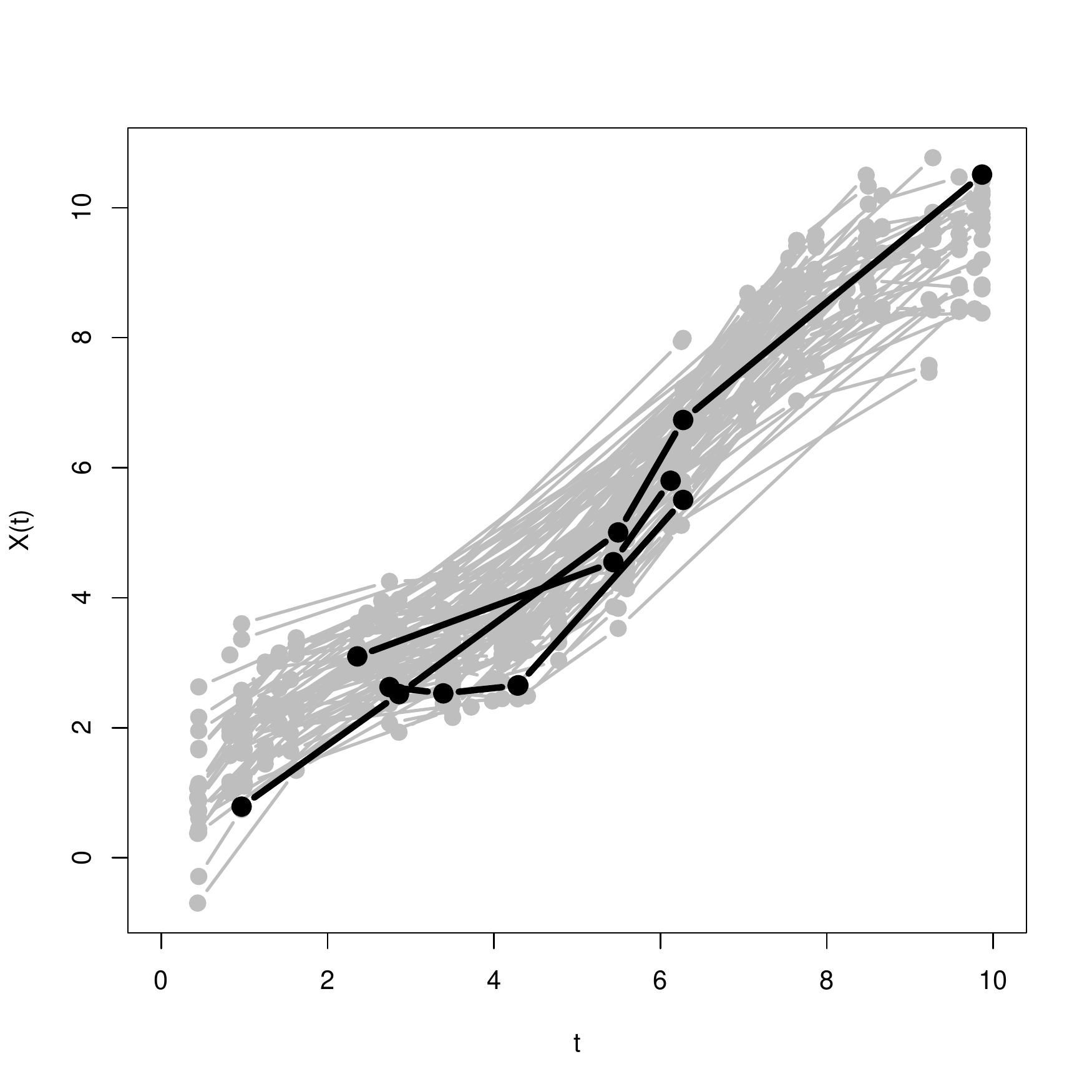}  &
  \includegraphics[scale=0.4]{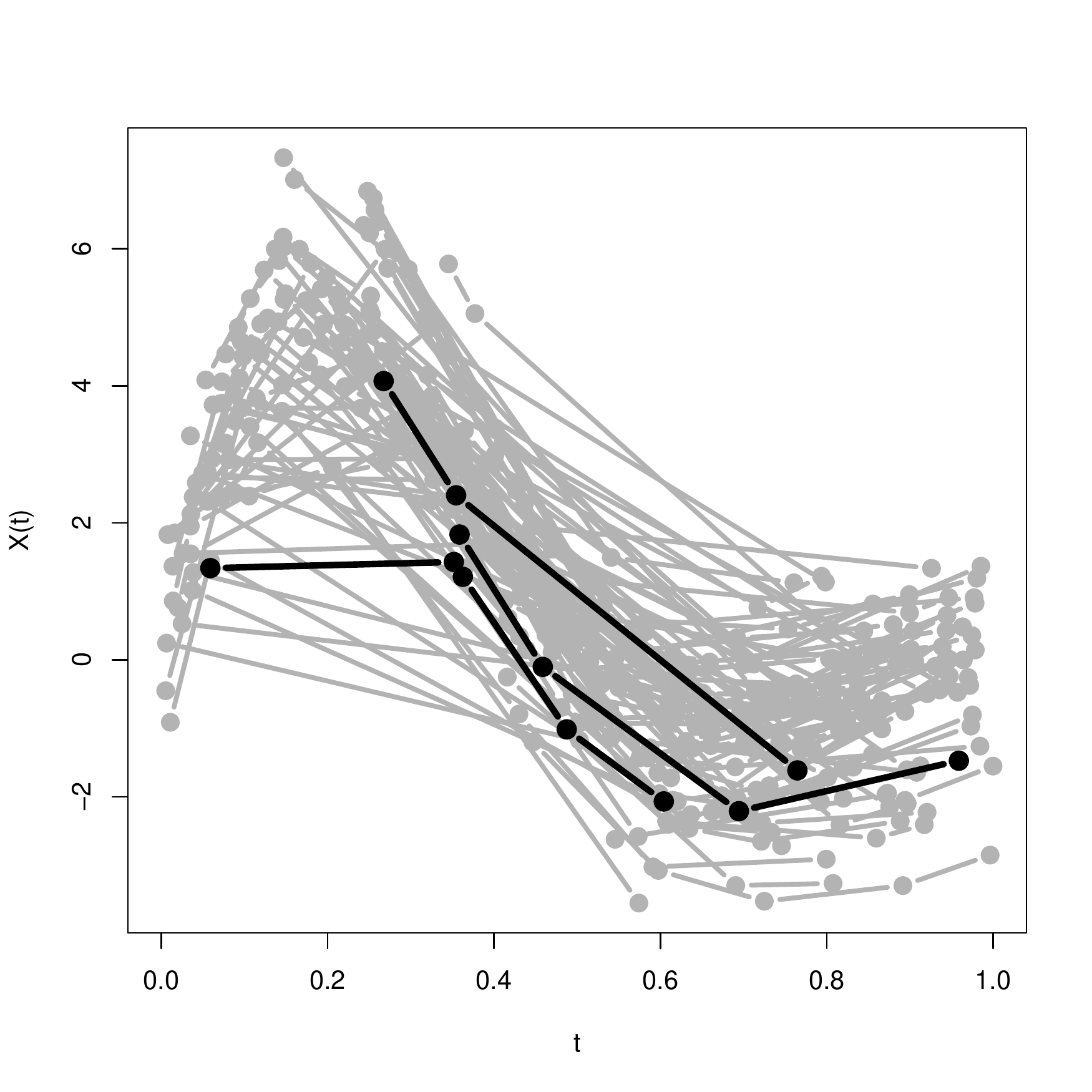}   
    \end{tabular}
\caption{\small \label{fig:datoseps0}  One typical sample from each Model, 
with 3 randomly chosen trajectories highlighted. Filled 
circles correspond to available observations.}
\end{center}
\end{figure}

Figure \ref{fig:datoseps0} illustrates typical samples obtained 
under each Model. We highlighted 3 randomly chosen
trajectories in each sample. The filled circles 
correspond to the observations.
 
\subsection{Outliers}

Atypical observations were introduced to 
alter the order among the principal directions
so that the estimated eigenfunctions and eigenvalues might be
affected. 
Outliers were introduced using a Bernoulli random variable  
$B_i \sim \itB(1, \epsilon)$, $i = 1, \ldots, N$, 
with $\epsilon = 0.05$ or $0.10$, which correspond
to 5\% and 10\% of outliers, respectively.  
When $B_i = 0$ the trajectory $X_i$ is
generated as in \eqref{eq:clean-simu}. 
When $B_i = 1$ the curve is contaminated
as follows: 
\begin{itemize}
\item \textbf{Model 1}: The score $Z_{i, 2}$ for the second 
direction is sampled from
$ Z_{i,2}  \sim  \itN \left( 12, 1 \right)$; 
\item \textbf{Model 2}: The scores 
for the second and third direction are sampled from the
following bivariate distribution:
$$
\left(\begin{array}{c}
Z_{i,2} \\
 Z_{i,3}\end{array}\right)
\sim   \itN \left(\bmu_c,  \bSi_c \right) 
 \quad \mbox{with} \quad
\bmu_c=\left(\begin{array}{r}
20 \\
25\end{array}\right)\quad \mbox{and}\quad \bSi_c=\left(\begin{array}{cc}
1/16 & 0 \\
0 & 1/16\end{array}\right) \, .$$ 
%%%%%%%%%%%%%%%%%%%%%%%%%%%%%%%%%%%%%%%%%%%%%%%
% con epsilon =0.05 esta contaminacion manda la primera autofuncion se convierta en la tercera
% con epsilon =0.1 la primera se convierte en la segunda
%%%%%%%%%%%%%%%%%%%%%%%%%%%%%%%%%%%%%%%%%%%%%%%
 \end{itemize}

\begin{figure}[ht!] 
\begin{center}
%   \newcolumntype{G}{>{\centering\arraybackslash}m{\dimexpr.36\linewidth-1\tabcolsep}}
\begin{tabular}{cc }
Model 1 & Model 2 \\
 \includegraphics[scale=0.4]{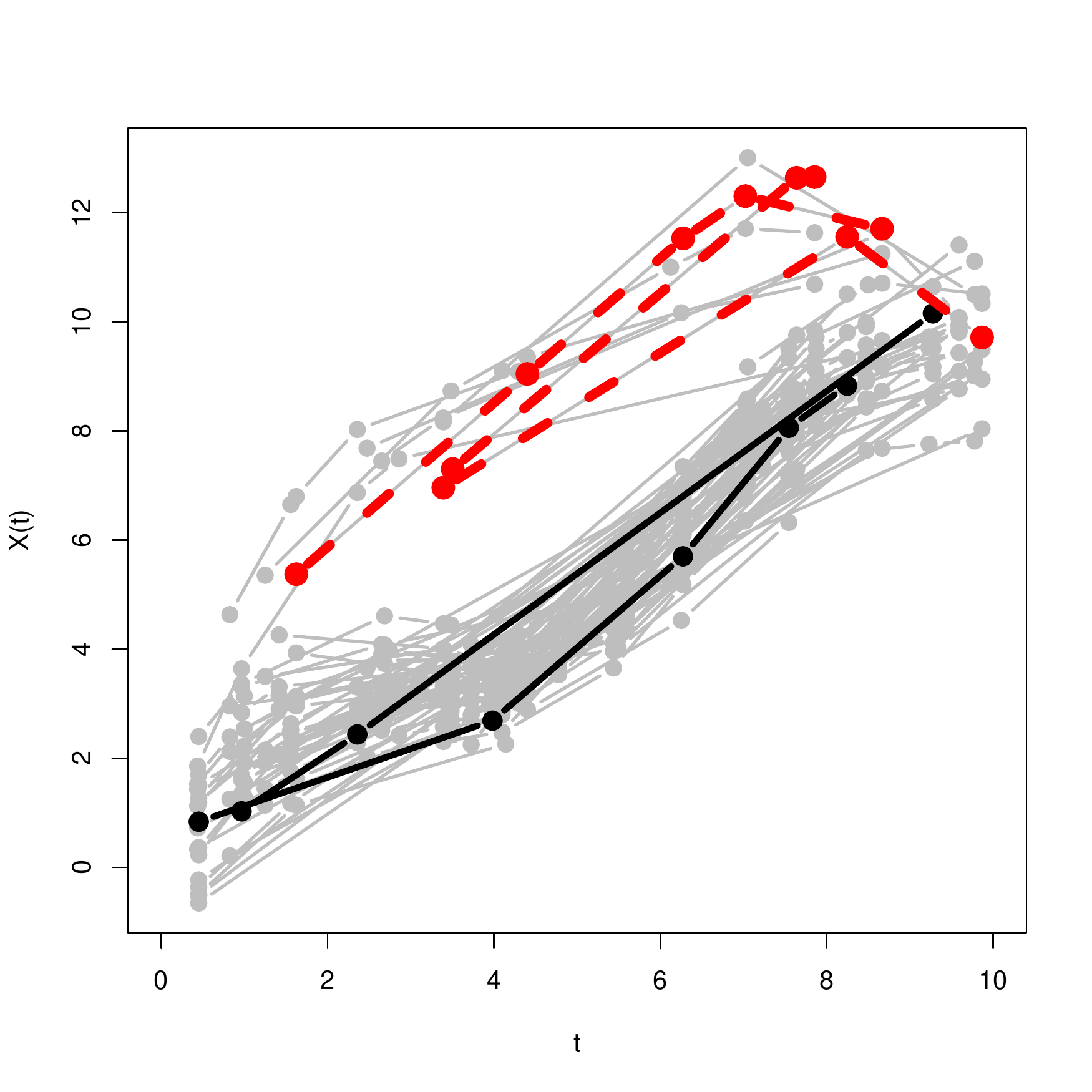}  &
  \includegraphics[scale=0.4]{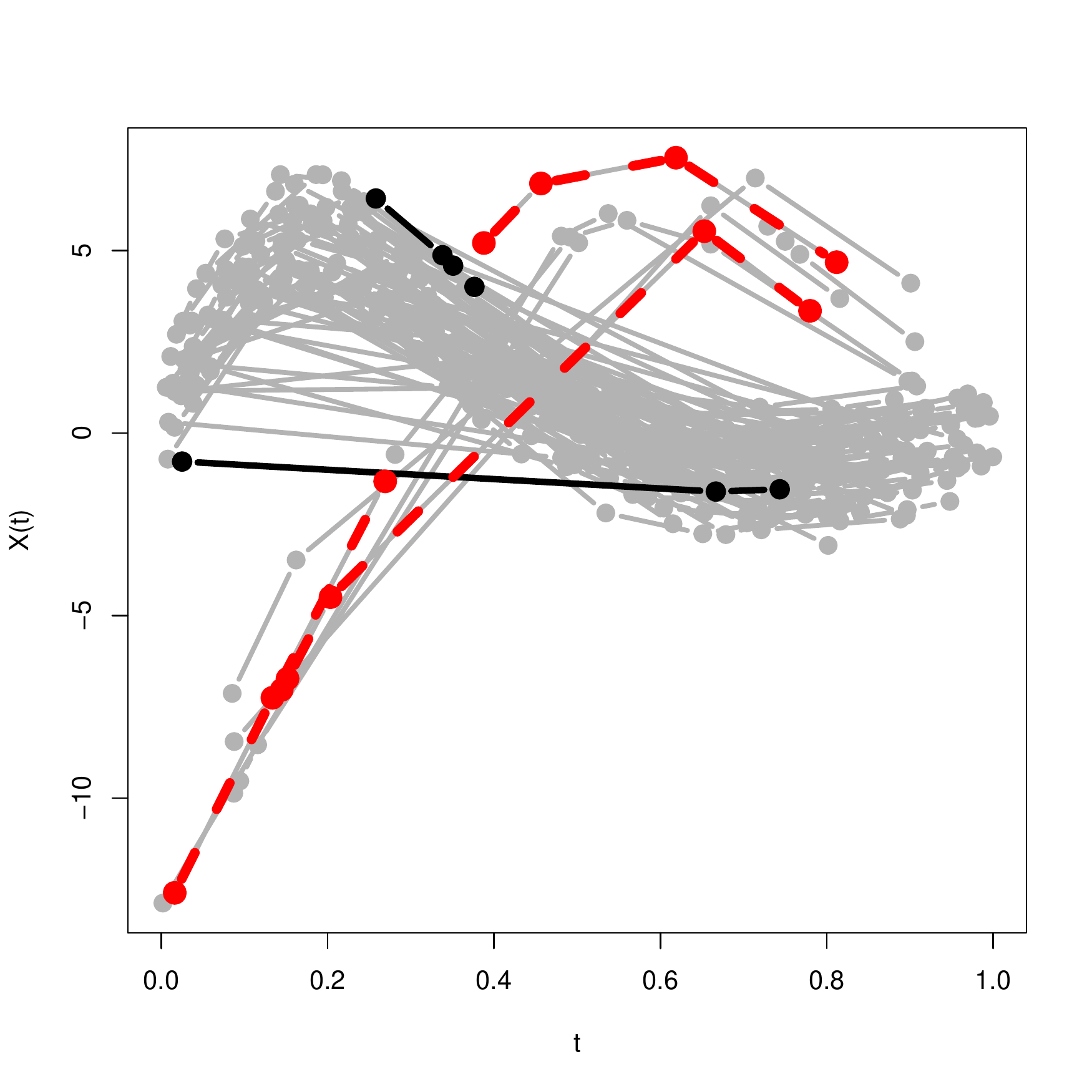}    
    \end{tabular}
\caption{\small \label{fig:datoseps01}  Contaminated sample from Models 1 to 3 with $\epsilon=0.10$. The filled points correspond to the observed trajectories.  Black solid lines correspond to two randomly chosen 
uncontaminated observations, and dashed red lines indicate 3 randomly 
selected outlying trajectories (color figures are available on the web version of this article).}
\end{center}
\end{figure}

Figure \ref{fig:datoseps01} shows how the introduced contaminations modify the pattern of the clean data when $\epsilon=0.10$. 
 Black solid lines correspond to two randomly chosen 
uncontaminated observations, and dashed red lines indicate 3 randomly 
selected outlying trajectories.

\subsection{The estimators}{\label{sec:estmonte}}

We considered three estimators: two of them 
are based on non-resistant procedures and the third one is our proposal 
in Section  \ref{sec:proposal}. Of the two non-robust methods, 
one is PACE (Yao \textsl{et al.} \cite{YMW}), and the other
is the non-robust version of our method as in Section \ref{sec:non-robust}. 
All computations were done using \texttt{R}
(R Core Team \cite{R}). The PACE 
estimator was computed using the package \texttt{fdapace}
(Chen \textsl{et al.} \cite{fdapace}). 
Our estimators were computed using \texttt{R} 
code in the package \texttt{sparseFPCA}
publicly available at \url{https://github.com/msalibian/sparseFPCA}. 

The tuning parameters for the robust FPCA estimator
were chosen as follows. To estimate $\mu(t)$ we use
a Huber function $\rho$ with constant $1.345$.
In Steps 2 and 3 we use $\rho$-functions in 
Tukey's biweight family. To estimate the 
diagonal elements of the scatter function $\gamma(t,t)$ 
the tuning constants were $c=1.54764$ and  
$b=1/2$. In Step 3,    to obtain a more efficient estimator for $\beta(t, s)$, we used a larger tuning constant. More specifically, we selected $c = 3.44369$, which in linear regression models leads to   estimators with breakdown point $1/2$  and asymptotic efficiency $0.85$. 

The kernel function in \eqref{eq:weights} and \eqref{eq:wbeta}
was an Epanechnikov 
kernel, $\itK(u)=0.75 \, (1-u^{2}) \indica_{[-1,1]}(u)$.   The two--dimensional 
smoother used in Step 3 was a thin plate regression spline, as described in Section 5.5.1 of Wood \cite{wood}, 
which does not require selecting knot locations.  The bandwidth 
parameters  were selected using a robust 5-fold cross-validation 
(holding 20\% of the curves for each fold). 
The cross-validation criterion for the robust estimator 
was  based on  a 50\% breakdown point 
$M$-scale of the residuals with respect to the fits in equations  \eqref{eq:Mpolylocalmuest} and \eqref{eq:wbeta}, while for
the other two approaches we used the mean 
squared prediction error. 

We chose the number $K$ of principal directions to estimate
with the goal of explaining at least 90\% of the total variability. This
corresponds to $K=2$ for both Models above.

\subsection{Simulation results}{\label{sec:result}}

To compare the different approaches we look at the corresponding scatter operators $\wGamma$, 
their associated eigenvalues and eigenfunctions and 
the predicted scores. We report summaries of their
performance computed over
500 samples of $N = 100$ trajectories each, 
generated under each Model using 
clean and contaminated data. 

To quantify the discrepancy between the estimated $\wGamma$
and the true $\Gamma$ 
scatter operators we use the following approximation to the 
norm of their
difference. We compute the operators over a 
fixed grid of 50 equidistant points in $\itI$, and calculate 
the squared Frobenius norm of the difference matrix. 
This is a discrete version of the corresponding 
operator norm for Hilbert-Schmidt 
operators $\Upsilon$ given by 
$\|\Upsilon\|_{\itF}^2 = \mbox{trace} \left[\Upsilon^{*}\Upsilon\right] = \sum_{j=1}^\infty \|\Upsilon u_j
\|^2$, with $\{u_j :  j \ge 1\}$ any orthonormal basis of $L^2(\itI)$.

To compare the different 
eigenvalue estimators we consider two loss measures:
$$\frac 1{500} \sum_{\ell=1}^{500} \left\{\log\left(\frac{\wlam_{\ell,k}}{\lambda_k}\right)\right\}^2 \, , \qquad \mbox{and} \qquad \frac 1{500} \sum_{\ell=1}^{500} \left\{ \left(\frac{\wlam_{\ell,k}}{\lambda_k}-1\right)\right\}^2 \, ,  $$
where $\wlam_{\ell,k}$ is the estimated $k$-th eigenvalue obtained with the $\ell$-th sample.

The accuracy of the eigenfunction estimators was measured using the cosine of the 
angle between the
estimated and true eigenfunctions.  Values close to 1 
correspond to a small angle and thus to a good estimated eigenfunction, whereas values
close to zero indicate estimated eigenfunctions that are close to being orthogonal to 
the true ones. We report
$$
\frac 1{500} \sum_{\ell=1}^{500} \cos( | \langle \wphi_{\ell, k}, \phi_k \rangle | ) \, ,
$$
where $\wphi_{\ell, k}$
and $\phi_k$ are the estimated $k$-th eigenfunction computed with the $\ell$-th 
sample, and the true $k$-th eigenfunction, respectively. 

Finally, to measure the discrepancy between  the true and predicted scores, 
for each $k$, we calculate the average of the scores distances over 
the observations in each sample:
$$ 
\mbox{MSE}_{\ell}( \xi_k )= (1/N) \, \sum_{i=1}^N   \left( \wxi_{i,k}^{(\ell)} - \xi_{i,k}^{(\ell)} \right)^2 
\, , \quad \ell = 1, \ldots, 500 \, ,$$
where $\xi_{i,k}^{(\ell)}$ and $\wxi_{i,k}^{(\ell)}$ denote the 
true and the estimated $k$-th score for the $i$-th trajectory of the $\ell$-th sample, $1\le k\le K$, 
respectively.  For each $k = 1, 2$, we report the 
mean and  median of $\mbox{MSE}_{\ell}( \xi_k )$
over the $\ell = 1, \ldots, 500$ replications.

Note that, for a fixed $  k$,  the predicted scores $\wxi_{i,k}$ defined in \eqref{eq:wxi} may be multiplied by -1 when  the $k-$th estimated eigenfunction is multiplied by -1, without changing the predicted trajectory. For that reason, in order to ensure that both the true and predicted scores have the same \textsl{orientation}, for each replication $\ell$ and each $k$, we compute the Spearman correlation $s_k^{(\ell)}$, between $\wxi_{i,k}^{(\ell)}$ and $\xi_{i,k}^{(\ell)}$, $1\le i\le N$. When $s_k^{(\ell)}<0$, we redefine the scores $\wxi_{i,k}^{(\ell)}$, $1\le i\le N$,  multiplying by -1 those originally obtained. 

In what follows the 
method proposed by Yao \textsl{et al.} \cite{YMW}
will be denoted \textsc{PACE}, our robust method method 
will be labeled \textsc{ROB}, and our non-robust variant 
(Section \ref{sec:non-robust}) \textsc{LS}. 
Results for experiments without outliers will be 
indicated with the label $C_0$,
while those corresponding to the contamination schemes described in  Section \ref{sec:settings} will
be indicated with $C_{0.05}$ and $C_{0.10}$ when $\epsilon=0.05$ and $0.10$, respectively. 

\begin{figure}[ht!]
\begin{tabular}{cc}
$C_0$ & \textsc{rob}\\
\includegraphics[scale=0.4]{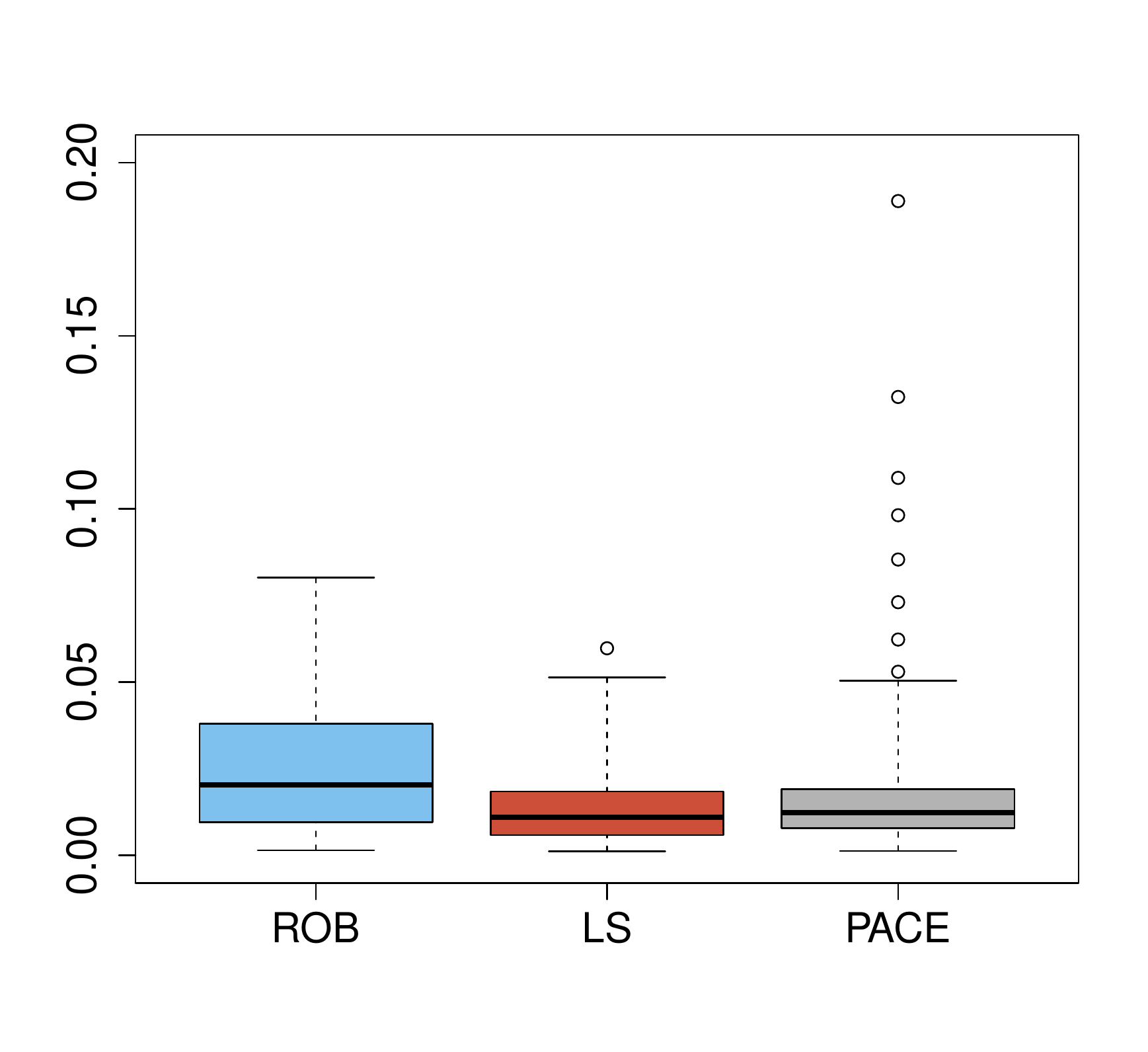} & 
\includegraphics[scale=0.4]{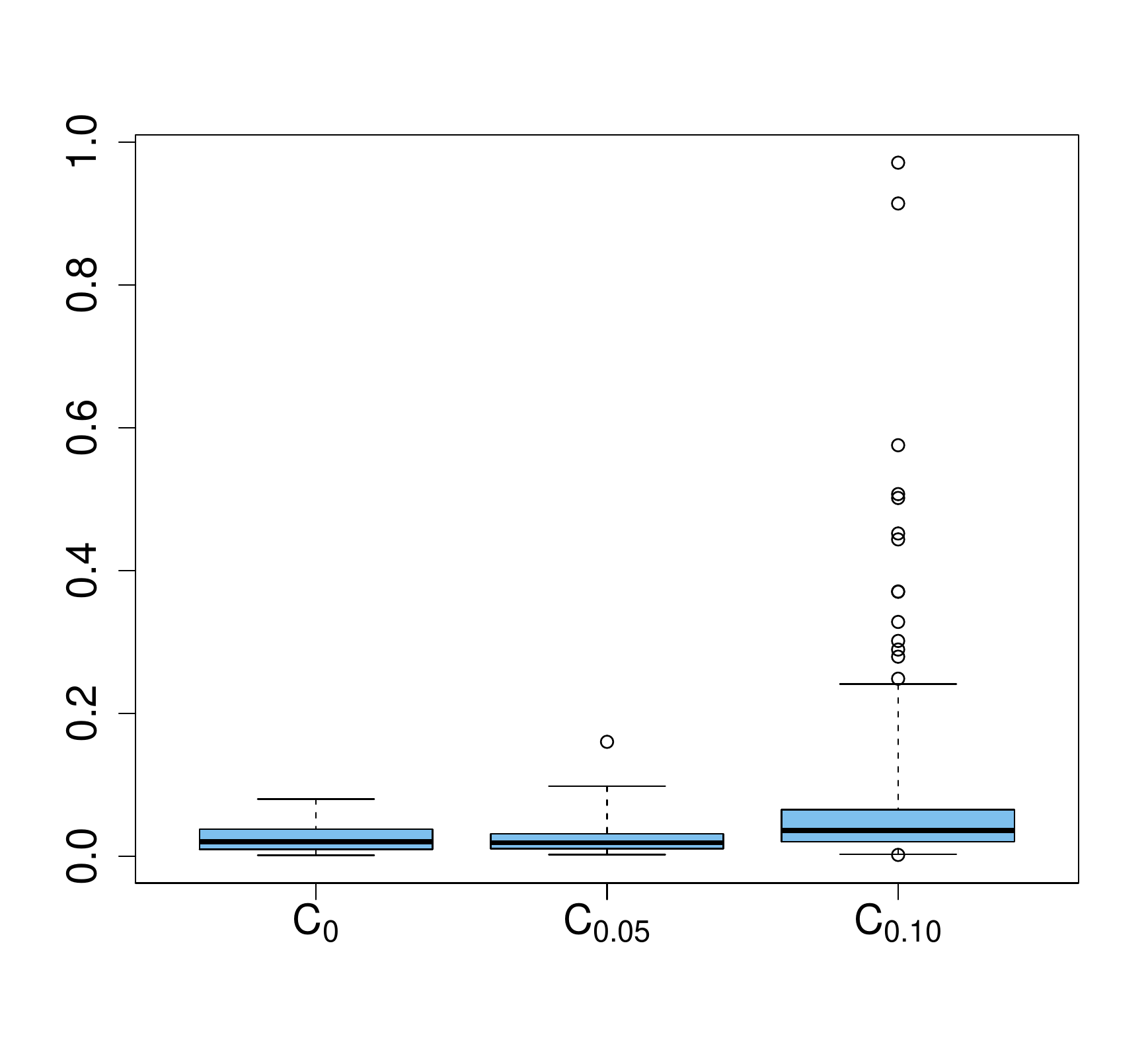}\\
\textsc{ls}& \textsc{pace}\\
\includegraphics[scale=0.4]{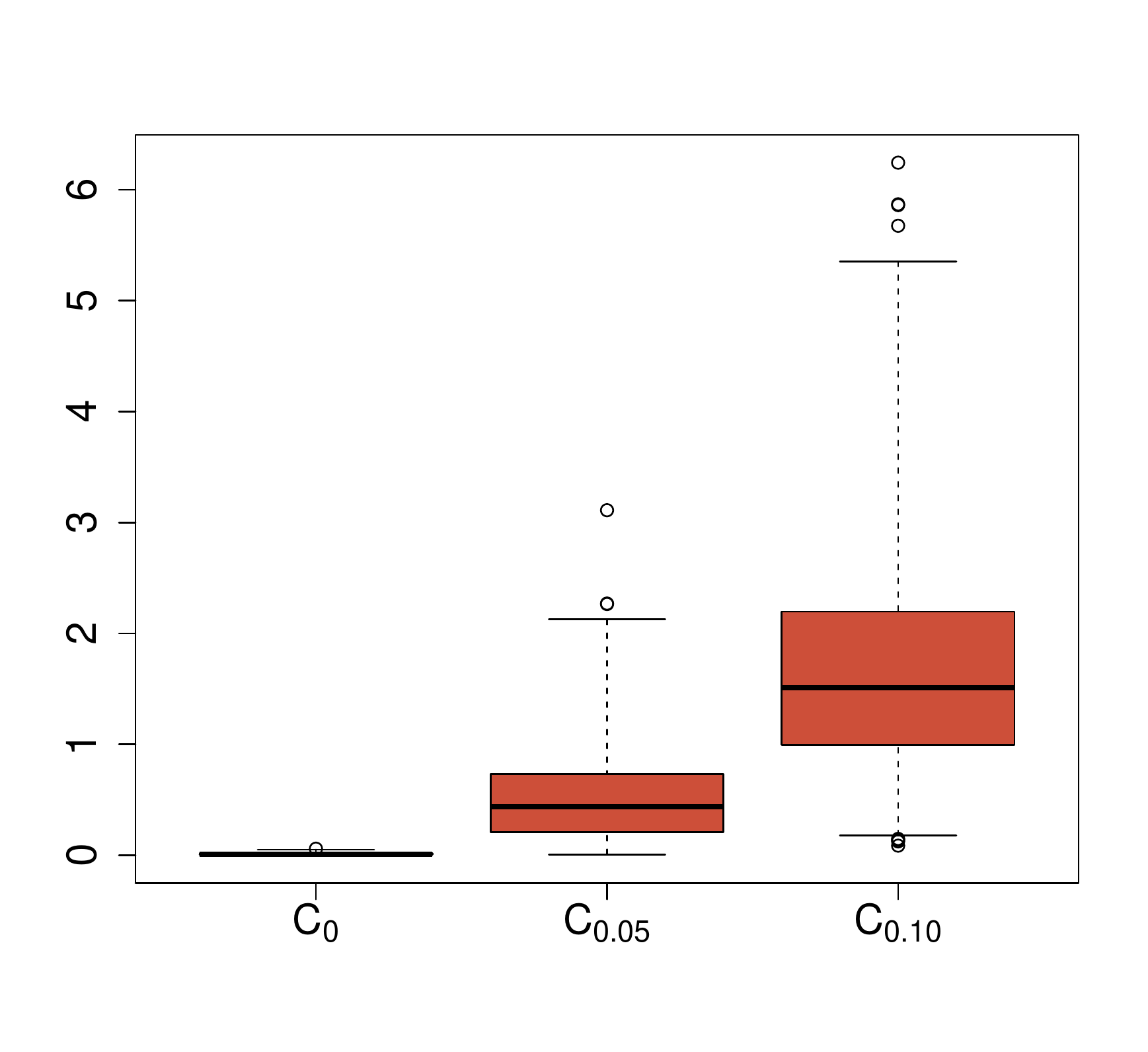} & 
\includegraphics[scale=0.4]{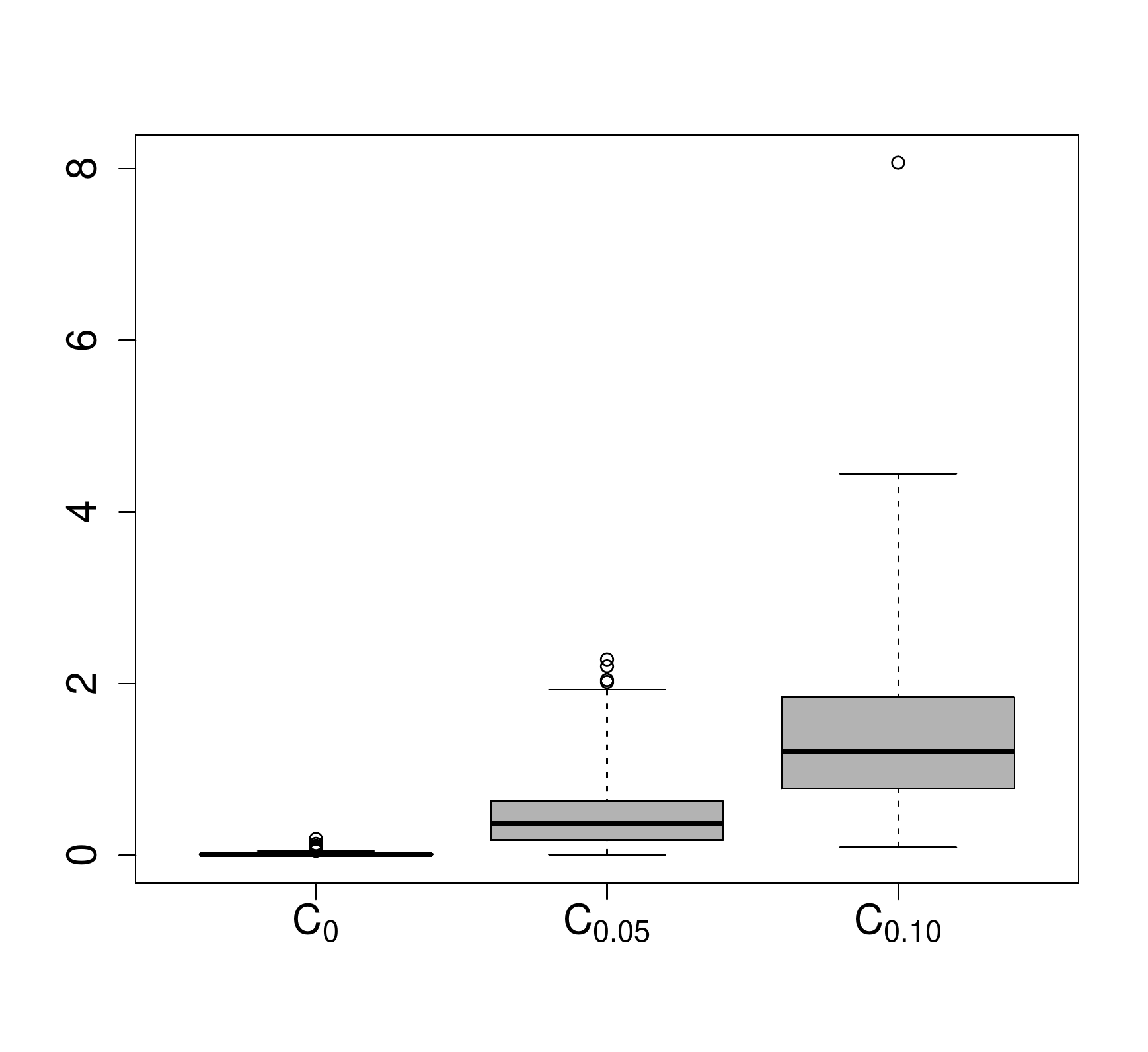}
\end{tabular}
\caption{\label{fig:operator-norm-model-1} Adjusted boxplots of $\|\wGamma-\Gamma\|_{\itF}^2 $, under Model 1.}
\end{figure}

\begin{figure}[ht!]
\begin{tabular}{cc}
$C_0$ & \textsc{rob}\\
\includegraphics[scale=0.4]{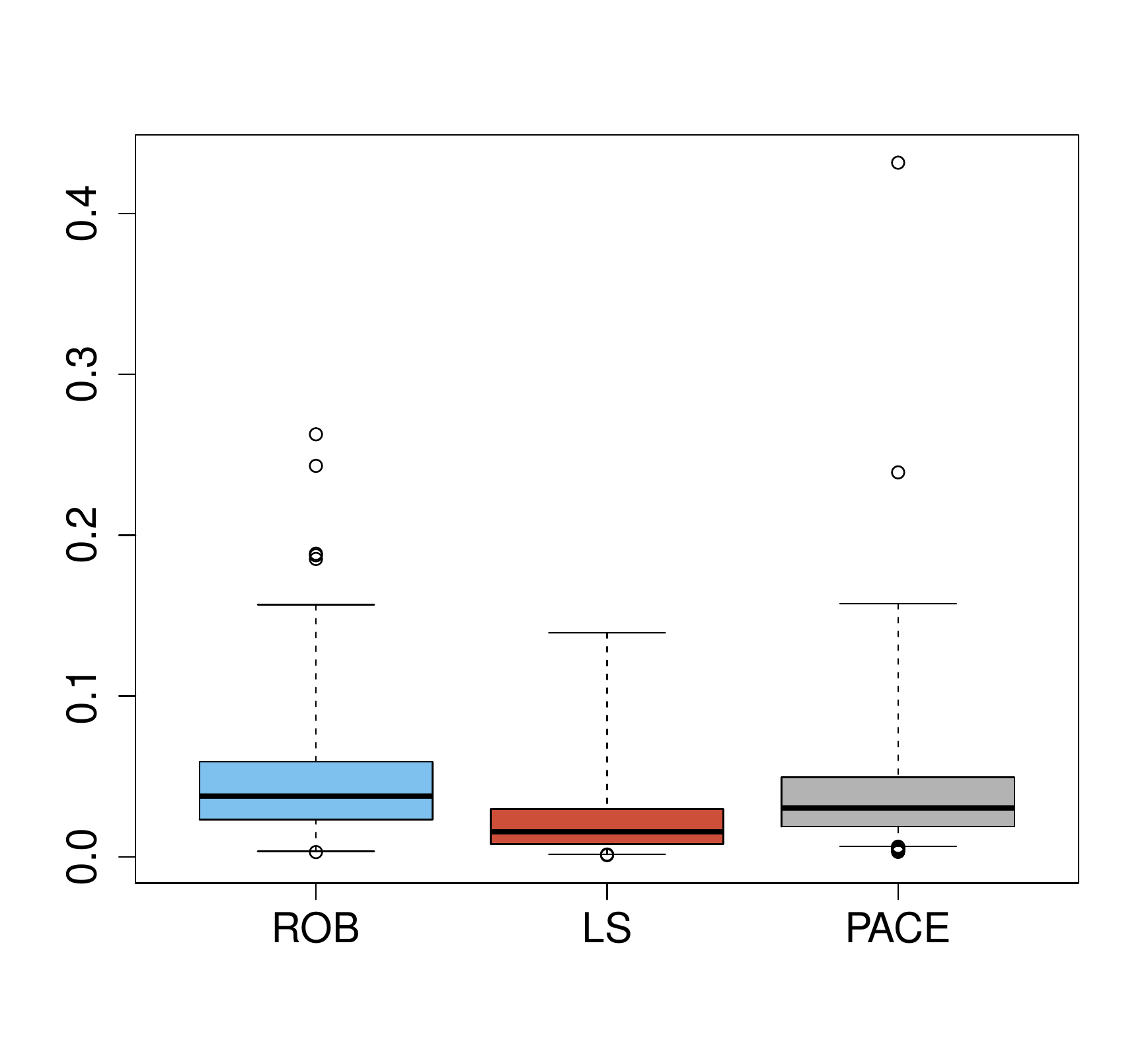} & 
\includegraphics[scale=0.4]{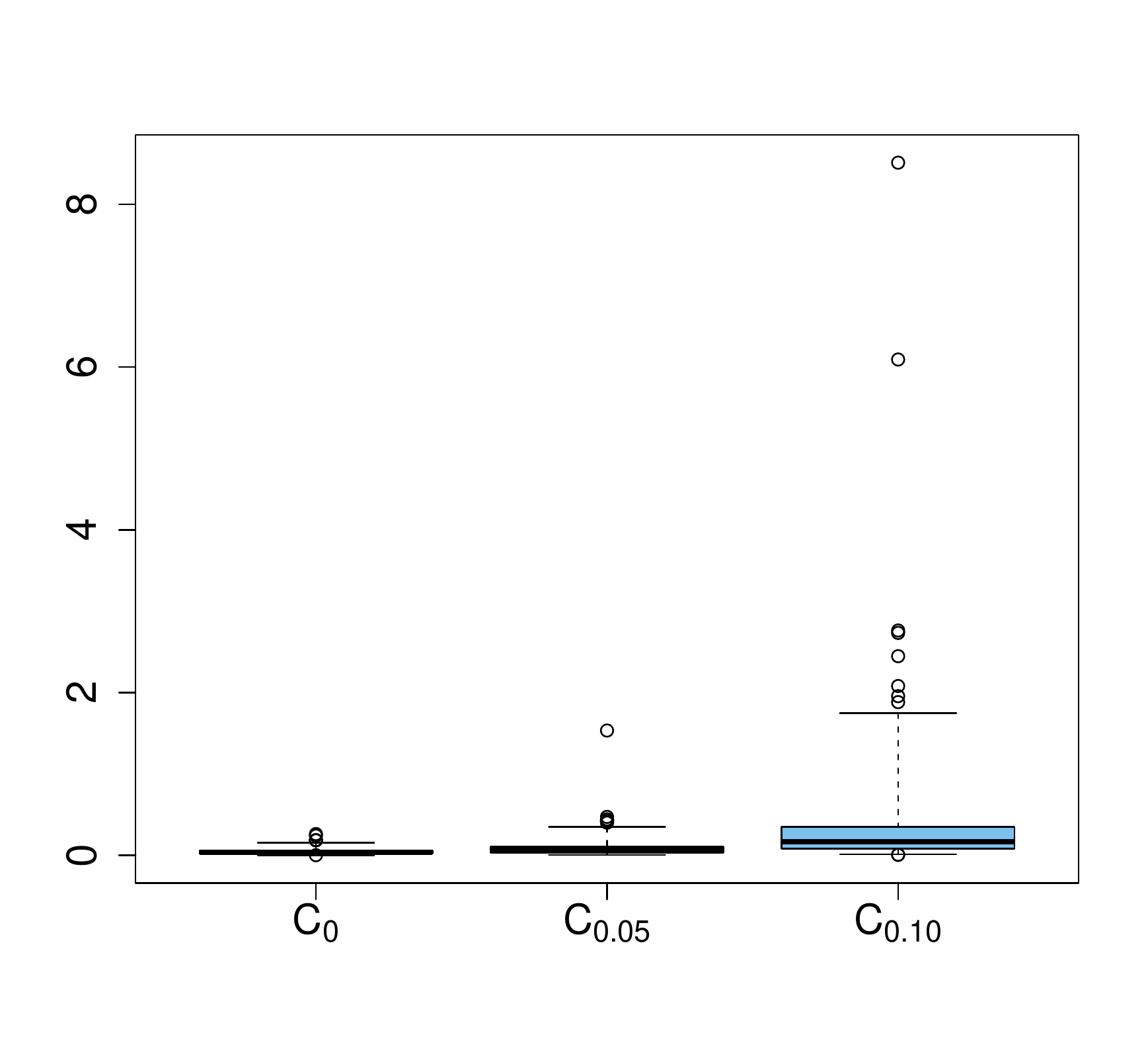}\\
\textsc{ls}& \textsc{pace}\\
\includegraphics[scale=0.4]{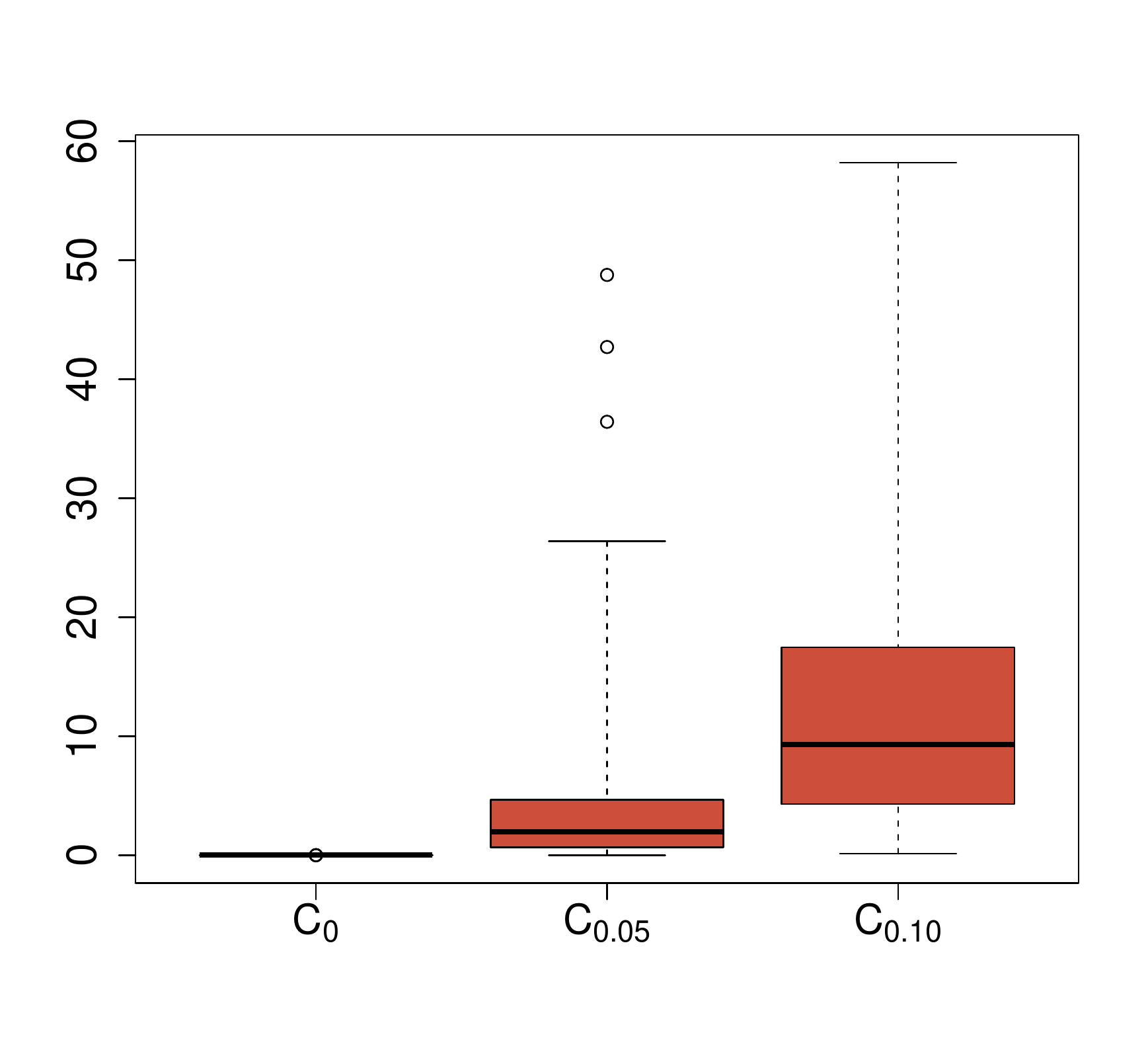} & 
\includegraphics[scale=0.4]{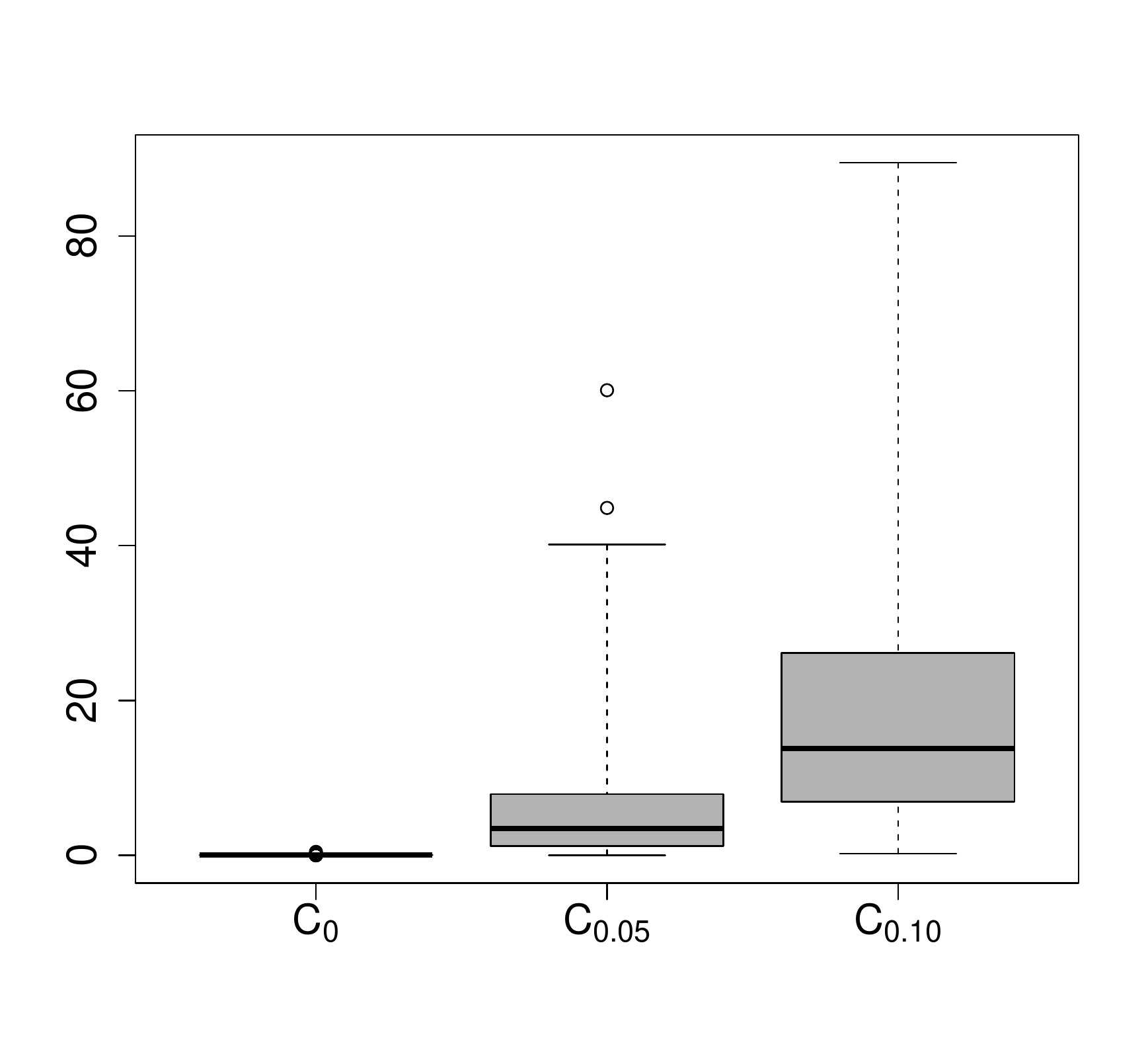}
\end{tabular}
\caption{\label{fig:operator-norm-model-2}Adjusted boxplots of $\|\wGamma-\Gamma\|_{\itF}^2$, under Model 2.}
\end{figure}

Taking into account that the squared operator norm and the mean scores distances 
$\mbox{MSE}_{\ell}( \xi_k )$ are non-negative and expected to have a 
skewed distribution, 
we use skewed-adjusted boxplots (Hubert and Vandervieren \cite{HV}) to
display our results. These can be found in 
Figures \ref{fig:operator-norm-model-1}, \ref{fig:operator-norm-model-2}  and 
\ref{fig:scores-12-model-123-sigcor}. 
%Figures \ref{fig:operator-norm-model-1} to \ref{fig:operator-norm-model-2} and 
%Figures \ref{fig:scores-12-model-123-sigcor} to \ref{fig:M2-12-model-123-sigcor}.

As expected, when no outliers are present all procedures are comparable, 
with the robust procedure performing slightly worse than the non-robust methods
when estimating the scatter operator, in particular under Model 1
(Table \ref{tab:operator-norm-model-123}).
Note that for both Models the non-robust version of our approach (\textsc{LS})
shows the best performance. Figures
\ref{fig:operator-norm-model-1} and \ref{fig:operator-norm-model-2} 
confirm this observation. 
The stability and advantage of \textsc{ROB} over \textsc{PACE} and \textsc{LS}
when outliers are present in the data can be seen in 
Table \ref{tab:operator-norm-model-123}, and  Figures \ref{fig:operator-norm-model-1} and \ref{fig:operator-norm-model-2}.

Table \ref{tab:cos-angle-model-123}
contains the average of the cosine of the angles between the estimated and true eigenfunctions. 
Here again \textsc{LS} has the best performance for clean data, although the difference 
between the three estimators is small. For contaminated data sets 
(even with only 5\% of outliers) the estimated eigenfunctions obtained with \textsc{PACE} and 
\textsc{LS} are heavily affected by the atypical observations, while the robust estimator 
remains stable.

Tables \ref{tab:Logs-lambdas-model-123} and
\ref{tab:Cuad-lambdas-model-123} show the 
results for the eigenvalue estimators. Note that under Model 1, even without 
contamination, all estimators for $\lambda_1$ are biased 
(this can be observed in Figure  \ref{fig:lambdas-12-model-123}). 
When the samples are contaminated the robust eigenvalue estimators 
perform better than the non-robust ones.  
The results for Model 2 are as one would expect: very little difference
when no outliers are present in the data, and a noticeable advantage of the robust
estimator in the contaminated settings.

The predicted scores of all estimators with clean samples behave very similar to each other
(Tables \ref{tab:mse-model-123} and \ref{tab:medse-model-123}). 
Note that the larger values of the mean and median of $ \mbox{MSE}( \xi_k )$, under Model 1, may be explained by
the fact that $\var(\xi_{ik})=\lambda_k$, and the eigenvalues of Model 1 are larger than those for
Model 2. When no outliers are present the \textsc{LS} estimator performs best. 

Noting that score estimates 
can be highly influenced by atypical trajectories,  
to evaluate the fit over the uncontaminated curves, 
we also considered the average over replications, $M_2( \xi_k )$,  
of the following quantity:
 \begin{eqnarray*}
M_{2,\ell}( \xi_k ) \, &=&  \frac{1}{\sum_{i=1}^N (1-B_i)} \, \sum_{i=1}^N (1-B_i) \left( \wxi_{i,k}^{(\ell)} - \xi_{i,k}^{(\ell)} \right)^2 \,.
\end{eqnarray*}
We expect a good procedure to fit well most of the uncontaminated data, resulting
in small values of $M_{2,\ell}( \xi_k )$. Note that, for clean samples,  $M_{2,\ell}(\xi_k)$ equals 
 $\mbox{MSE}_{\ell}( \xi_k )$.
Figure  \ref{fig:M2-12-model-123-sigcor} and Table \ref{tab:M2-model-123} show that   classical procedures
result in a compromise between outlying and non-outlying trajectories, leading to bad predictions for the 
uncontaminated data. Under the contamination settings in this study, the robust estimator has 
the best performance overall. The  scores mean squared error of the robust approach is the smallest, 
sometimes by a factor larger than 6 (see Table \ref{tab:mse-model-123}). 
Table \ref{tab:M2-model-123} reveals that the robust procedure also provides  better fits to the 
non-contaminated samples. For $\epsilon=0.05$, the values of $M_2$ for the robust method are 
similar to those obtained under $C_0$, while for  $\epsilon=0.10$ they increase by a factor
less than 2.5.  The values of $M_2$ for \textsc{PACE} and \textsc{LS}, 
when the data contain outliers, can be more than 10 times higher than 
their values under $C_0$. 
%Regarding the behaviour of $M2$ when \textsc{pace} or \textsc{ls} are considered, they present 
%values which are more than 10 times those obtained under $C_0$, in particular for $k=1$ 
%the values show an increment of more than 200\%.

\begin{table}[ht!]
\centering
\begin{tabular}{c|ccc|ccc|}
  \hline
  & \multicolumn{3}{c|}{Model 1}& \multicolumn{3}{c|}{Model 2} \\
  \hline
 & \textsc{rob} & \textsc{ls} & \textsc{pace}  & \textsc{rob} & \textsc{ls} & \textsc{pace}  \\ 
  \hline
$C_0$ & 0.0256 & 0.0133 & 0.0154  & 0.0468 & 0.0230 & 0.0398   \\ 
  $C_{0.05}$ & 0.0235 & 0.5344 & 0.4668  & 0.0887 & 3.5896 & 6.0193  \\ 
 $ C_{0.10}$  & 0.0584 & 1.7041 & 1.3851 & 0.3228 & 12.3002 & 17.5266    \\ 
    \hline
\end{tabular}
\caption{\label{tab:operator-norm-model-123} Average of $\|\wGamma-\Gamma\|_{\itF}^2$ over 
500 samples.}
\end{table}

\begin{table}[ht!]
\centering
 \setlength{\tabcolsep}{3pt} 
  \begin{tabular}{c|c|ccc|ccc |}
  \hline
 & & \multicolumn{3}{c|}{$\wphi_1$} & \multicolumn{3}{c|}{$\wphi_2$} \\ 
  \hline
  & & \textsc{rob} & \textsc{ls} & \textsc{pace}  & \textsc{rob} & \textsc{ls} & \textsc{pace}   \\ 
  \hline 
          & $C_0$  & 0.9622 & 0.9923 & 0.9872 & 0.9483 & 0.9812 & 0.9594   \\ 
Model 1 & $C_{0.05}$  & 0.9486 & 0.2304 & 0.3015 & 0.9388 & 0.2297 & 0.3022  \\ 
        & $C_{0.10}$ & 0.8521 & 0.0898 & 0.1189 & 0.8468 & 0.0905 & 0.1215  \\  
\hline 
        & $C_0$ & 0.9919 & 0.9977 & 0.9938 & 0.9198 & 0.9776 & 0.8586   \\ 
Model 2 & $C_{0.05}$  & 0.9878 & 0.4388 & 0.3940 & 0.8821 & 0.3867 & 0.3396   \\ 
        & $C_{0.10}$ & 0.9390 & 0.1889 & 0.1982 & 0.8165 & 0.1725 & 0.1881   \\ 
\hline
\end{tabular}
\caption{\label{tab:cos-angle-model-123} Average of
$\cos( | \langle \wphi_{\ell, k}, \phi_k \rangle |) $ over $\ell = 1, \ldots, 500$, for $k = 1, 2$.}  
\end{table}

\begin{table}[ht!]
\centering
 \setlength{\tabcolsep}{3pt} 
\small
\begin{tabular}{c|c|ccc|ccc|}
  \hline
  & & \multicolumn{3}{c|}{$k=1$} & \multicolumn{3}{c|}{$k=2$} \\ 
  \hline
  & & \textsc{rob} & \textsc{ls} & \textsc{pace}  & \textsc{rob} & \textsc{ls} & \textsc{pace}   \\ 
  \hline 
 & $C_{0}$ &  0.2652 & 0.1187 & 0.0628 & 0.0685 & 0.0931 & 0.0530  \\ 
Model 1 & $C_{0.05}$ & 0.1228 & 0.4714 & 0.4058 & 0.2141 & 1.1298 & 1.3880 \\ 
 & $C_{0.10}$ &  0.0786 & 1.4122 & 1.1931 & 0.6719 & 1.4483 & 1.6548 \\  
    \hline
        & $C_0$ & 0.0376 & 0.0259 & 0.0339 & 0.1059 & 0.1525 & 0.1772   \\ 
Model 2 & $C_{0.05}$ & 0.0505 & 0.6590 & 1.0158 & 0.5273 & 4.3406 & 4.5832   \\ 
        & $C_{0.10}$  & 0.0934 & 1.8684 & 2.3644 & 2.0295 & 5.2132 & 5.6738   \\ 
       \hline
     \hline
\end{tabular}
\caption{\label{tab:Logs-lambdas-model-123} 
Average of
$( \log( \wlam_{\ell,k} / \lambda_k ) )^2$ over $\ell = 1, \ldots, 500$, for $k = 1, 2$.} 
\end{table}

\begin{table}[ht!]
\centering
 \setlength{\tabcolsep}{2pt} 
\small
\begin{tabular}{c|c|ccc|ccc|}
  \hline
 & & \multicolumn{3}{c|}{$k=1$} & \multicolumn{3}{c|}{$k=2$} \\ 
  \hline
  & & \textsc{rob} & \textsc{ls} & \textsc{pace}  & \textsc{rob} & \textsc{ls} & \textsc{pace}   \\ 
  \hline 
 & $C_{0}$ & 0.1390 & 0.0760 & 0.0493 & 0.0838 & 0.0586 & 0.0569  \\ 
Model 1 & $C_{0.05}$ & 0.0763 & 1.2663 & 1.0439 & 0.4461 & 4.1994 & 5.8317 \\ 
 & $C_{0.10}$ & 0.0776 & 6.1153 & 4.6775 & 2.0962 & 6.2134 & 7.7069 \\ 
  \hline
        & $C_0$ & 0.0344 & 0.0233 & 0.0283 & 0.1662 & 0.0913 & 0.2062   \\ 
Model 2 & $C_{0.05}$ & 0.0659 & 2.6433 & 5.0494 & 1.9130 & 66.5294 & 76.7619   \\ 
        & $C_{0.10}$  & 0.1608 & 12.0135 & 17.9939 & 17.6268 & 95.6142 & 124.1025   \\ 
    \hline
\end{tabular}
\caption{\label{tab:Cuad-lambdas-model-123} Average of
$( ( \wlam_{\ell,k} / \lambda_k ) - 1 )^2$ over $\ell = 1, \ldots, 500$, for $k = 1, 2$. } 
\end{table}

\begin{table}[ht!]
\centering
 \setlength{\tabcolsep}{3pt} 
\small
\begin{tabular}{c|c|ccc|ccc|}
  \hline
 & & \multicolumn{3}{c|}{$k=1$} & \multicolumn{3}{c|}{$k=2$} \\ 
  \hline
  & & \textsc{rob} & \textsc{ls} & \textsc{pace}  & \textsc{rob} & \textsc{ls} & \textsc{pace}   \\ 
  \hline
        & $C_0$  & 0.2787 & 0.1715 & 0.3828 & 0.2746 & 0.1149 & 0.3011  \\ 
Model 1 & $C_{0.05}$ & 0.9717 & 8.7177 & 8.3245 & 0.8070 & 9.3234 & 8.6898  \\ 
        & $C_{0.10}$  & 3.6692 & 14.8339 & 15.1670 & 3.1779 & 16.9066 & 16.6332  \\ 
           \hline
        & $C_0$  & 0.0461 & 0.0386 & 0.0427 & 0.0510 & 0.0401 & 0.0413 \\ 
Model 2 & $C_{0.05}$ & 0.3319 & 1.9847 & 1.9123 & 0.4021 & 1.5955 & 1.5153 \\ 
        & $C_{0.10}$ & 0.8653 & 4.2007 & 3.8671 & 0.8770 & 3.3106 & 3.0065   \\ 
   \hline
 \end{tabular}
\caption{\label{tab:mse-model-123}  
Average of $\mbox{MSE}_{\ell}( \xi_k )$ over $\ell = 1, \ldots, 500$, for $k = 1, 2$. } 
\end{table}

\begin{table}[ht!]
\centering
 \setlength{\tabcolsep}{3pt} 
\small
\begin{tabular}{c|c|ccc|ccc|}
  \hline
 & & \multicolumn{3}{c|}{$k=1$} & \multicolumn{3}{c|}{$k=2$} \\ 
  \hline
  & & \textsc{rob} & \textsc{ls} & \textsc{pace}  & \textsc{rob} & \textsc{ls} & \textsc{pace}   \\ 
  \hline
           & $C_0$  & 0.1864 & 0.1560 & 0.3640 & 0.1315 & 0.1004 & 0.2823  \\ 
Model 1 & $C_{0.05}$ & 0.3960 & 9.2000 & 8.7213 & 0.3944 & 9.6801 & 8.8836   \\ 
        & $C_{0.10}$ & 0.8955 & 14.8595 & 15.2054 & 0.9741 & 16.6023 & 16.3510  \\ 
   \hline
           & $C_0$  & 0.0425 & 0.0352 & 0.0397 & 0.0470 & 0.0385 & 0.0367  \\ 
Model 2 & $C_{0.05}$ & 0.2421 & 1.9905 & 1.9149 & 0.2965 & 1.5463 & 1.4494   \\ 
        & $C_{0.10}$ & 0.5115 & 4.3153 & 3.8731 & 0.6043 & 3.3486 & 2.9270  \\ 
   \hline
 \end{tabular}
\caption{\label{tab:medse-model-123} 
Median of $\mbox{MSE}_{\ell}( \xi_k )$ over $\ell = 1, \ldots, 500$, for $k = 1, 2$. 
} 
\end{table}
    
\begin{table}[ht!]
\centering
 \setlength{\tabcolsep}{3pt} 
\small
\begin{tabular}{c|c|ccc|ccc|}
  \hline
 & & \multicolumn{3}{c|}{$k=1$} & \multicolumn{3}{c|}{$k=2$} \\  
  \hline
  & & \textsc{rob} & \textsc{ls} & \textsc{pace}  & \textsc{rob} & \textsc{ls} & \textsc{pace}   \\ 
  \hline
        & $C_{0}$    & 0.2787 & 0.1715 & 0.3828 & 0.2746 & 0.1149 & 0.3011  \\ 
Model 1 & $C_{0.05}$ & 0.3259 & 3.8781 & 3.6919 & 0.3866 & 3.2214 & 3.1439  \\ 
        & $C_{0.10}$ & 0.7847 & 5.6363 & 5.6406 & 0.9642 & 3.4723 & 3.8266  \\ 
   \hline
        & $C_{0}$    & 0.0461 & 0.0386 & 0.0427 & 0.0510 & 0.0401 & 0.0413   \\ 
Model 2 & $C_{0.05}$ & 0.0578 & 0.6877 & 0.7245 & 0.0876 & 0.5497 & 0.4676   \\ 
        & $C_{0.10}$ & 0.1226 & 1.4081 & 1.3645 & 0.1936 & 0.6895 & 0.5456   \\ 
   \hline
  \end{tabular}
\caption{\label{tab:M2-model-123} Summary measure $M_2$ for  the scores.} 
\end{table}

\begin{figure}[ht!]
 \begin{center}
 \footnotesize
 \renewcommand{\arraystretch}{0.4}
  \newcolumntype{M}{>{\centering\arraybackslash}m{\dimexpr.1\linewidth-1\tabcolsep}}
     \newcolumntype{G}{>{\centering\arraybackslash}m{\dimexpr.5\linewidth-1\tabcolsep}}
\begin{tabular}{M GG}
 & Model 1 & Model 2   \\
$C_0$ &
\includegraphics[scale=0.4]{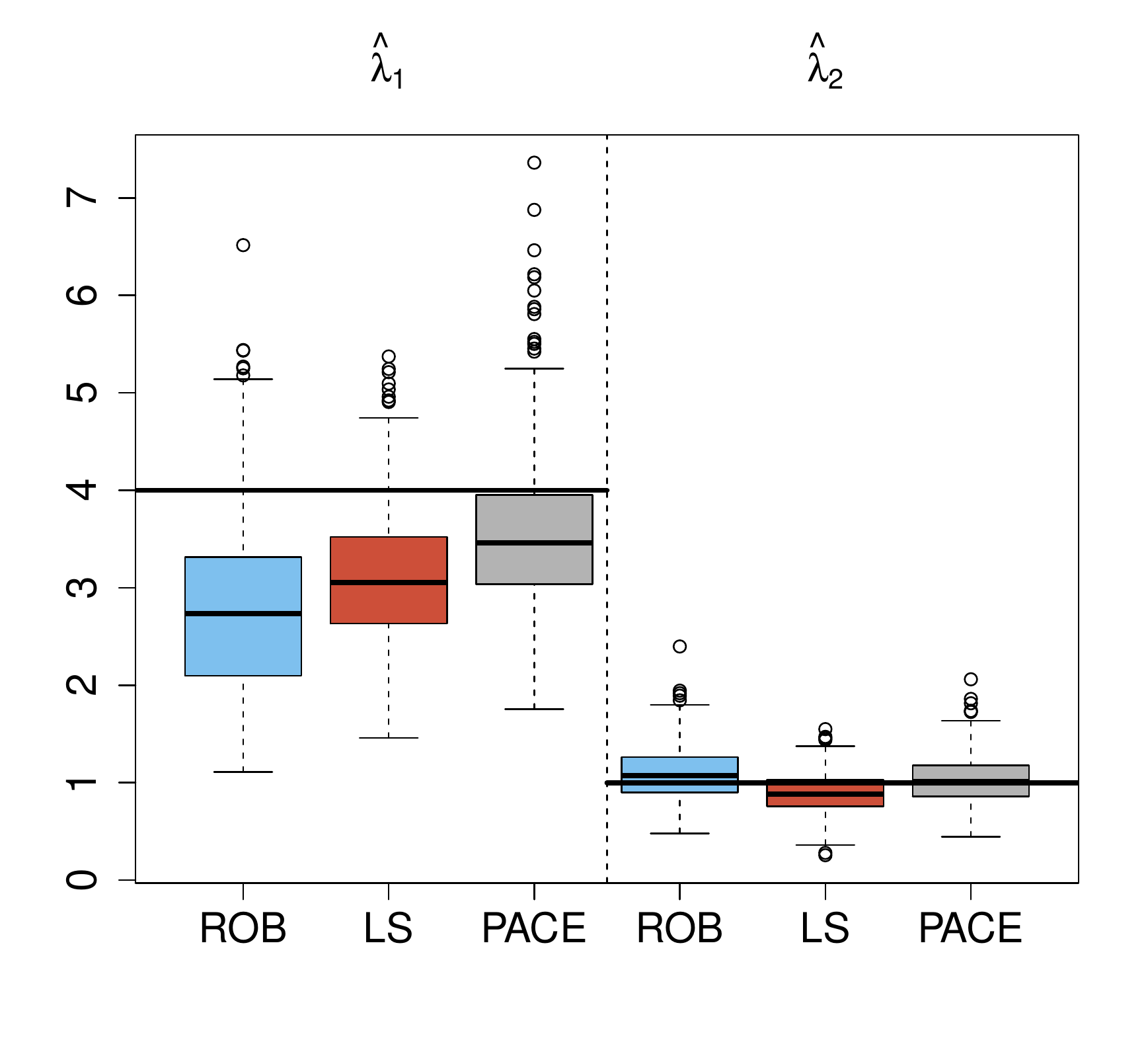} & 
\includegraphics[scale=0.4]{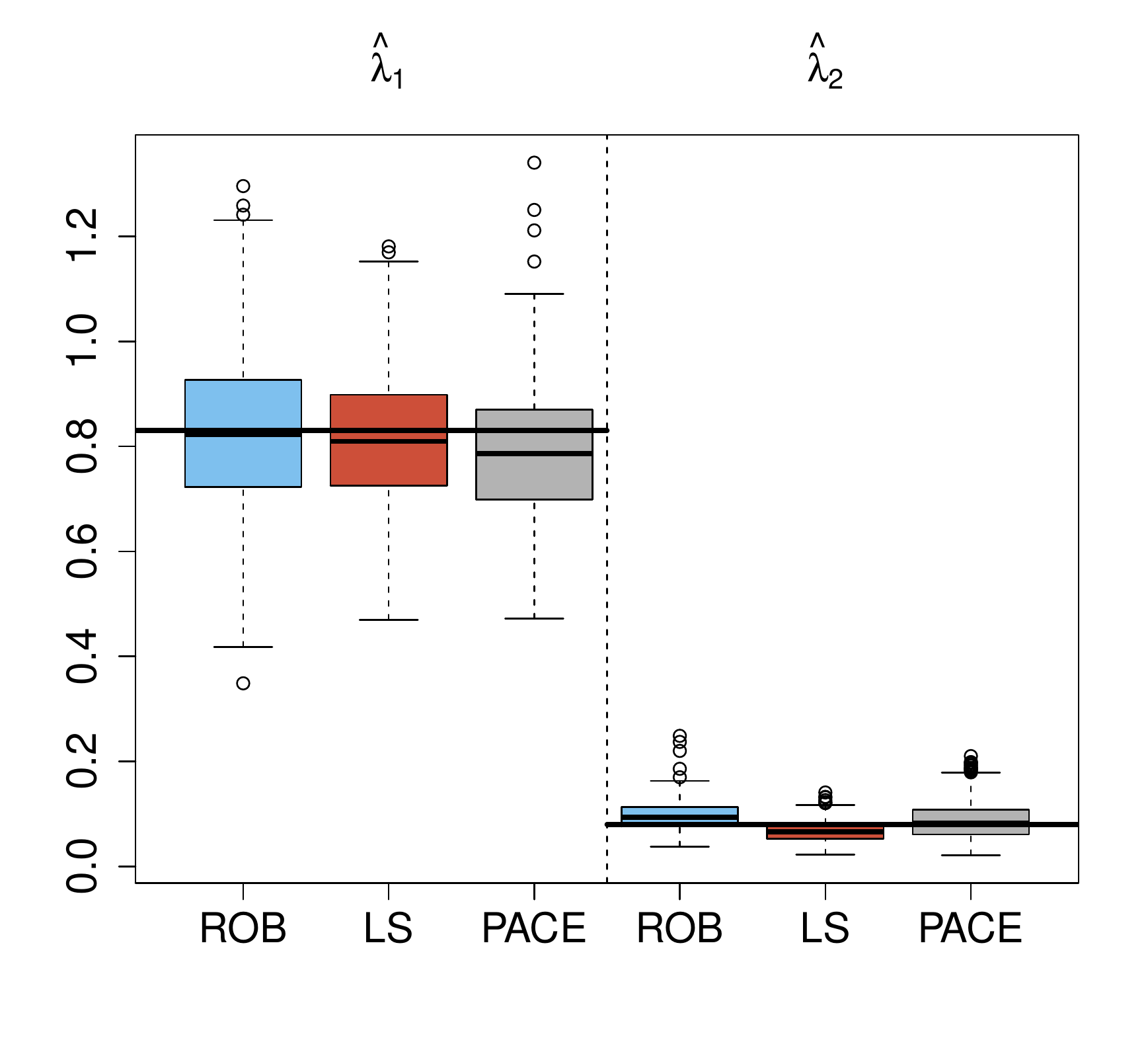}  \\
$C_{0.05}$ & 
\includegraphics[scale=0.4]{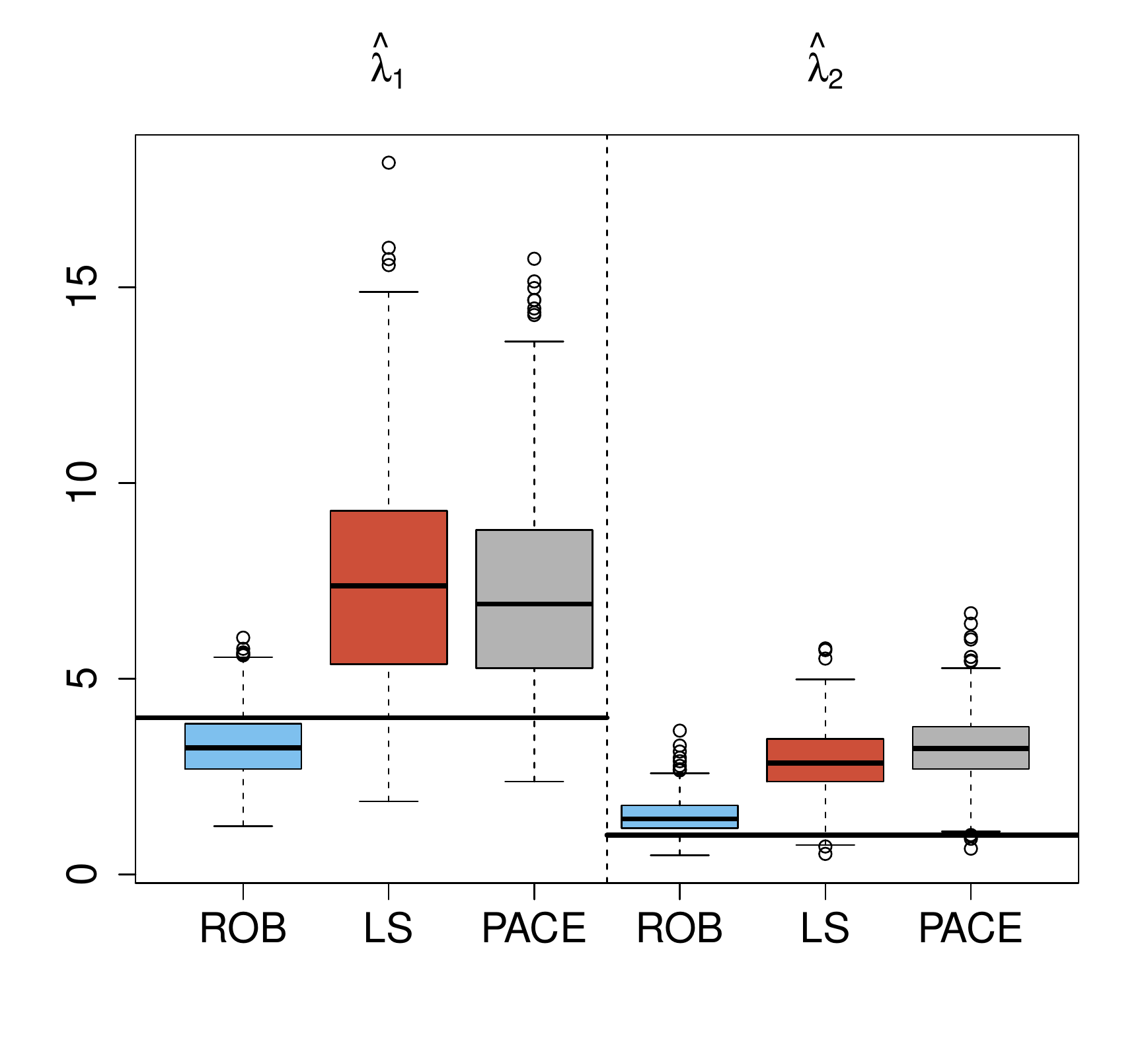} & 
\includegraphics[scale=0.4]{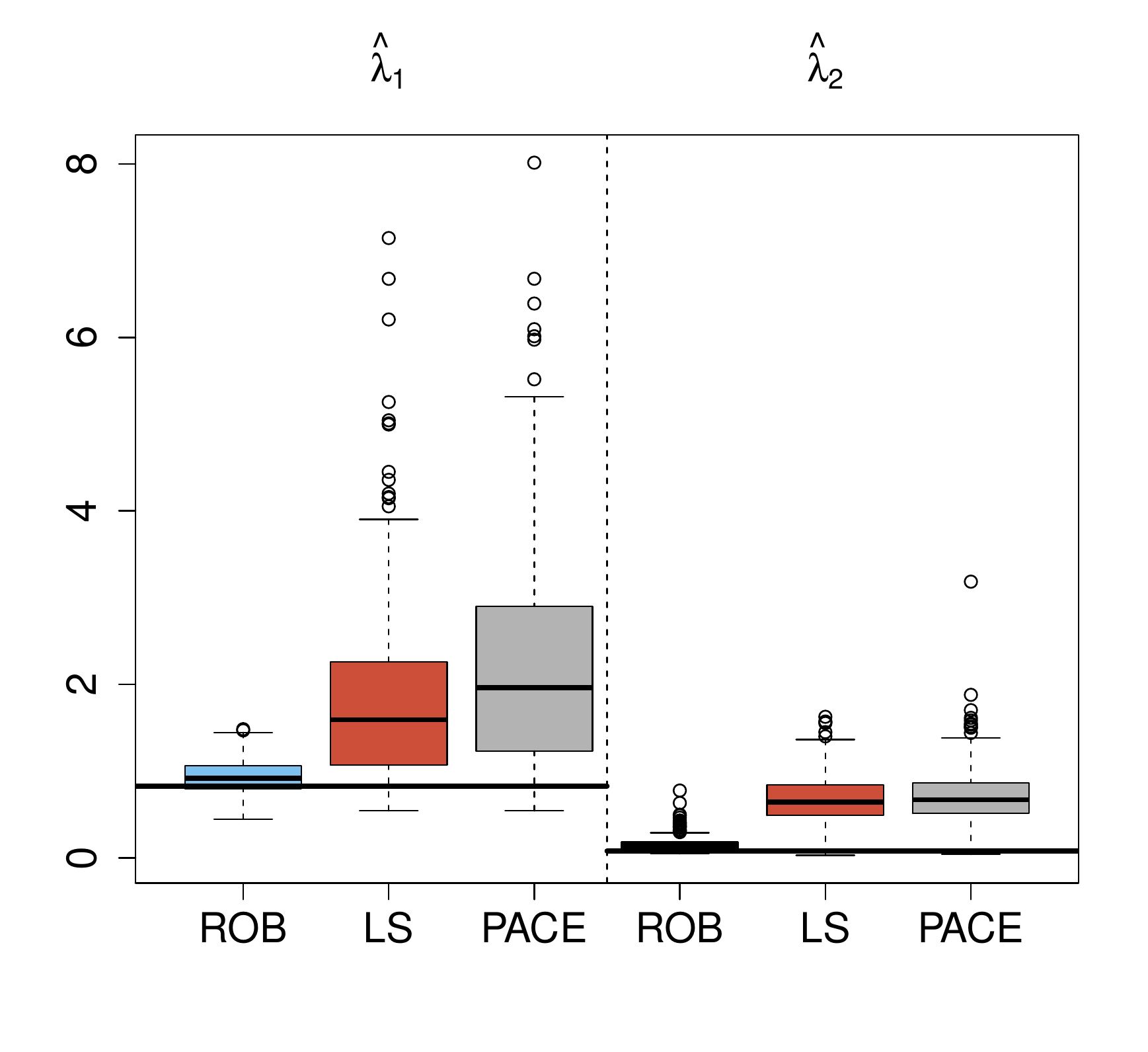}   \\
$C_{0.10}$ & 
\includegraphics[scale=0.4]{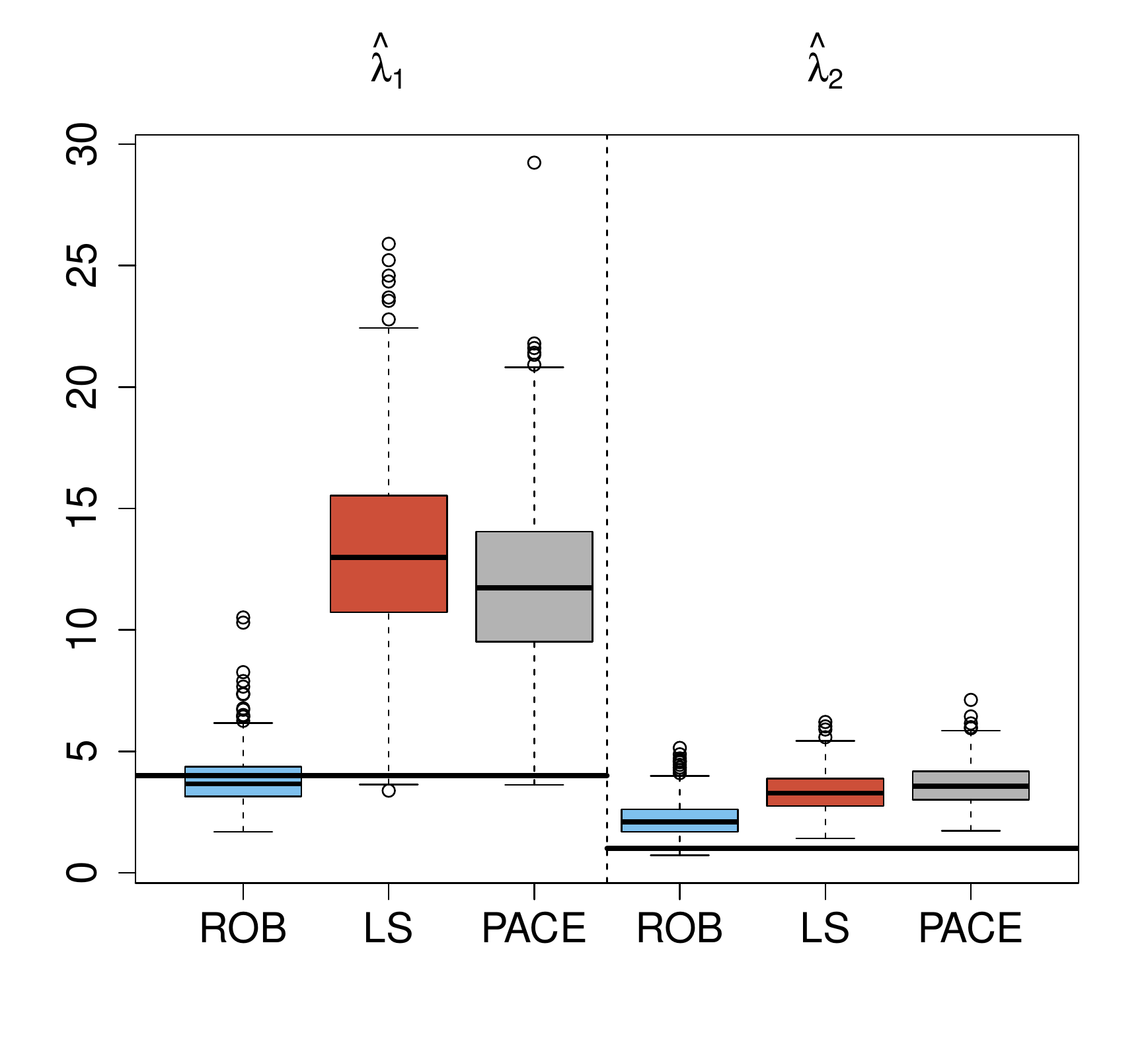}&
\includegraphics[scale=0.4]{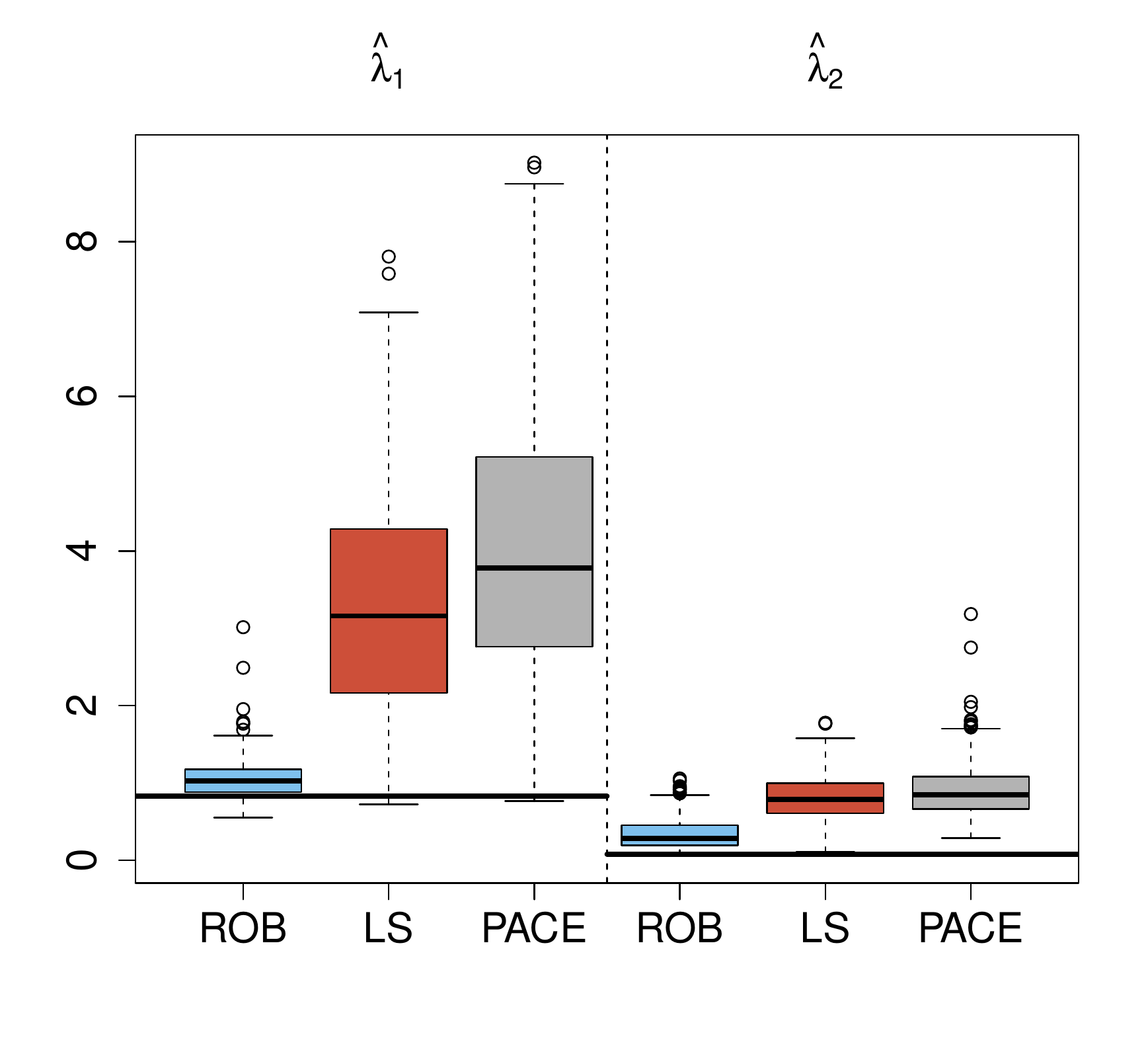}   
\end{tabular}
\caption{\label{fig:lambdas-12-model-123} Boxplots of $ \wlam_j $, for $j=1,2$.  The
horizontal lines indicate the value of the true eigenvalues $\lambda_j$. }
\end{center}
\end{figure}

\begin{figure}[ht!]
 \begin{center}
 \footnotesize
 \renewcommand{\arraystretch}{0.4}
\newcolumntype{M}{>{\centering\arraybackslash}m{\dimexpr.1\linewidth-1\tabcolsep}}
     \newcolumntype{G}{>{\centering\arraybackslash}m{\dimexpr.5\linewidth-1\tabcolsep}}
\begin{tabular}{M GG}
 & Model 1 & Model 2  \\
$C_0$ &
\includegraphics[scale=0.4]{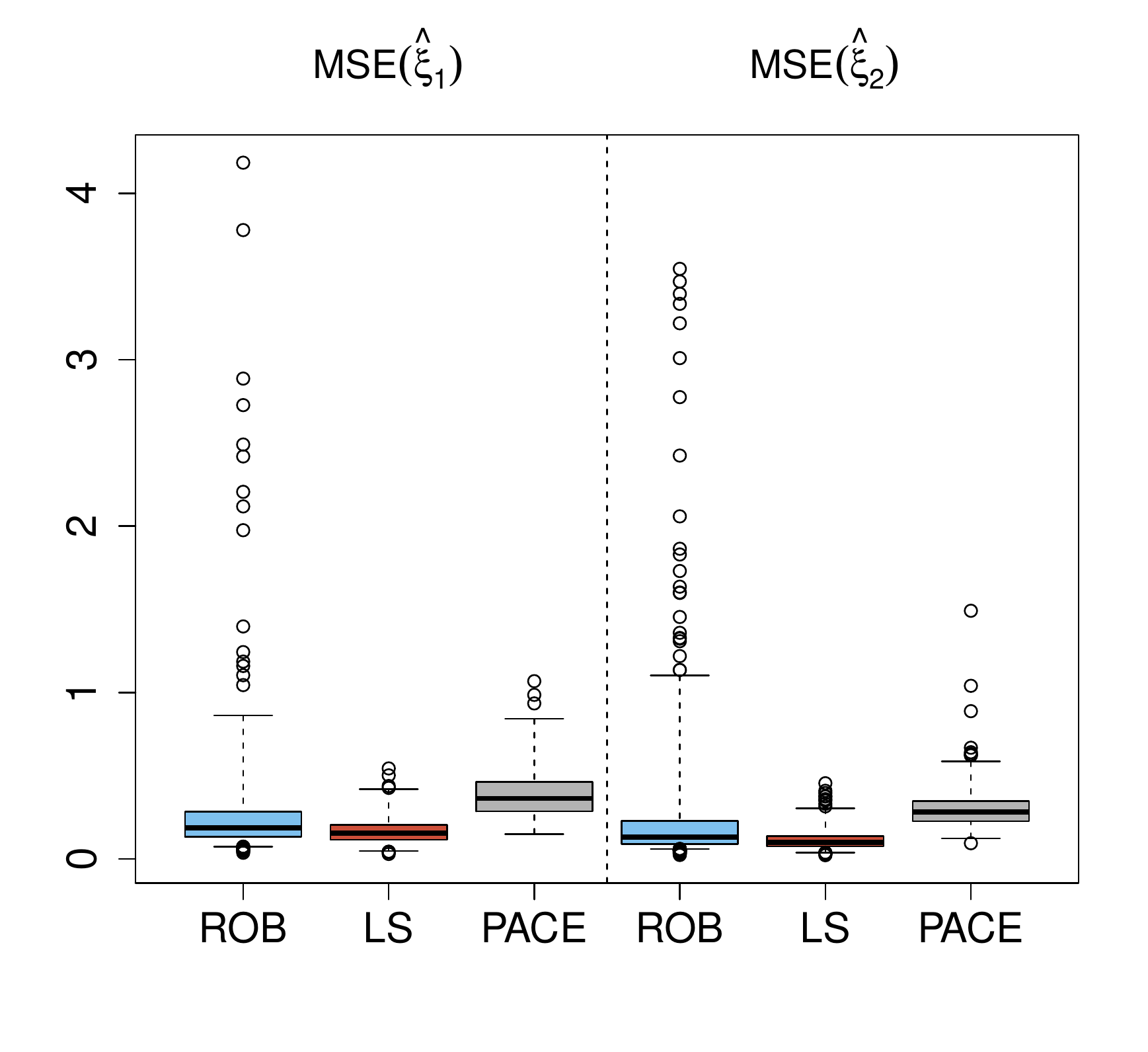}& 
\includegraphics[scale=0.4]{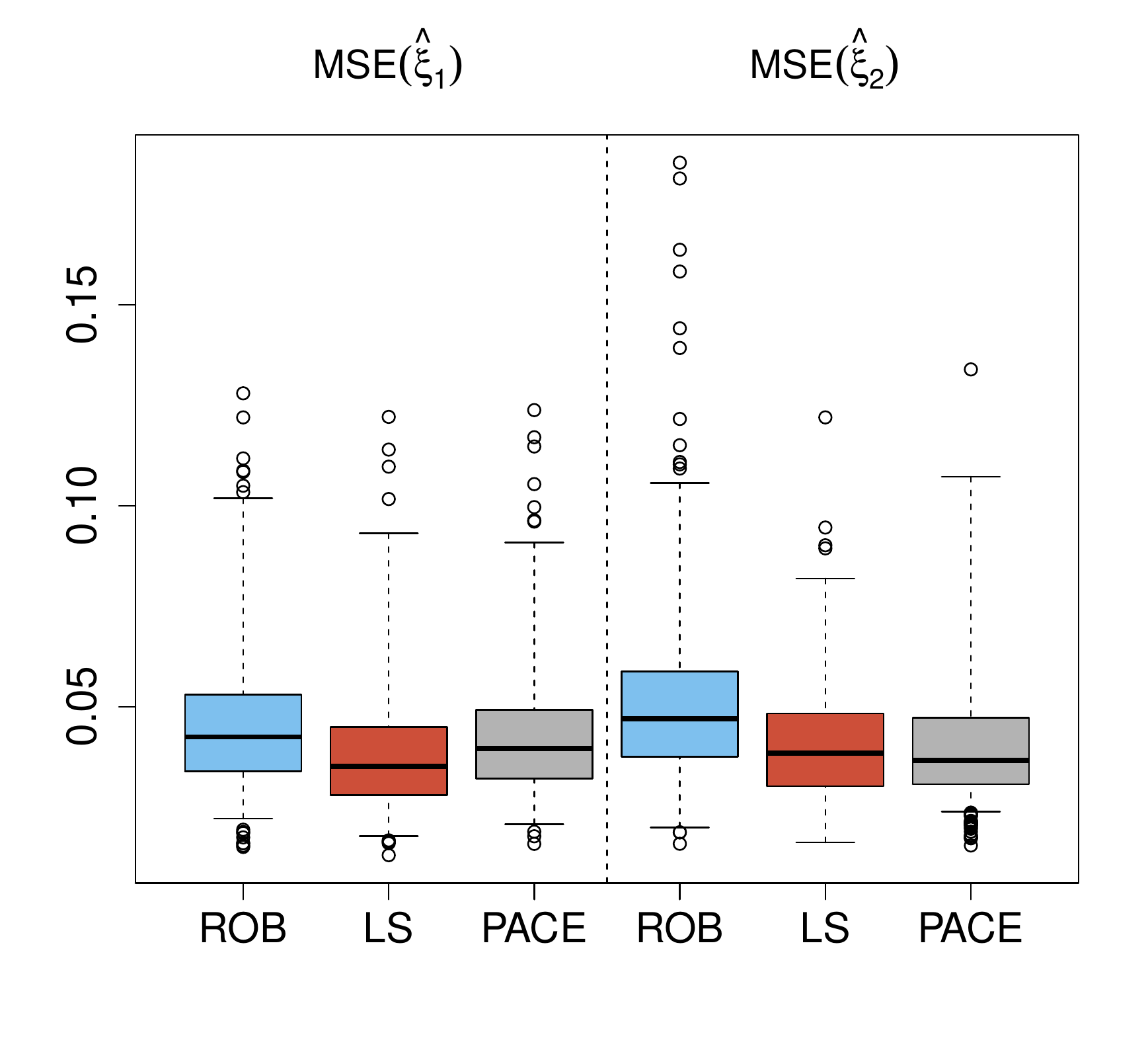}   
\\
$C_{0.05}$ & 
\includegraphics[scale=0.4]{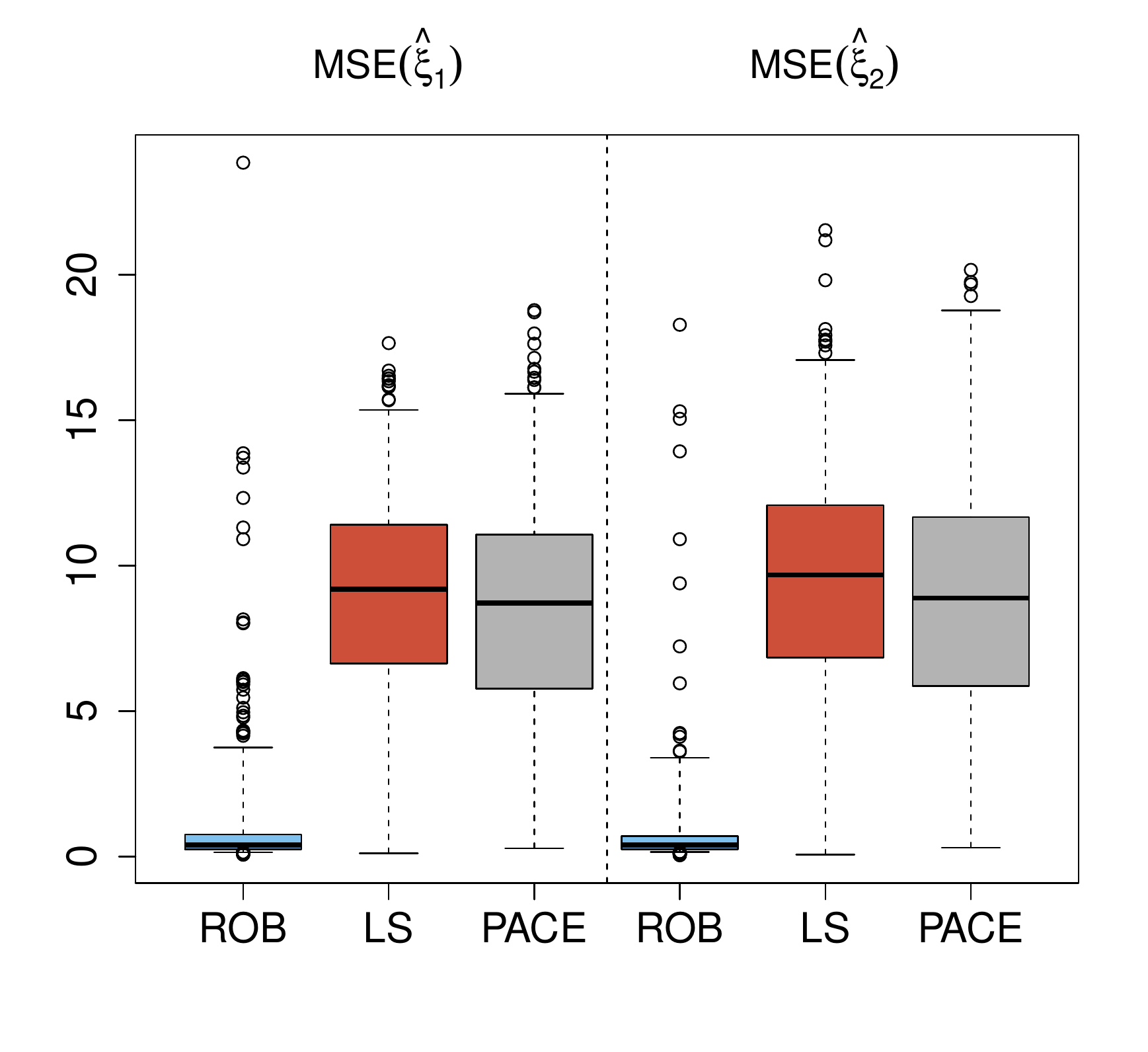} & 
\includegraphics[scale=0.4]{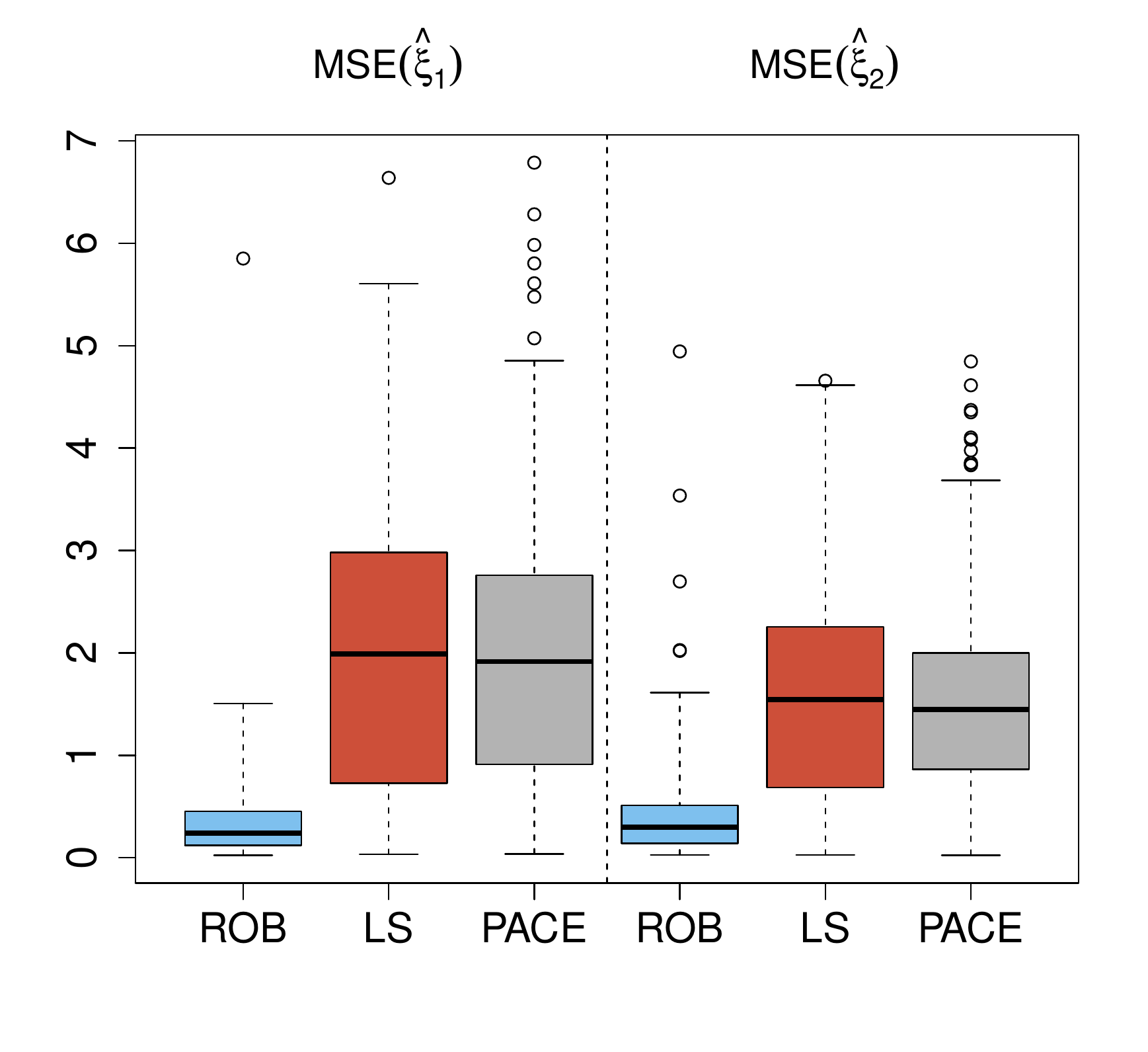}   
\\
$C_{0.10}$ & 
\includegraphics[scale=0.4]{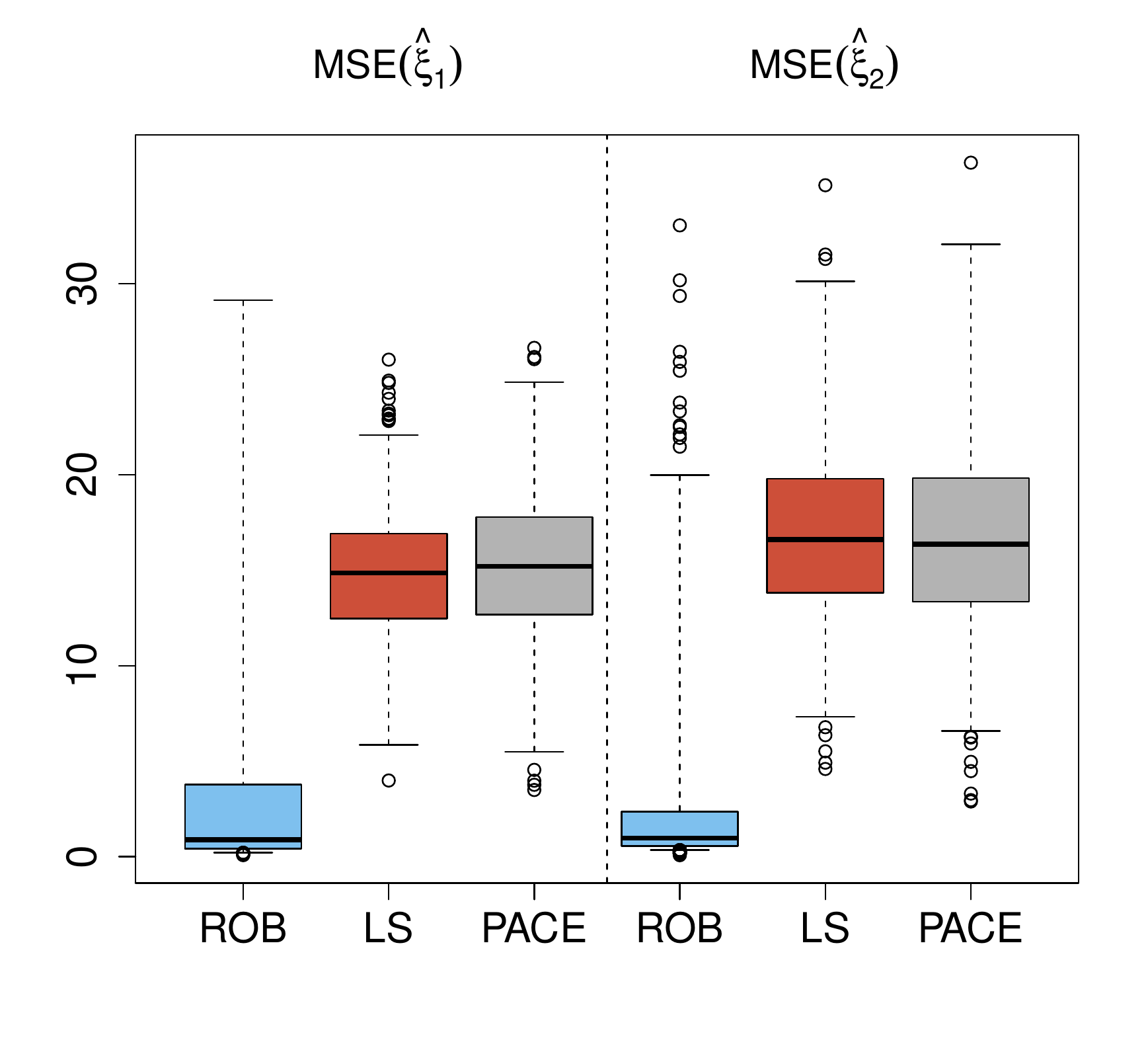}&
\includegraphics[scale=0.4]{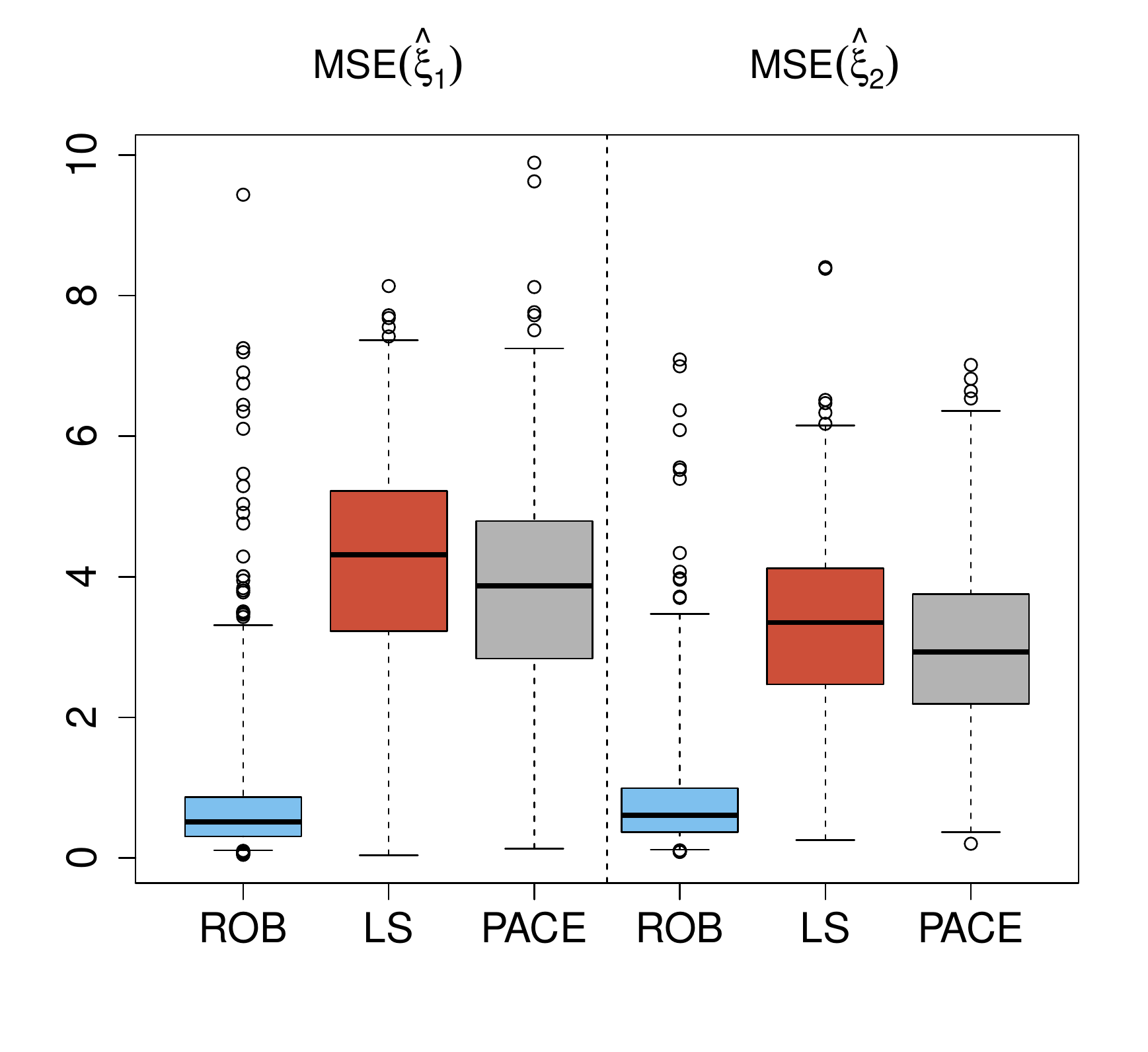}  

\end{tabular}
\caption{\label{fig:scores-12-model-123-sigcor} Adjusted boxplots of the mean scores distance  $\mbox{MSE}_{\ell}( \xi_k )$, $k=1,2$.}

\end{center}
\end{figure}

\begin{figure}[ht!]
 \begin{center}
 \footnotesize
 \renewcommand{\arraystretch}{0.4}
\newcolumntype{M}{>{\centering\arraybackslash}m{\dimexpr.1\linewidth-1\tabcolsep}}
     \newcolumntype{G}{>{\centering\arraybackslash}m{\dimexpr.5\linewidth-1\tabcolsep}}
\begin{tabular}{M GG}
 & Model 1 & Model 2  \\
$C_0$ &
\includegraphics[scale=0.4]{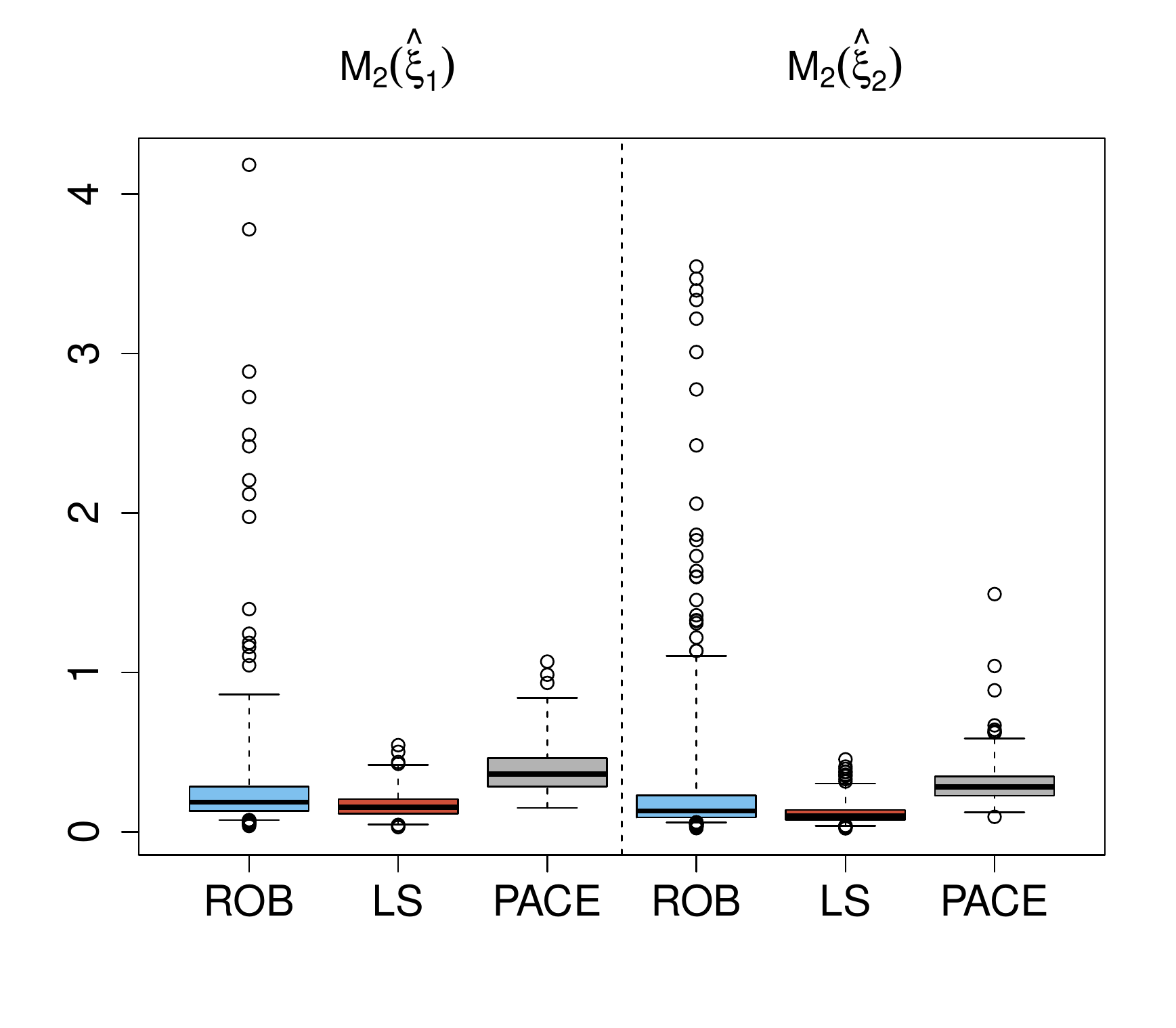} & 
\includegraphics[scale=0.4]{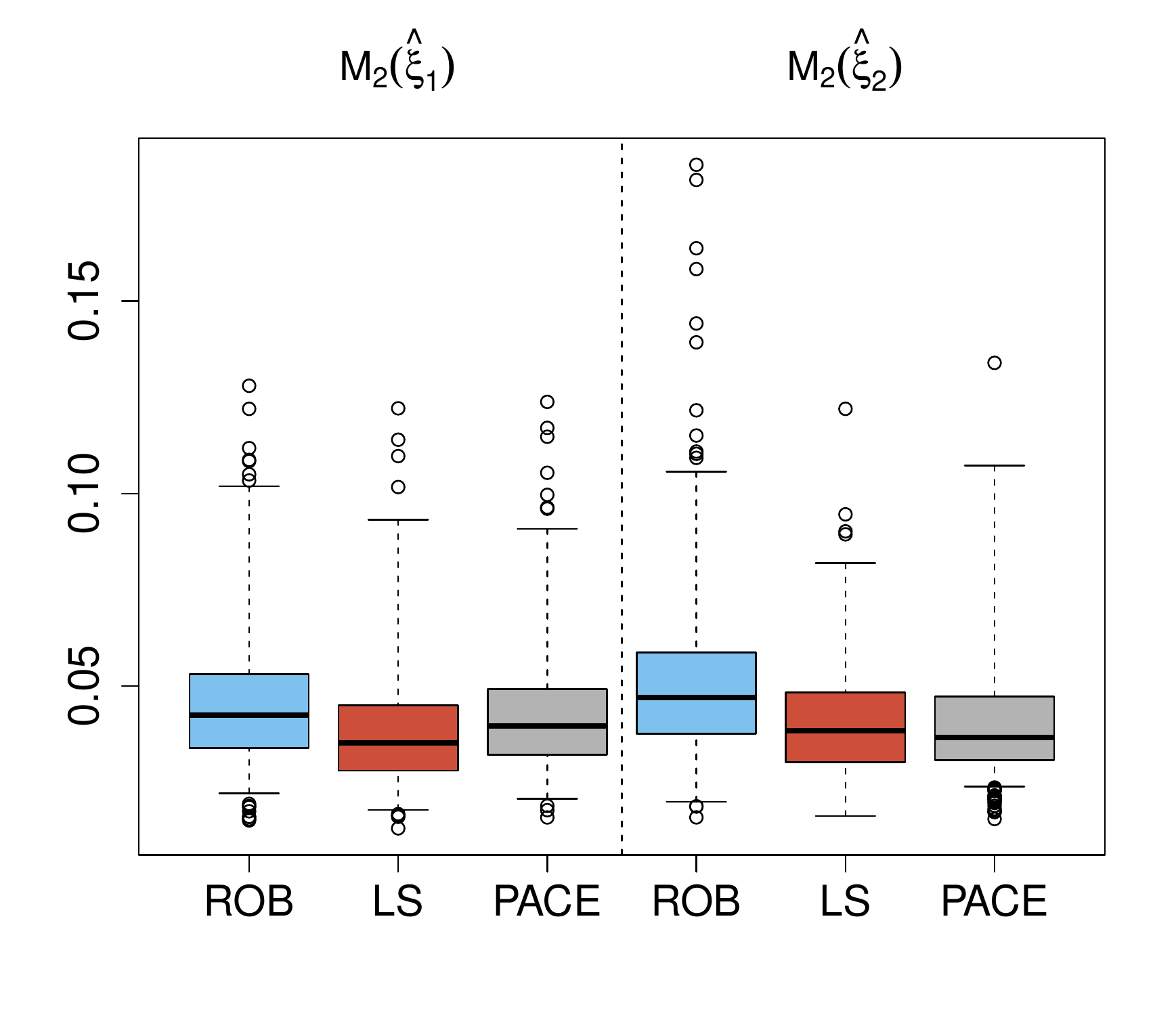}  
\\
$C_{0.05}$ & 
\includegraphics[scale=0.4]{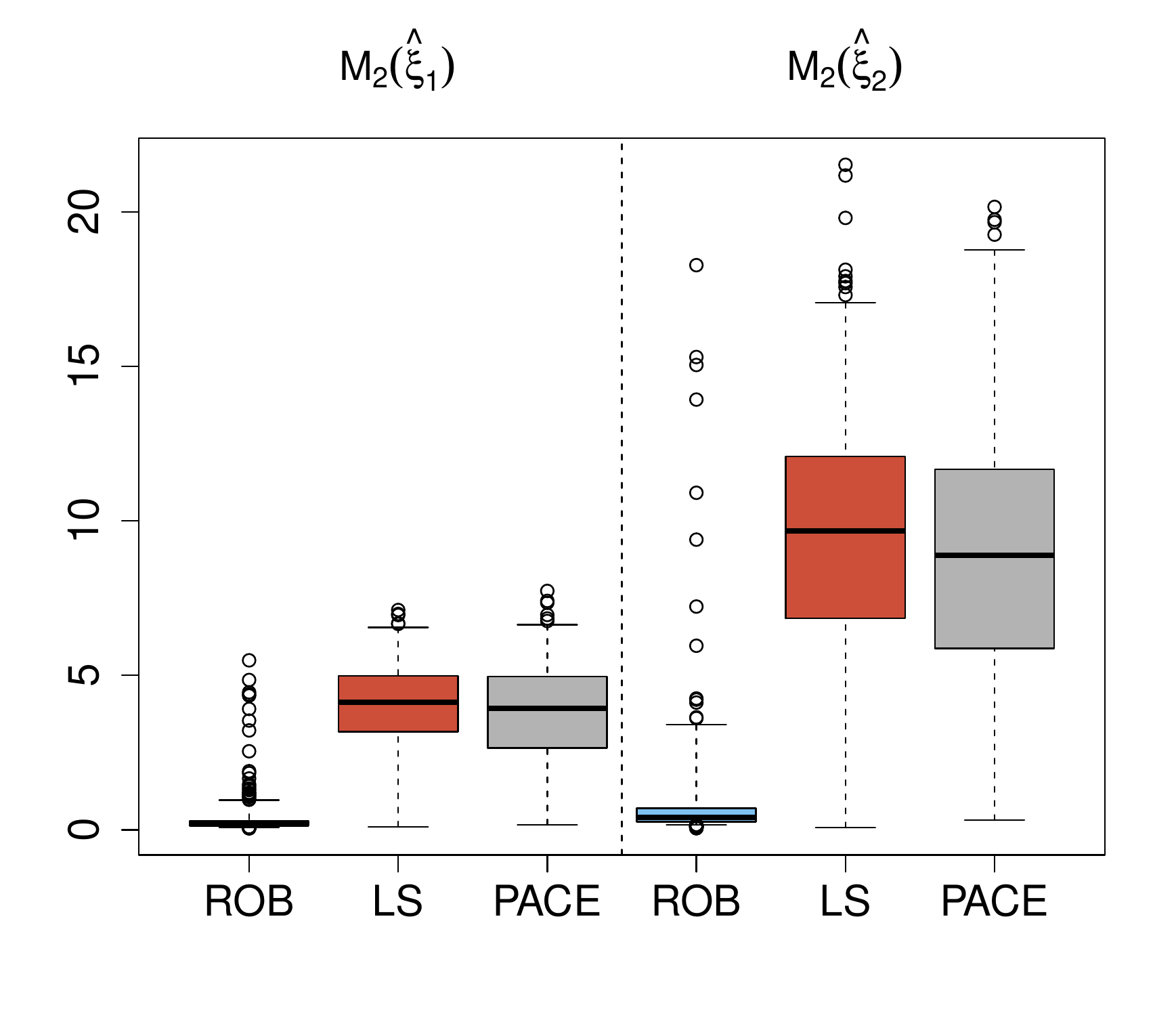}  & 
\includegraphics[scale=0.4]{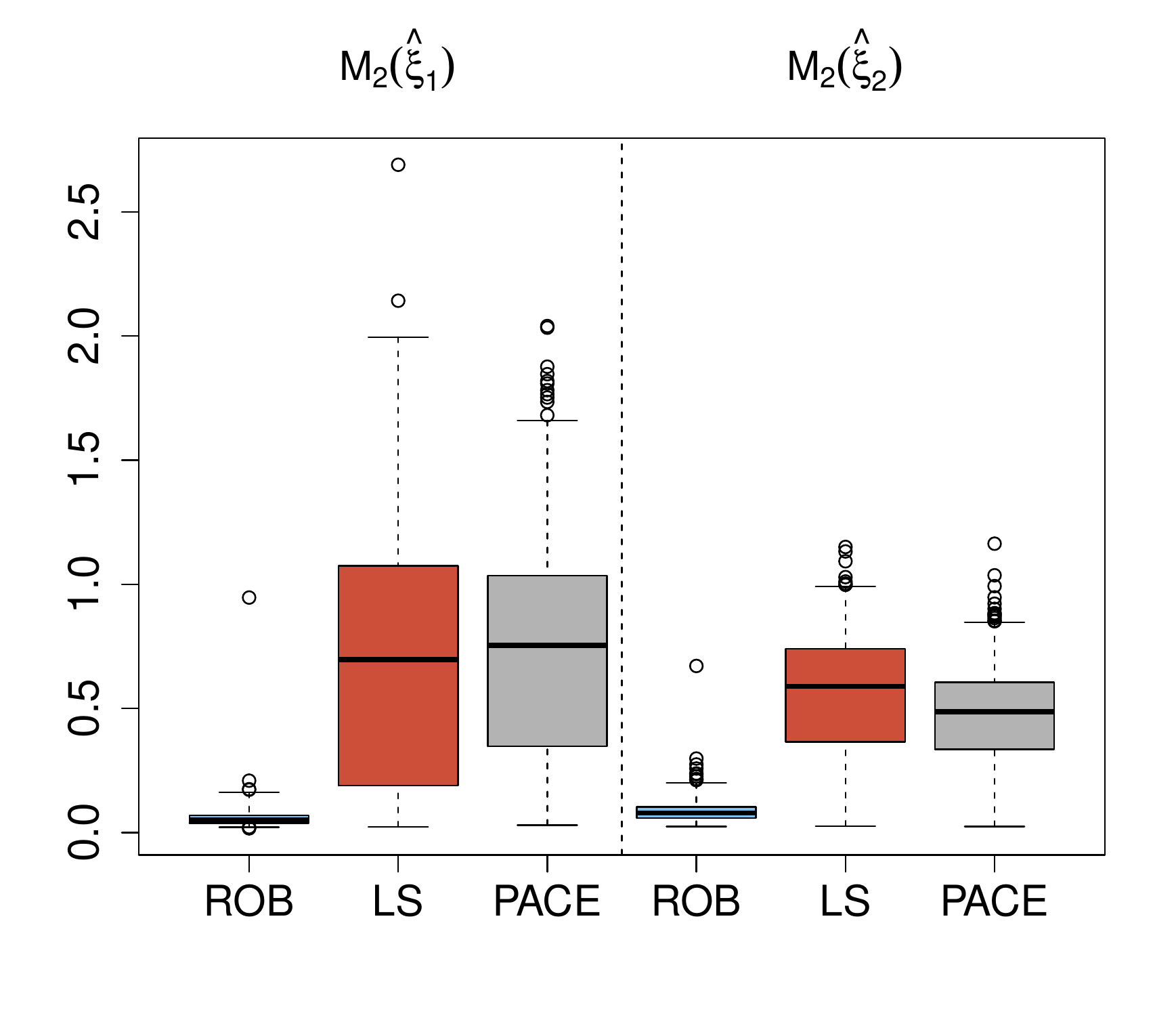}   
\\
$C_{0.10}$ & 
\includegraphics[scale=0.4]{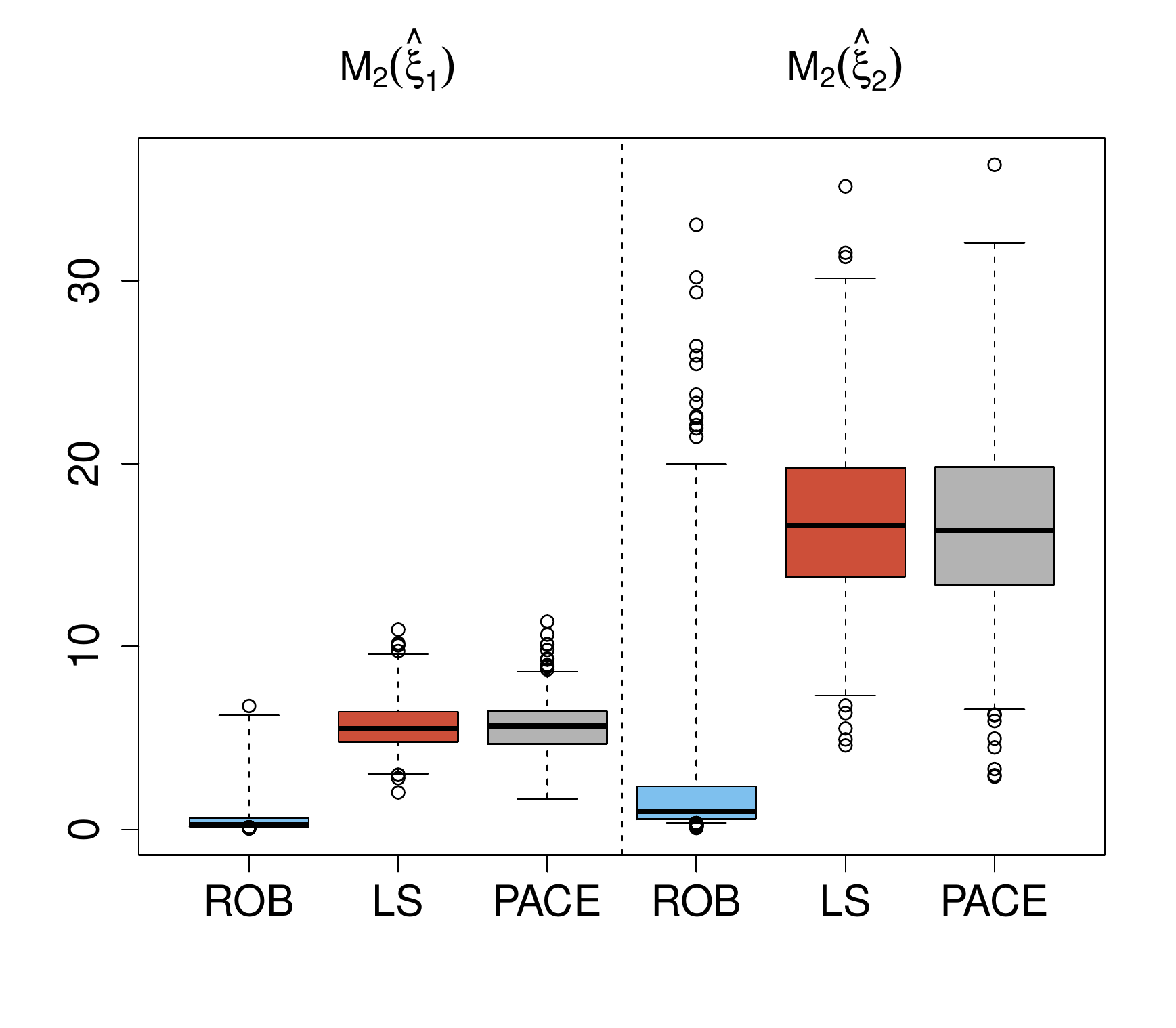}&
\includegraphics[scale=0.4]{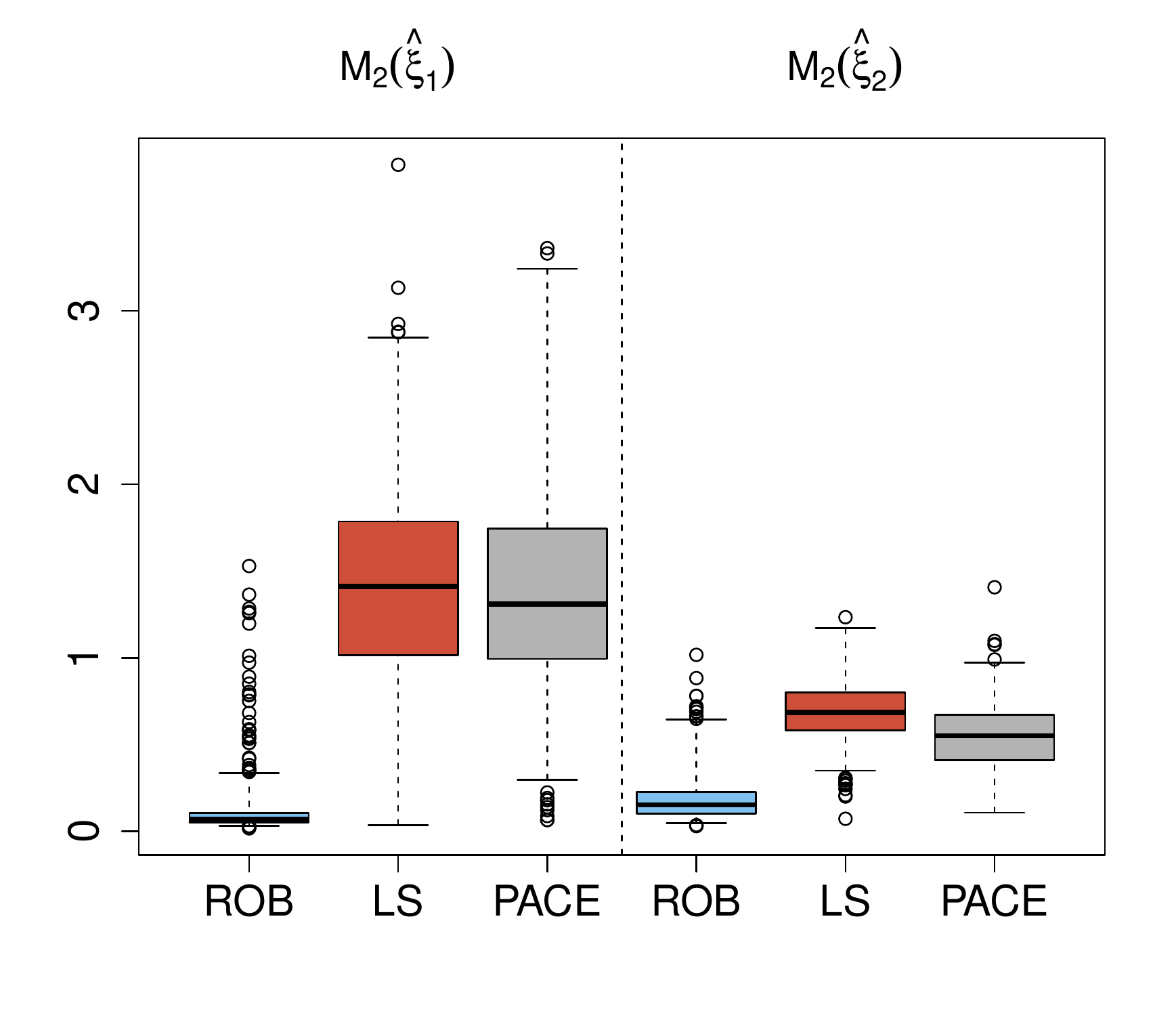}  
\end{tabular}
\caption{\label{fig:M2-12-model-123-sigcor}  Adjusted boxplots of $M_{2,\ell}(\xi_k)$ for $k=1,2$.}
\end{center}
\end{figure}

\clearpage
\section{A real data example}{\label{sec:realdata}}
 
\begin{figure}[ht!]
  \begin{center}
    \newcolumntype{G}{>{\centering\arraybackslash}m{\dimexpr.5\linewidth-1\tabcolsep}}
\begin{tabular}{GG}
(a) Data & (b) Most outlying curves \\
\includegraphics[scale=0.35]{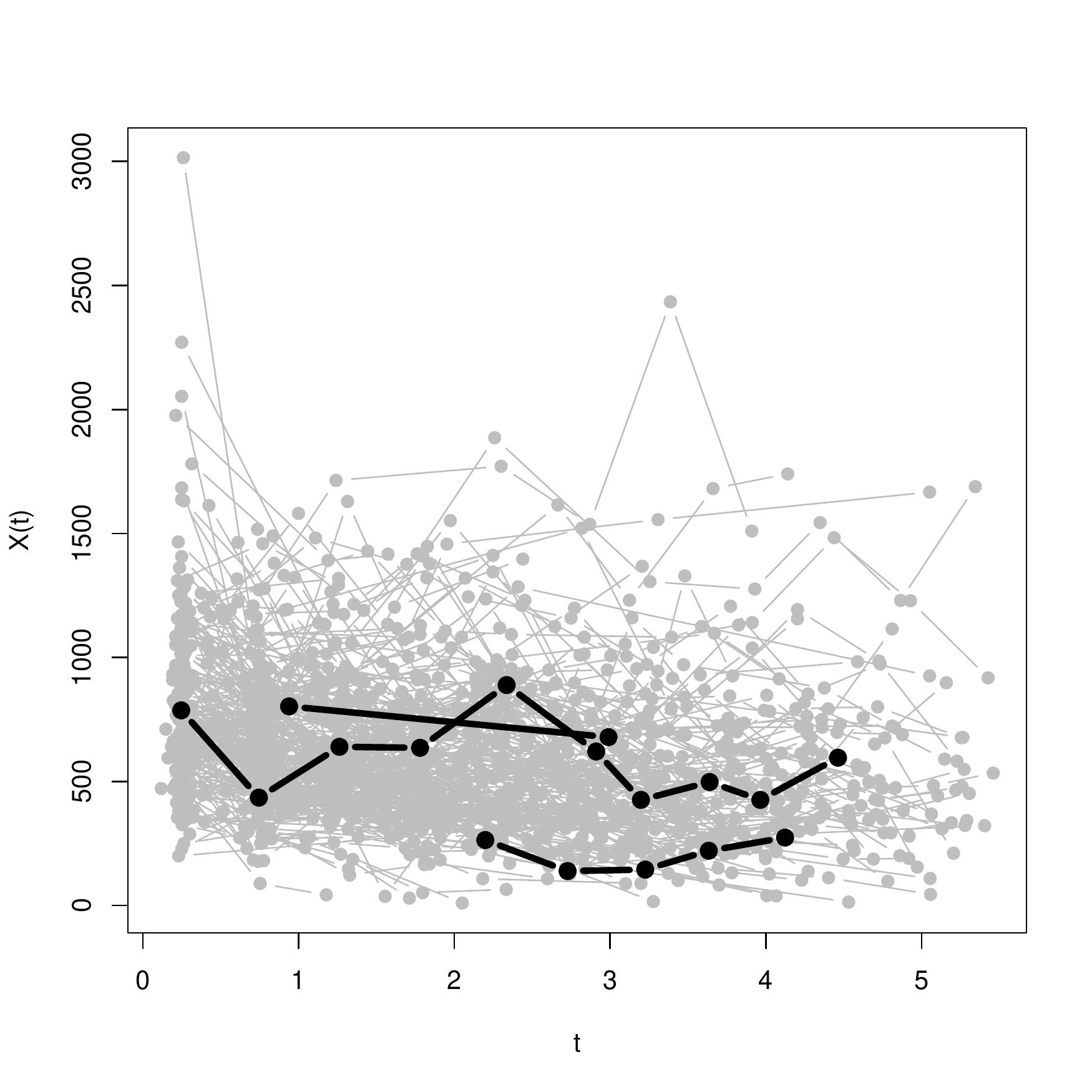} &
\includegraphics[scale=0.35]{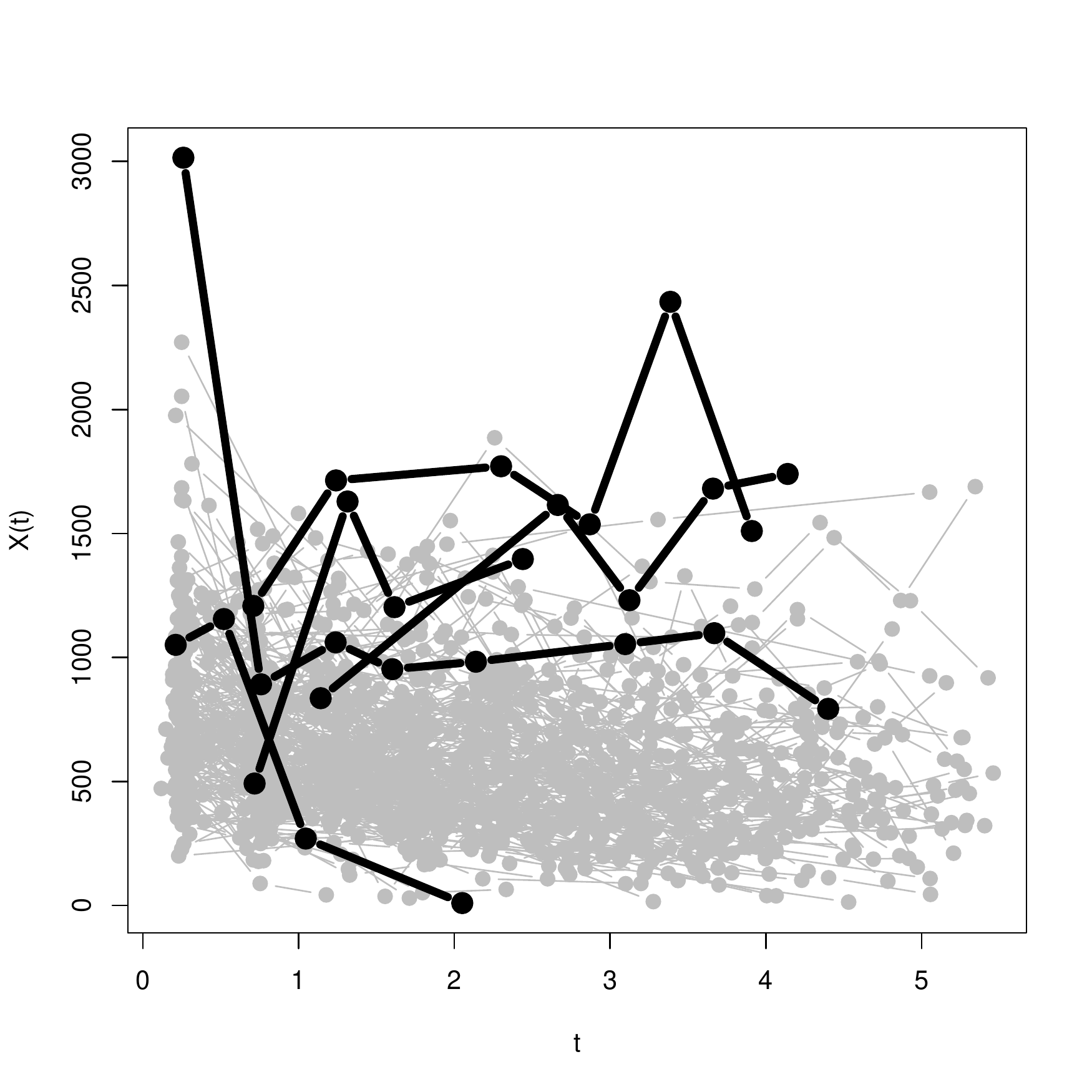} 
\end{tabular}
\caption{\label{fig:cd4data} CD4 counts for $N = 292$ 
patients where $t \ge 0$. Three 
randomly chosen curves are highlighed with solid black lines
on the left panel, while the right panel
shows the five most outlying trajectories.}
\end{center}
\end{figure}

We illustrate our method on the CD4 data, which 
is part of the Multicentre AIDS Cohort Study (Zeger and Diggle \cite{ZD}). 
The data consists of 2376 measurements of CD4 cell counts, taken on 369
men. The times are measured in years since seroconversion ($t = 0$). 
The whole data set is available from the 
\texttt{catdata} package for \texttt{R} (Schauberger and Tutz \cite{catdata}). 
To ensure that there are enough observations to estimate the 
covariance function at every pair of times $(s, t)$, we focus on the 
observations with $t \ge 0$, and on individuals with more than one 
measurement. This results in $N = 292$ curves, with the number 
of observations per individual ranging between 2 and 11 (with a median of 5).
The data set is shown in  Figure \ref{fig:cd4data}, 
with three randomly chosen trajectories highlighted with
solid black lines.

We compare the covariance function estimates obtained with our
robust FPCA method in Section \ref{sec:ourmethod}
(ROB), the corresponding non-robust 
variant (Section \ref{sec:non-robust}) (LS) and PACE
(Yao \textsl{et al.} \cite{YMW}). 
The tuning parameters and bandwidths for ROB and 
LS  were chosen as in the simulation study (Section \ref{sec:estmonte}) using
10-fold cross-validation,  while  the 
smoothing parameters for PACE were set as described in
Wang \textsl{et. al} \cite{WCM}.
 
\begin{figure}[ht!]
  \begin{center}
% \footnotesize
    \newcolumntype{G}{>{\centering\arraybackslash}m{\dimexpr.3\linewidth-1\tabcolsep}}
\begin{tabular}{GGG}
(a) ROB & (b) LS & (c) PACE \\
\includegraphics[scale=0.3]{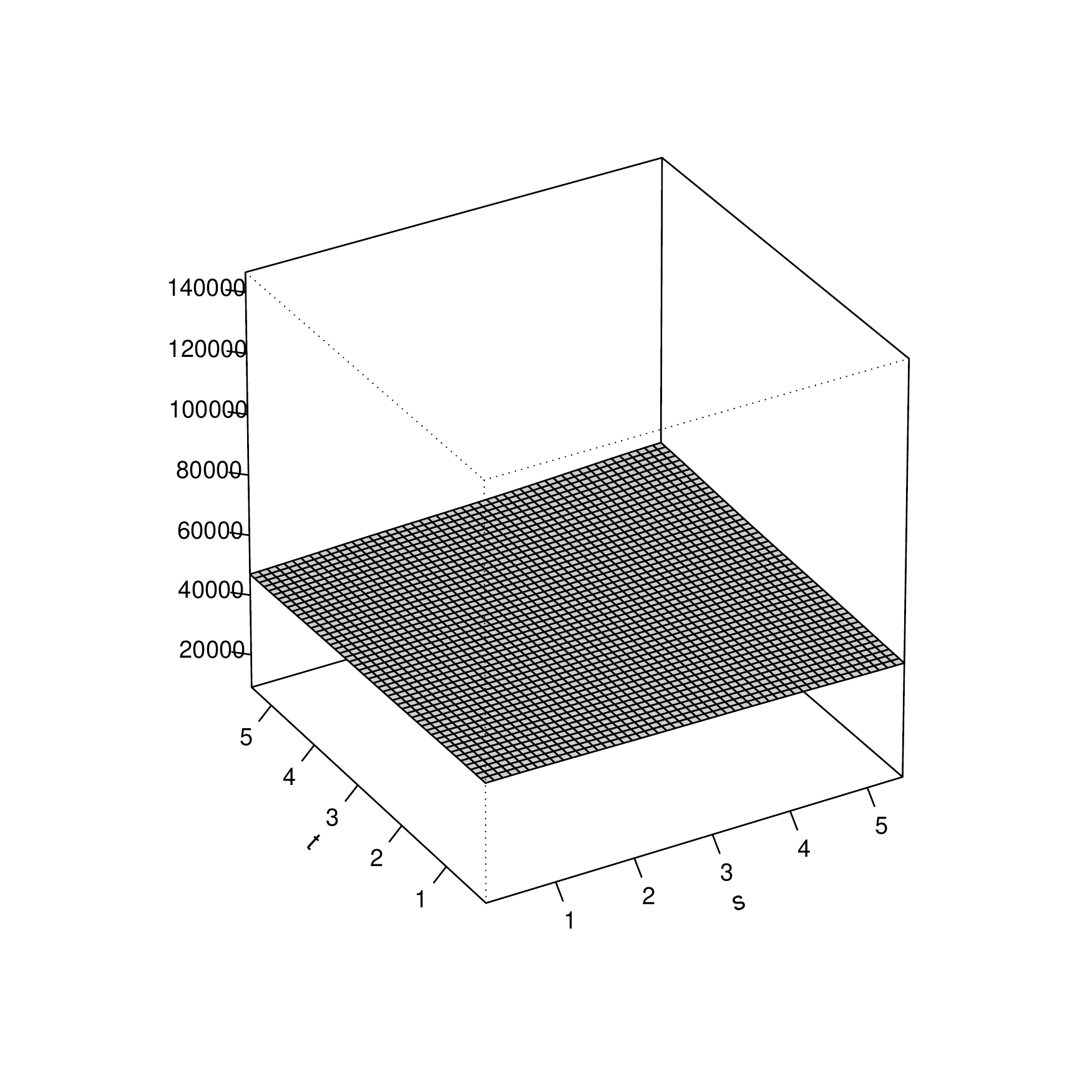} &
\includegraphics[scale=0.3]{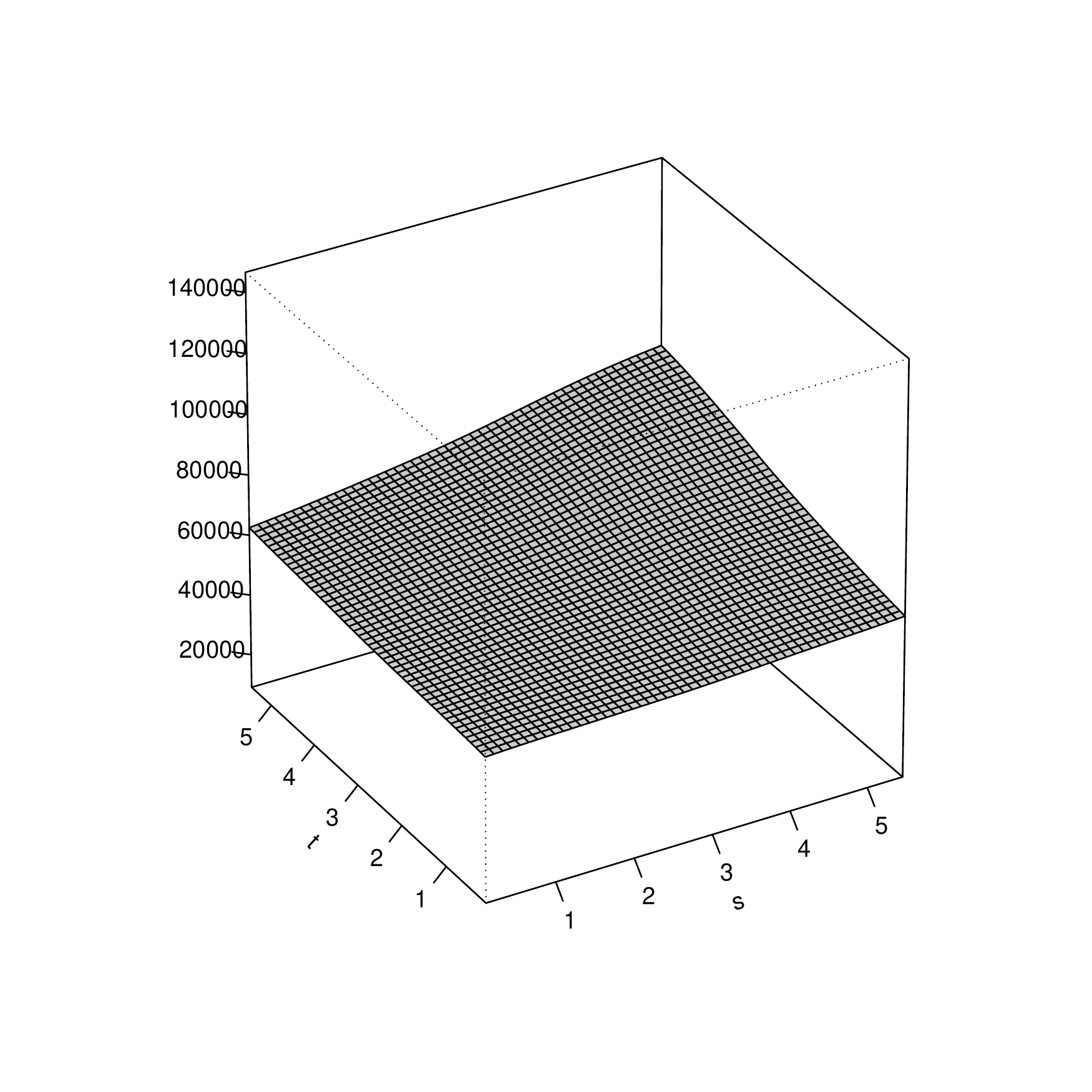} &
\includegraphics[scale=0.3]{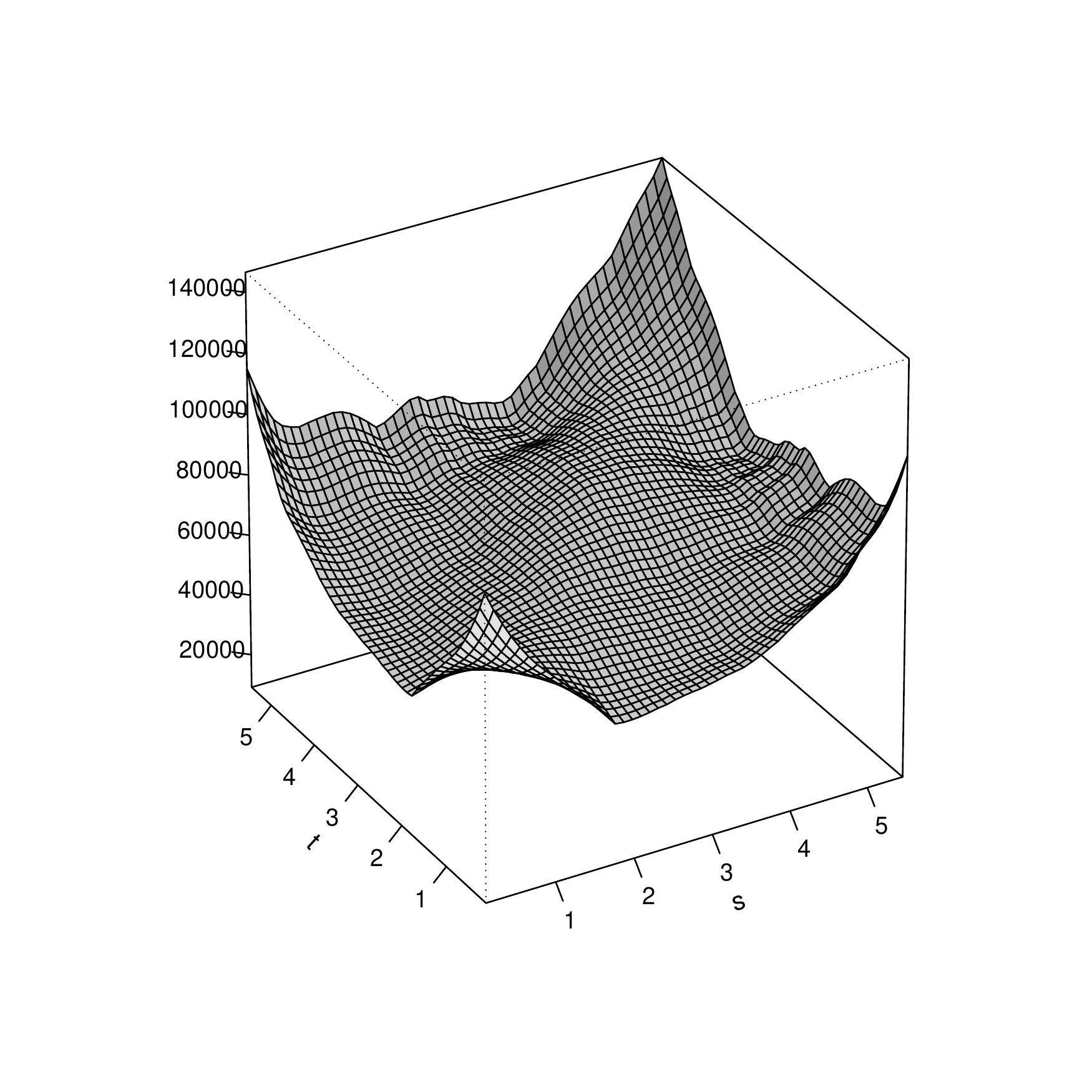}
\end{tabular}
\caption{\label{fig:covs} Estimated covariance functions for the three estimators (ROB, LS and PACE).}
\end{center}
\end{figure}

\begin{figure}[ht]
  \begin{center}
 \begin{tabular}{cc}
(a) First eigenfunctions & (b) Second eigenfunctions \\
{\includegraphics[scale=0.35]{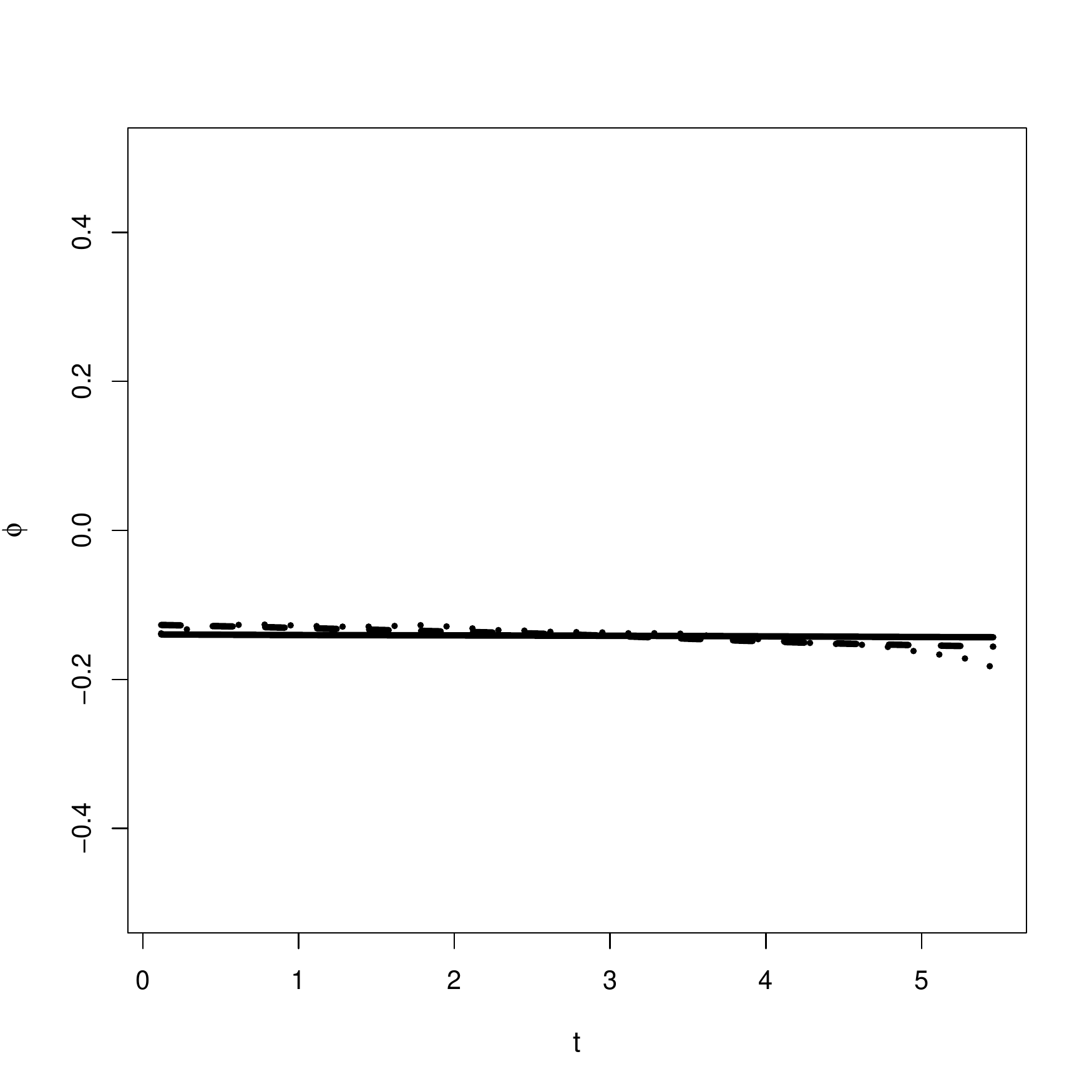}} &
{\includegraphics[scale=0.35]{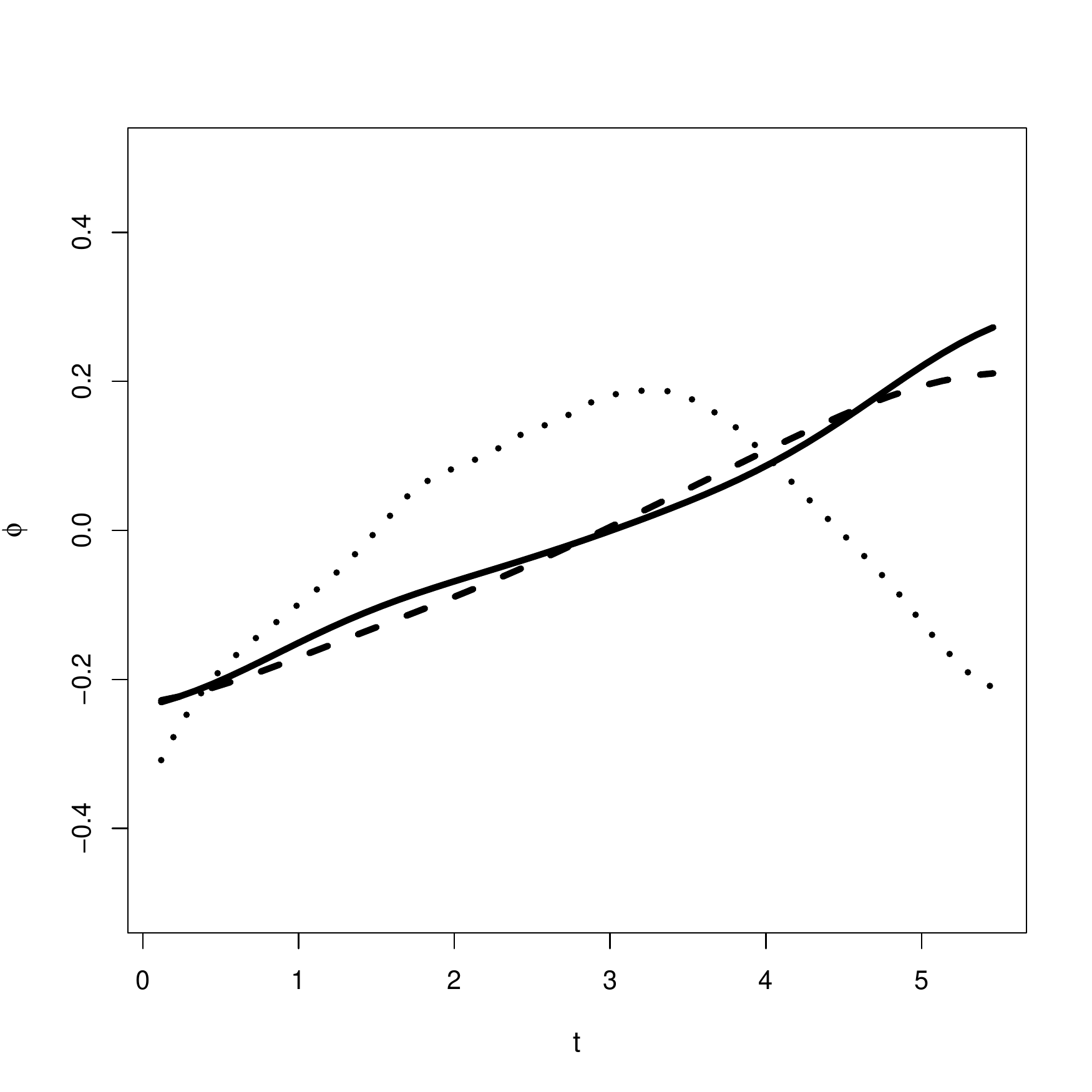}}
\end{tabular}
\caption{Estimated eigenfunctions 
associated with the two largest eigenvalues of the
corresponding covariance function estimate. 
Robust estimates (ROB) are displayed 
with solid lines, their non-robust alternatives (LS) use dashed lines, while PACE is
shown with dotted curves.} \label{fig:eigenfun} 
\end{center}
\end{figure}

The first two
principal components 
estimated with ROB and LS account for over 99\% of the 
total variability, while for PACE they reach over 94\%. 
Figures \ref{fig:covs} and \ref{fig:eigenfun} show the 
estimated covariance functions and 
eigenfunctions, respectively. 

\begin{figure}[ht!]
  \begin{center}
% \footnotesize
    \newcolumntype{G}{>{\centering\arraybackslash}m{\dimexpr.3\linewidth-1\tabcolsep}}
\begin{tabular}{GGG}
(a) ROB & (b) LS & (c) PACE\\
\includegraphics[scale=0.3]{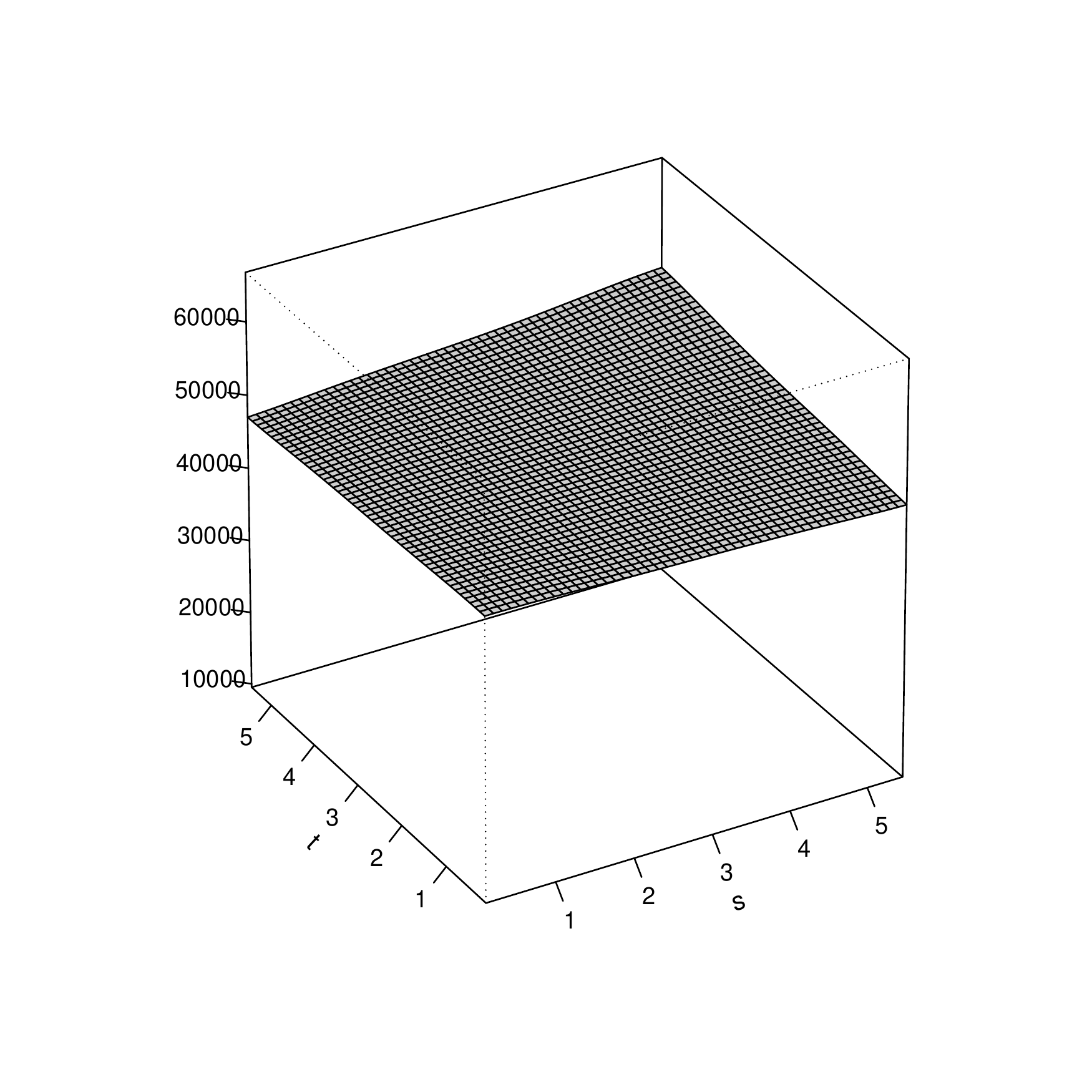} &
\includegraphics[scale=0.3]{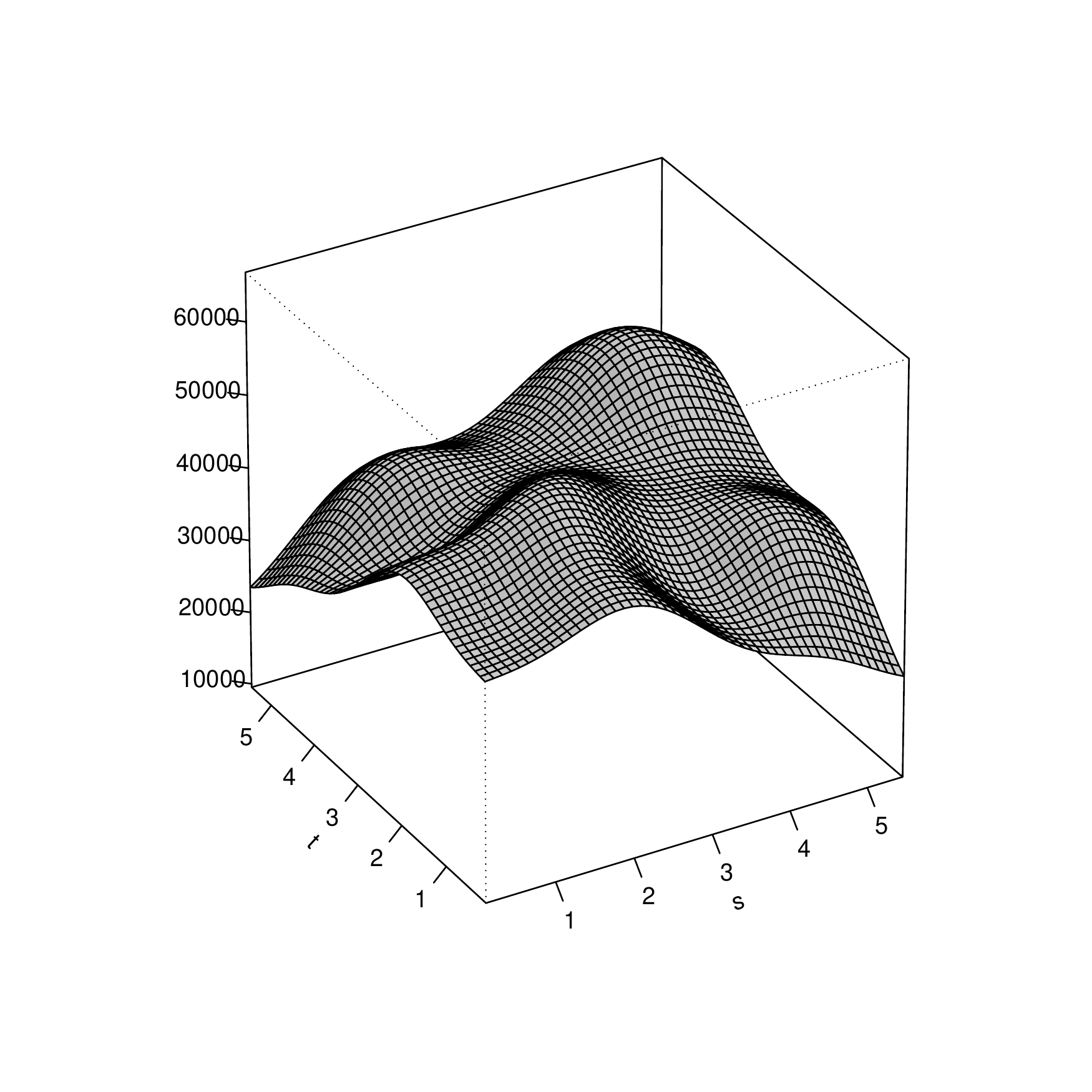} &
\includegraphics[scale=0.3]{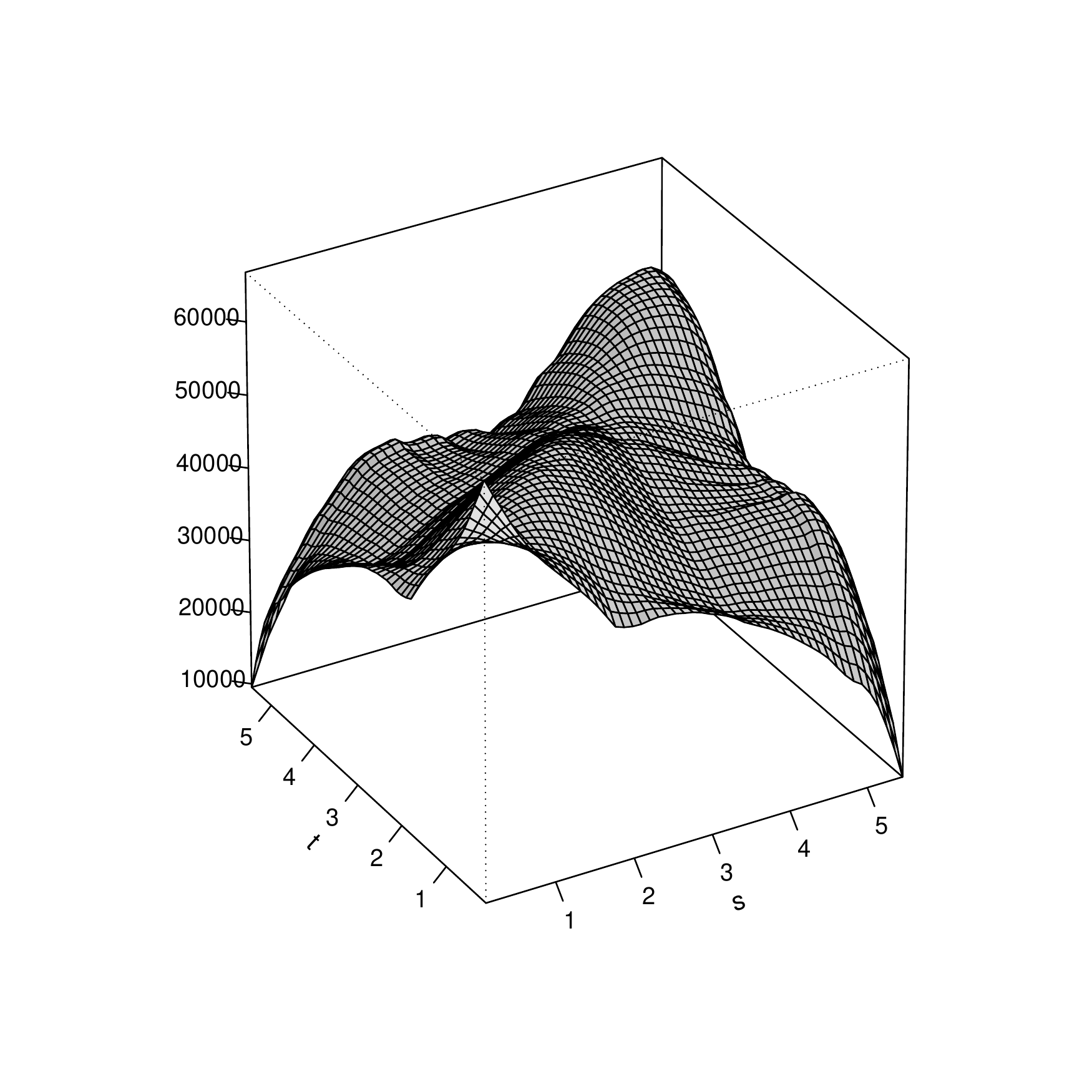}
\end{tabular}
\caption{\label{fig:covs-clean} Estimated covariance functions with LS and PACE 
after removing trajectories that were flagged as possible outliers together with the robust estimated covariance function using all the data.}
\end{center}
\end{figure}

\begin{figure}[ht]
  \begin{center}
 \begin{tabular}{cc}
(a) First eigenfunctions & (b) Second eigenfunctions \\
  {\includegraphics[scale=0.35]{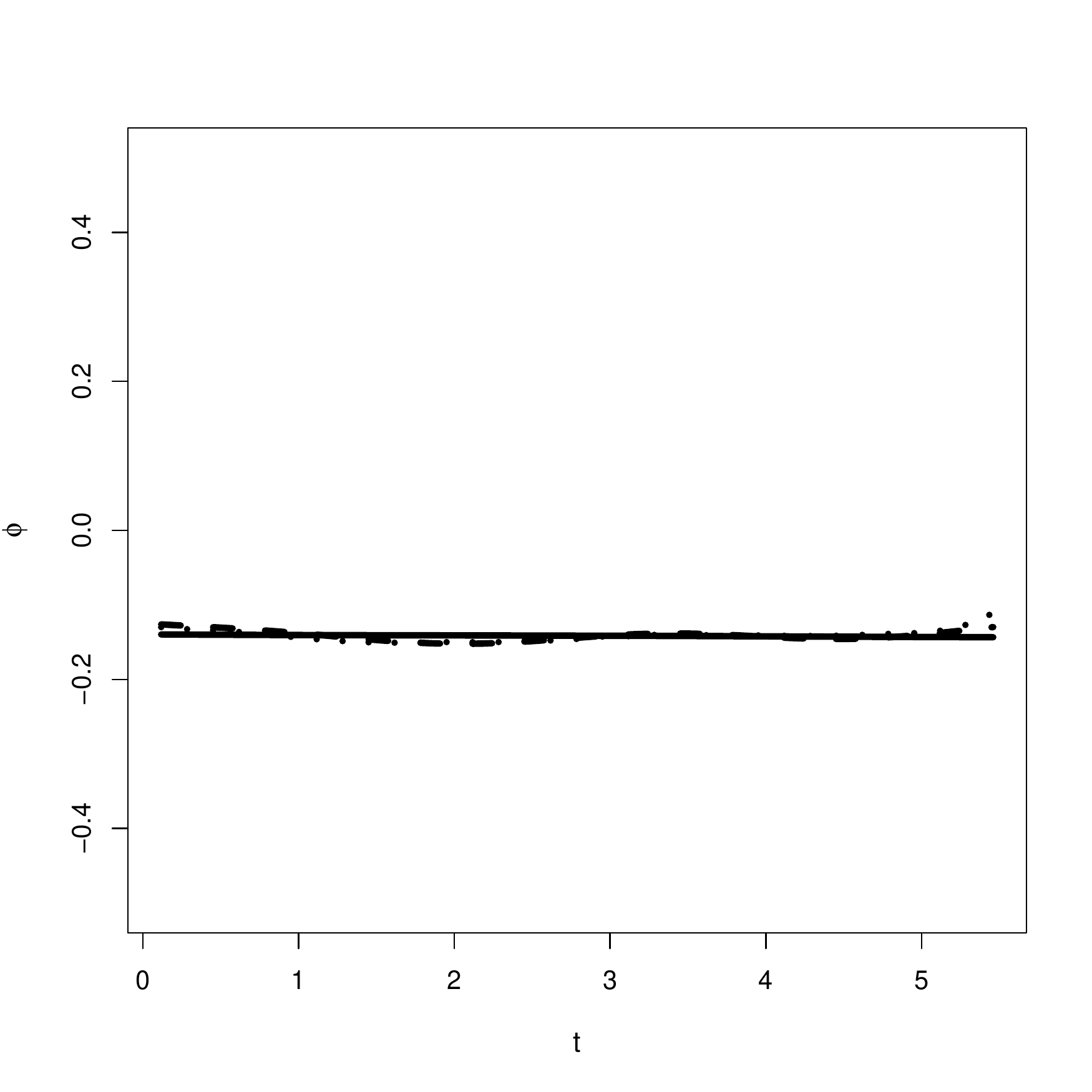}} &
 {\includegraphics[scale=0.35]{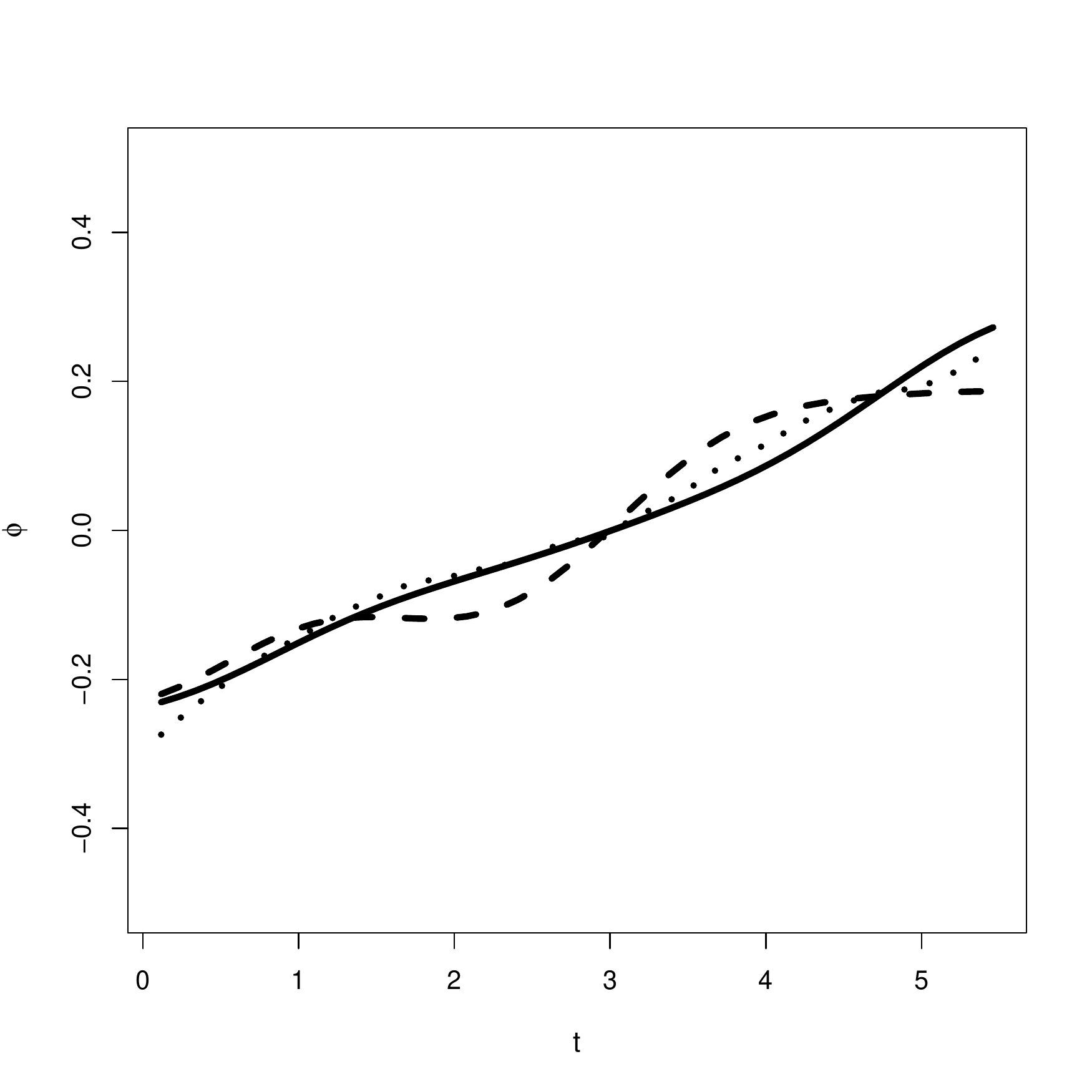}}
\end{tabular}
\caption{\label{fig:eigenfun-clean} Estimated eigenfunctions
after removing trajectories that were flagged as possible outliers.
The robust estimates (ROB) are displayed 
with solid lines, the non-robust version (LS) uses dashed lines, while PACE is
shown with dotted curves.}
\end{center}
\end{figure}

Although the estimated first principal directions 
are similar for the three methods, the PACE estimate for the
second eigenfunction is notably different. To explore the 
possibility that this difference is due to the
effect of a few potentially atypical observations, we 
identify outliers using the scores 
$(\hat{\xi}_{i1}, \hat{\xi}_{i2})$, $i=1, \ldots, 292$, 
of the curves  based  on the first 2  robustly estimated  eigenfunctions.
To decide which score 
vectors may be outliers, we compute their robust Mahalanobis distance $D_i$ 
using an $MM-$location and scatter estimator. We flag 
as outliers those vectors $(\hat{\xi}_{i1}, \hat{\xi}_{i2})$ with $D_i^2$  larger than the 99.5\% quantile of a $\chi^2_2$ distribution. 
 This resulted in 18 trajectories being identified as
potentially atypical. The five most outlying ones are shown in the right panel of
Figure \ref{fig:cd4data}. Note that these curves appear to either decrease too rapidly (with 
respect to the rest),
or to remain at high values over time. The other outlying curves also show one of these 
these two main patterns. More details about this analysis,
including documented \texttt{R} code reproducing these
analyses is publicly available on-line at \url{https://github.com/msalibian/sparseFPCA}.
We now re-compute the 
non-robust estimators (LS and PACE) 
after removing the outliers. 
It is interesting to note that on this ``clean'' data set, 
the PACE and ROB estimators are qualitatively similar
(see Figures \ref{fig:covs-clean} and \ref{fig:eigenfun-clean}).

We further compare these fits in terms of
their prediction ability by randomly splitting the data 
into a training set and a test set (with 80\% and 20\% of
the curves, respectively). We estimate the mean 
and covariance functions using the training set, and use
them to obtain predicted curves for the 20\% held-out
curves in the test set.
Figure \ref{fig:predictions} below displays four 
curves in the test set (shown in gray), with 
the 3 predicted trajectories. 
\begin{figure}[ht!]
\centering
\subfigure[Predictions]{\includegraphics[scale=0.25]{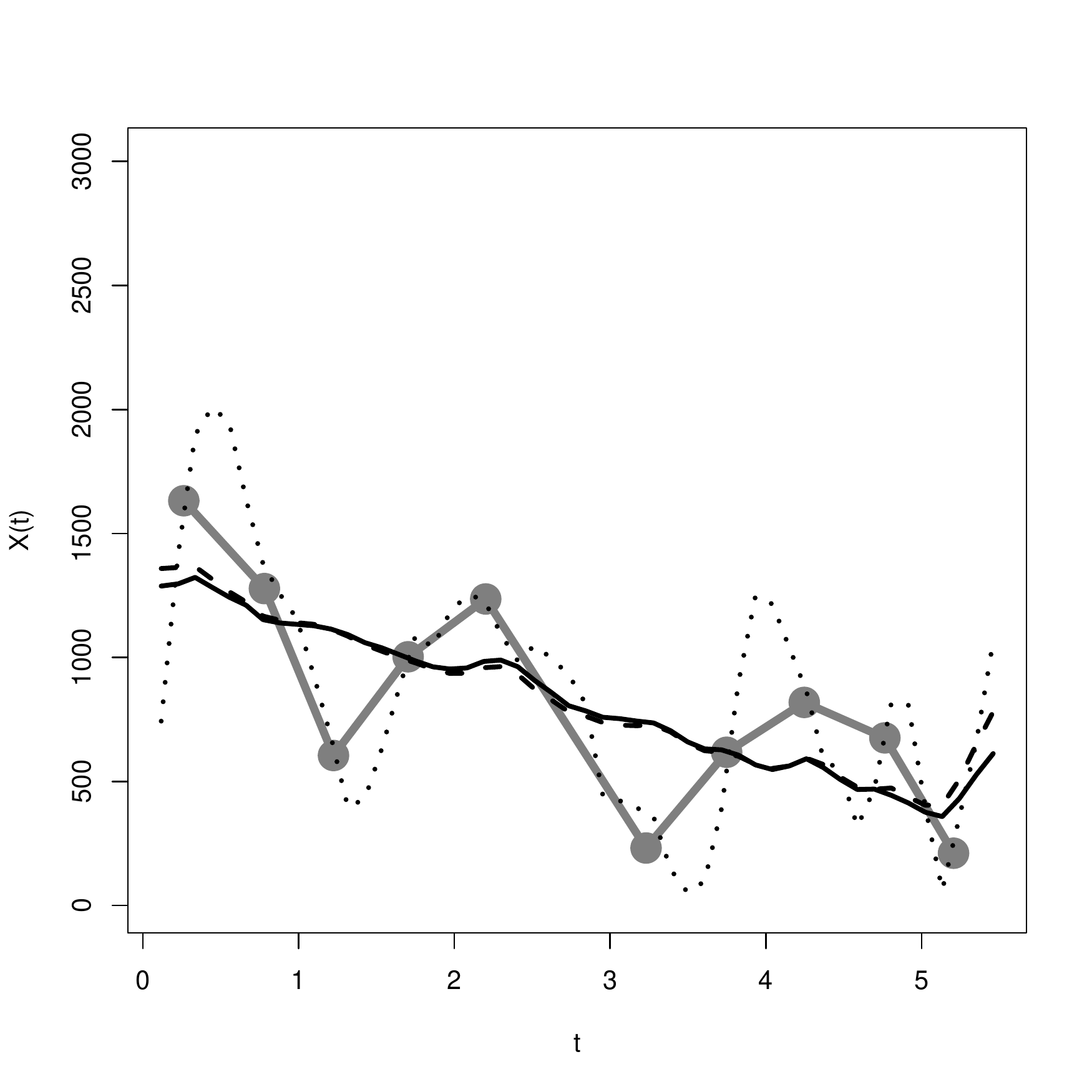}} 
\subfigure[Predictions]{\includegraphics[scale=0.25]{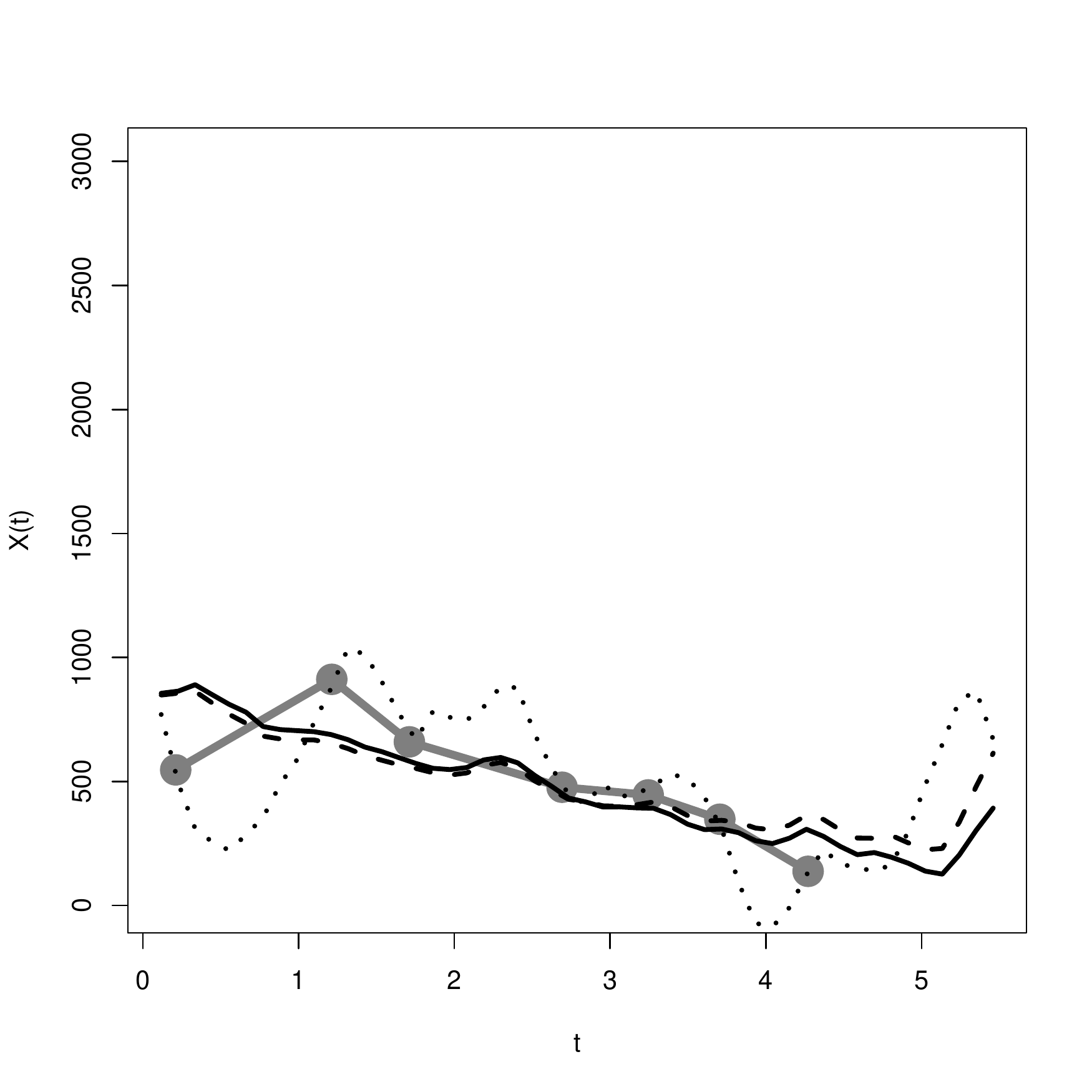}} 
\subfigure[Predictions]{\includegraphics[scale=0.25]{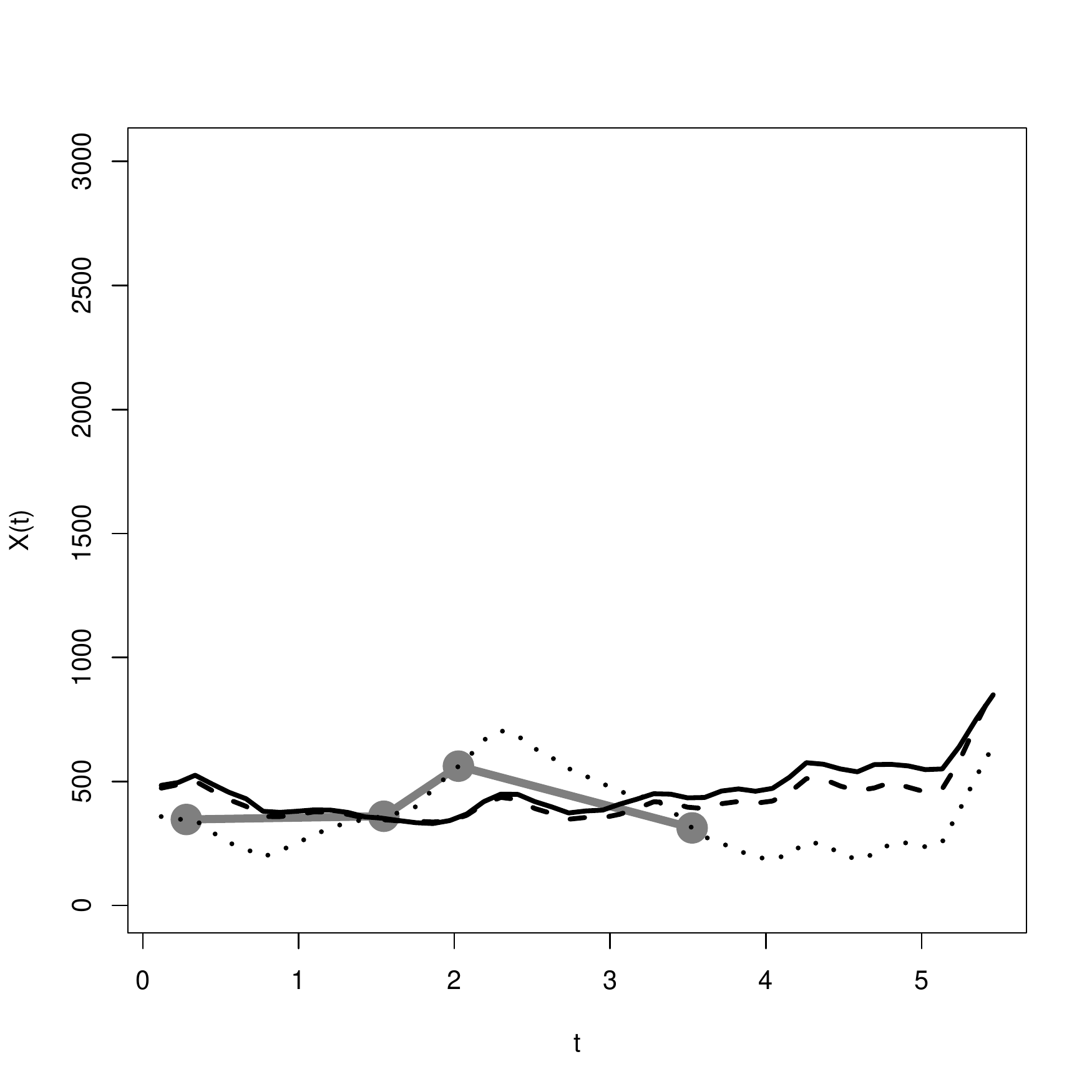}} 
\subfigure[Predictions]{\includegraphics[scale=0.25]{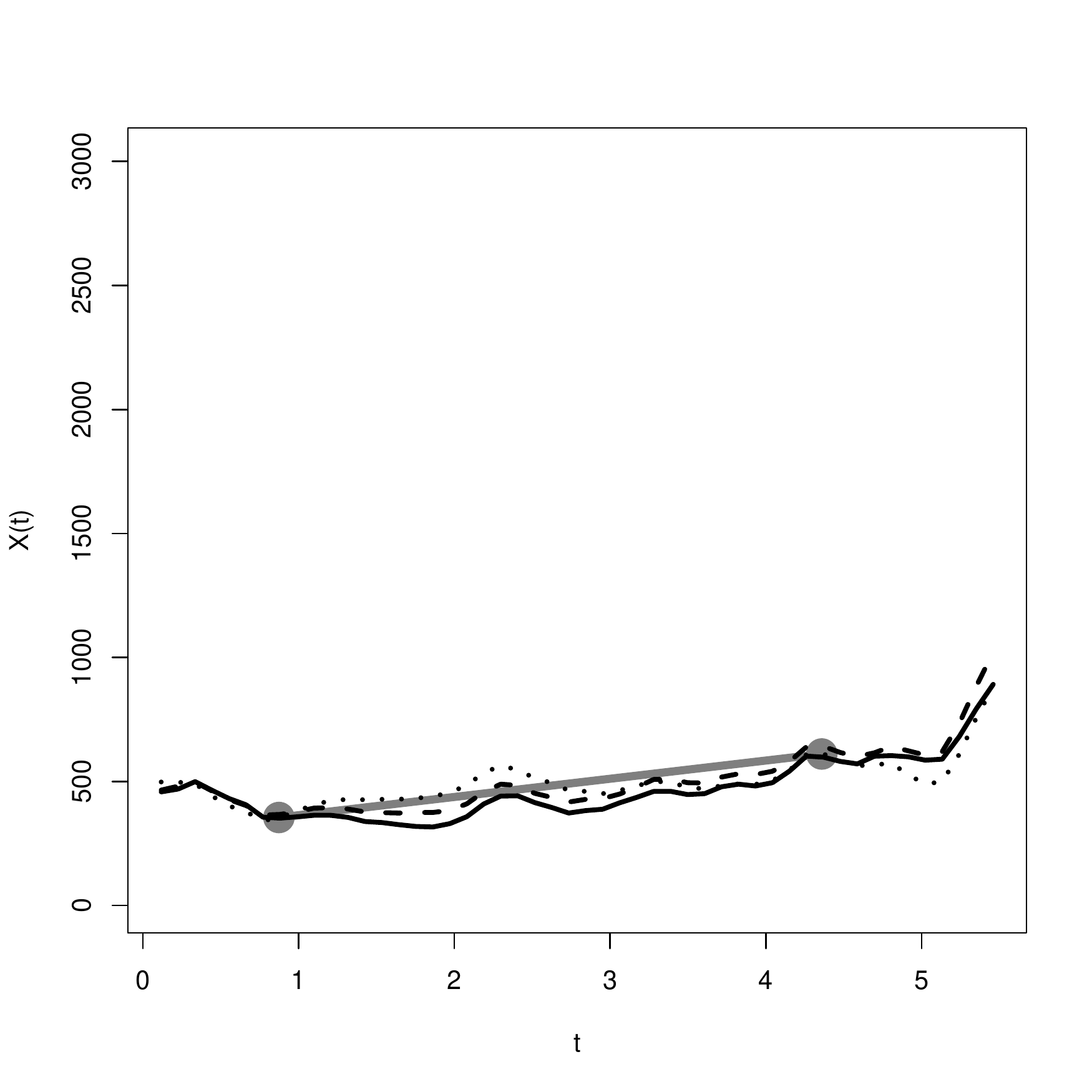}} 
\caption{Predicted trajectories for 
four curves in the test set. 
The predictions based on the ROB estimator are shown 
with solid lines, the non-robust ones based on LS use dashed lines, while 
those based on PACE are
shown with dotted lines.} \label{fig:predictions} 
\end{figure}

\section{Conclusion}{\label{sec:conclusion}}

We propose a novel robust functional principal 
components analysis (FPCA) method that is appropriate
for applications in which only a few observations 
per unit or trajectory are available.
Such longitudinal data sets
with few points per curve (possibly recorded at irregular 
intervals) are relatively 
common in applications, and it is often natural to assume
an underlying functional structure 
(of smooth trajectories, for example). 
Although our motivation was the development of 
a robust FPCA approach, the proposed method 
can easily be extended to obtain an alternative to 
PACE (Yao \textsl{et al.} \cite{YMW}) for cases where 
no atypical observations are present in the data. 
Our method only assumes that the underlying 
random process has an elliptical distribution, and thus
it is applicable even with heavy-tailed data (for example, 
without finite moments). Our simulation studies confirm
that the robust version of this approach remains
informative when the data contains outliers (which need not
be extreme values), and behave similarly to the non-robust
alternatives when the data are clean. Furthermore, in both our
numerical experiments and the example the non-robust
variant of our proposal compares favourably to the existing
methods in the literature. Our methodology could, in principle, also be
used to derive robust estimators for functional principal components
regression models (see Febrero-Bande \textsl{et al.} \cite{fpcaregression}
for a recent review) when only a few observations per curve
are available. We will explore this idea in future work.

\section*{Acknowledgements} 

The authors wish to thank two anonymous referees for valuable comments which led to an improved version of
the original paper. They also thank Prof.\! Jane-Ling Wang for 
helpful discussions during the early stages of our work. 
This research was partially supported by Grants    20020170100022BA from the Universidad de Buenos Aires and \textsc{pict} 2018-00740 from \textsc{anpcyt} at Buenos Aires, Argentina and also by the Spanish Project {MTM2016-76969P} from the Ministry of Economy, Industry and Competitiveness  (MINECO/AEI/FEDER, UE) (Graciela Boente) and by Discovery Grant RGPIN-2016-04288 of the Natural Sciences and Engineering Research Council of Canada (M. Salibi\'an Barrera). 

\section{Appendix - Proof of Proposition \ref{lemma:proposal}} \label{sec:appendix}

Although the proof of Proposition \ref{lemma:proposal}  
is immediate when $\Gamma$ has finite rank, 
we include it here for completeness. The 
infinite-dimensional case is proved using 
the representation $ X \sim  \mu + S \,  V $ 
for elliptical random elements (Boente \textsl{et al.} \cite{BST}), 
where $S \ge 0$ is a random variable independent 
from the zero-mean Gaussian  random element $V$
% with covariance operator $\Gamma$, see 
%Proposition 2.1 in Boente \textsl{et al.} \cite{BST}. 

%\vskip0.1in

\paragraph{Proof of Proposition \ref{lemma:proposal}.}  
First note that it is sufficient to prove the result when $\mu=0$. 
Specifically, let $\wtX = X - \mu \sim \, \itE\left(0, \Gamma, \varphi \right)$, 
$\bX_m = (X(t_1),\dots, X(t_m))\trasp$ and 
$\widetilde{\bX}_m = (\wtX(t_1), \ldots, \wtX(t_m))\trasp = 
\bX_m - (\mu(t_1), \ldots, \mu(t_m))\trasp$. 
It will then follow that $\widetilde{\bX}_m$ 
has a multivariate elliptical distribution with location vector ${\mathbf 0}$, 
and hence that $\bX_m$ has a multivariate elliptical distribution with location 
$(\mu(t_1), \ldots, \mu(t_m))\trasp$ and the same scatter matrix as $\widetilde{\bX}_m $.  

In what follows assume that $\mu=0$.  Consider first 
the case where there are only finitely many
non-zero eigenvalues, $\lambda_1 \ge \lambda_2 \ge \cdots \ge \lambda_q$, 
$\lambda_\ell = 0$ for $\ell > p$, and let
$\phi_1$, \ldots, $\phi_q$ be the corresponding eigenfunctions.  Define the operator 
$A : L^2({\itI}) \to \real^q$ as $A f = \left( \langle f, \phi_1 \rangle,   
\ldots, \langle f, \phi_q \rangle \right)\trasp$. This operator is linear, and also bounded
since by the Cauchy-Schwartz inequality we have
$$
\| A f \|_{\real^p}^2 = \sum_{i=1}^q{\left( \int{f(t)\phi_i(t)dt}\right)^2} \leq 
\sum_{i=1}^q \left(\|f \|_{L^2}\|\phi_i\|_{L^2}\right)^2 =\, q \, \|f \|_{L^2}^2 \, .
$$
Using that $A$ is a linear and bounded operator and the definition of elliptical elements in $L^2({\itI})$, 
we have that $AX$ is an elliptical random vector. Furthermore, using \eqref{eq:repr}, we have
$$
AX  = (\langle X, \phi_1\rangle,\langle X,\phi_2 \rangle,\cdots,\langle X,\phi_q\rangle)\trasp  
 =  (\xi_1, \xi_2, \cdots, \xi_q)\trasp \; .
$$
Thus, $\bxi=(\xi_1, \xi_2, \cdots, \xi_q)\trasp$ is an elliptical random vector $\itE_q( \bcero_q, A \Gamma A^*, \varphi)$. 
It is easy to see that $A \Gamma A^*= \mbox{diag}(\lambda_1, \dots, \lambda_q)$, 
since $A^* \bu=\sum_{j=1}^q u_j \phi_j$, for any $\bu\in \real^q$.   
Hence $\bxi \sim \itE_q( \bcero_q, \Lambda_q, \varphi)$, 
where $\Lambda_q =  \mbox{diag}(\lambda_1, \dots, \lambda_q)$. 
Moreover, noticing that for any $s\ge 1$ and $\ell_1\le \ell_2\le \dots \le \ell_s$, 
such that   $\ell_1>q$, $\left(\langle X, \phi_{\ell_1}\rangle, \dots, \langle X, \phi_{\ell_s}\rangle\right) $ 
has also an elliptical distribution with location zero and null  scatter  matrix, that is, 
it equals $\bcero_s$ with probability one, we get that $X\sim \sum_{j=1}^q \xi_j \phi_j$.  
Let $\bB \in \real^{m \times q}$ with $(i,j)$-th entry 
$\bB_{ij}=\phi_j(t_i)$, $1 \le i \le m$, $1 \le j \le q$. Then, 
$$
\bX_m = (X(t_1),X(t_2), \cdots, X(t_m))\trasp = \bB \, \bxi \, \sim
\itE_m( \bcero_m, \bB \Lambda_q \bB\trasp, \varphi) \, .
$$ 
%has an elliptical distribution, since $\bX_m$ is linear transformation of 
%$\bxi$, and $\bX_m \sim \itE_m( \bcero_m, \bB \Lambda_q \bB\trasp, \varphi)$. 
Finally, note that the 
$(s,\ell)$ element of $\bB \Lambda_q \bB\trasp$ equals 
$\sum_{j=1}^q \lambda_j \phi_j(t_s)\phi_j(t_\ell)=\gamma(t_s,t_\ell)$, 
which concludes the proof of (a) for finite rank processes.

If infinitely many $\lambda_i$'s are non-zero, then using 
Proposition 2.1 in Boente \textsl{et al.} \cite{BST} we have 
$X \sim   S\;V$ where $S\geq 0$ is a random variable independent of the Gaussian element $V$. 
Hence 
$$
\bX_m \, = \, (X(t_1),X(t_2), \cdots, X(t_m))\trasp \, = \, S \; \bV_{p} \, ,
$$ 
where $S \ge 0$
is independent of $\bV_{p}=(V(t_1),\dots, V(t_m))\trasp$. Note that   the covariance operator of $V$ is a scalar multiple of $\Gamma$, without loss of generality, we may assume that $\Gamma_V=  \Gamma$. The fact that $V$ is a Gaussian process with continuous covariance kernel $ \gamma(s,t)$ entails that $\bV_{p}$ is an $m-$variate normally distributed vector.  Effectively, Theorem 1.5 in Bosq \cite{B} implies that
$$\lim_{p\to \infty}\sup_{t\in \itI}\esp \left(V(t)-\;\sum_{\ell=1}^p \,\eta_{\ell}\,\phi_{\ell}(t)\right)^2=0\,,$$
where $\eta_{\ell}=\langle V, \phi_{\ell}\rangle\sim N(0, \lambda_\ell)$ are independent. Let 
$V_p(t)= \sum_{\ell=1}^p \,\eta_{\ell}\,\phi_{\ell}(t)$, then $\bV_{p,m}=(V_p(t_1), \dots, V_p(t_m))\trasp \convdist \bV_{m}=(V(t_1), \dots, V(t_m))\trasp$, as $p\to \infty$. Note that $\bV_{p,m}$ has a multivariate normal distribution $N_m(\bcero_m, \bSi_{p,m})$ with $(s,\ell)$ element of $\bSi_{p,m}$ equal to $\sum_{j=1}^p \lambda_j\; \phi_{j}(t_s) \; \phi_{j}(t_\ell)$. Then, taking into account that $\sum_{\ell \ge 1 }\lambda_{\ell} \phi_{\ell}^2 (t_i)=\gamma(t_i,t_i)<\infty$, we obtain easily that $\bV_{m} \sim N(0, \bSi_{\bV_{m}}) $, where the $(s,\ell)$ element of $\bSi_{\bV_{m}}$ equals $\gamma(t_s,t_\ell)$. 
Hence, $\bX_m$ has an elliptical distribution. Taking into account that  $\Gamma_V= \Gamma$,  we obtain 
that $\bX_m \sim \itE_m(\bcero_m, \bSi , \varphi)$, where the $(s,\ell)-$th element of $\bSi$ equals $ \gamma(t_s,t_\ell)$, concluding the proof of (a).

Part (b) follows immediately from (a).

To prove (c), note again that it is enough to obtain the result when $\mu=0$. 
Let us first consider the finite-rank case, when $\lambda_\ell = 0$ for $\ell > q$. 
From (a), we have 
$\bxi=(\xi_1, \xi_2, \cdots, \xi_q)\trasp\sim \itE_q( \bcero_q, \mbox{diag}(\lambda_1, \dots, \lambda_q), \varphi)$
and $\bX_m = (X(t_1),  \cdots, X(t_m))\trasp = \bB \, \bxi$. Hence,
$$\bW=\left(\xi_k, X(t_1),X(t_2), \cdots, X(t_m)\right)\trasp = \left(\begin{array}{c}
\be_k\trasp\\
\bB
\end{array}\right) \bxi=\bM \bxi $$
with $\be_k$ the $k-$th canonical vector in $\real^q$. Thus, $\bW\sim \itE_{q+1}( \bcero_{q+1}, \bSi_{\bW}, \varphi)$, where $\bSi_{\bW}=\bM \; \mbox{diag}(\lambda_1, \dots, \lambda_q) \;\bM\trasp$ is given by
\begin{equation}
\bSi_{\bW}= \left(\begin{array}{ccc}
\lambda_k &  &\lambda_k \bphi_k\trasp \\
\lambda_k \; \bphi_k & & \bSi_{\bX_m}
\end{array}\right)
\label{eq:SIGMAu}
\end{equation}
with $\bSi_{\bX_m}$   and $\bphi_k$ as in the statement of the Proposition. The conclusion follows now immediately from properties of the conditional distribution of elliptical distributions, see, for instance, Corollary 8  in Frahm \cite{F}, where an expression for the characteristic generator is also given.  

When $\Gamma$ does not have finite rank, as in the proof of (a), from 
Proposition 2.1 in Boente \textsl{et al.} \cite{BST} we conclude that 
$X \sim   S\;V$,  where $S\geq 0$ is a random variable independent of the Gaussian element $V$ and the covariance operator of $V$ equals $\Gamma$. Then, $$\bW=\left(\xi_k, X(t_1),X(t_2), \cdots, X(t_m)\right) \trasp \, =  \ S \, \left(\eta_k, V(t_1),V(t_2), \cdots, V(t_m)\right)\trasp$$
with $\eta_{k}=\langle V, \phi_{k}\rangle\sim N(0, \lambda_k)$. Note that the fact that $\eta_k\sim N(0, \lambda_k)$ entails that $\xi_k=S\, \eta_k$ is finite almost surely. Moreover, 
with a similar argument as in the proof of part (a) and using 
the continuity of $\gamma$ it is easy to see that 
the random vector $\bV=\left(\eta_k, V(t_1),V(t_2), \cdots, V(t_m)\right)$ 
is normally distributed with zero mean and covariance matrix given in \eqref{eq:SIGMAu}. 
Hence $\bW \sim \itE_{p+1}( \bcero_{p+1}, \bSi_{\bW}, \varphi)$ 
and the result follows again from the properties of the multivariate elliptical distributions. $\qed$

\end{document}